\documentclass[aps,twocolumn,superscriptaddress,groupedaddress,nofootinbib]{emulateapj}  

\usepackage{amssymb,amsmath,amsfonts,amsbsy}
\usepackage{color}
\usepackage{enumerate}
\usepackage{graphicx}
\usepackage{pifont}
\RequirePackage[colorlinks,citecolor=blue,urlcolor=blue]{hyperref}

\usepackage{kotex}

\usepackage{hyperref}

\newcommand{\orcid}[1]{\href{https://orcid.org/#1}}

\newcommand{\parallelsum}{\mathbin{\|}}

\begin{document}

\title{Cosmological Parameter Estimation from the Two-Dimensional Genus Topology -- Measuring the Shape of the Matter Power Spectrum}

\author{Stephen Appleby}\email{stephen@kias.re.kr}
\affiliation{Quantum Universe Center, Korea Institute for Advanced Study, 85 Hoegiro, Dongdaemun-gu, Seoul 02455, Korea}
\author{Changbom Park}
\affiliation{School of Physics, Korea Institute for Advanced Study, 85
Hoegiro, Dongdaemun-gu, Seoul, 02455, Korea}
\author{Sungwook E. Hong  (홍성욱)}
\affiliation{Natural Science Research Institute, University of Seoul, 163
Seoulsiripdaero, Dongdaemun-gu, Seoul, 02504, Korea}
\author{Ho Seong Hwang}
\affiliation{School of Physics, Korea Institute for Advanced Study, 85
Hoegiro, Dongdaemun-gu, Seoul, 02455, Korea}
\affiliation{Korea Astronomy and Space Science Institute, 776 Daedeokdae-ro, Yuseong-gu, Daejeon 34055, Korea}
\author{Juhan Kim}
\affiliation{Center for Advanced Computation, Korea Institute for Advanced
Study, 85 Hoegiro, Dongdaemun-gu, Seoul, 02455, Korea}

\begin{abstract}
We present measurements of the two-dimensional genus of the SDSS-III BOSS catalogs to constrain cosmological parameters governing the shape of the matter power spectrum. The BOSS data are divided into twelve concentric shells over the redshift range $0.2 < z < 0.6$, and we extract the genus from the projected two-dimensional galaxy density fields. We compare the genus amplitudes to their Gaussian expectation values, exploiting the fact that this quantity is relatively insensitive to non-linear gravitational collapse. The genus amplitude provides a measure of the shape of the linear matter power spectrum, and is principally sensitive to $\Omega_{\rm c}h^{2}$ and scalar spectral index $n_{\rm s}$. A strong negative degeneracy between $\Omega_{\rm c}h^{2}$ and $n_{\rm s}$ is observed, as both can increase small scale power by shifting the peak and tilting the power spectrum respectively. We place a constraint on the particular combination $n_{\rm s}^{3/2} \Omega_{\rm c}h^{2}$ -- we find $n_{\rm s}^{3/2} \Omega_{\rm c}h^{2} = 0.1121 \pm  0.0043$ after combining the LOWZ and CMASS data sets, assuming a flat $\Lambda$CDM cosmology. This result is practically insensitive to reasonable variations of the power spectrum amplitude and linear galaxy bias. Our results are consistent with the Planck best fit $n_{\rm s}^{3/2}\Omega_{\rm c}h^{2} = 0.1139 \pm 0.0009$. 
\end{abstract}

\maketitle

\section{Introduction}

The Minkowski Functionals (MFs) are a class of statistics that describe the morphology and topology of excursion sets of a field. They have a long history of application within cosmology \citep{Gott:1989yj,1991ApJ...378..457P,Mecke:1994ax,Schmalzing:1997aj,Schmalzing:1997uc,1989ApJ...345..618M,1992ApJ...387....1P,2001ApJ...553...33P}. Early adopters measured the MFs of the CMB \citep{Gott:1989yj,1991ApJ...378..457P,Schmalzing:1997uc,Hikage:2006fe} and nascent large scale structure catalogs \citep{1989ApJ...345..618M,1992ApJ...387....1P,1992ApJ...385...26G}. More recent applications have involved the measurement of the MFs from modern cosmological data sets; see for example  \citet{Hikage:2002ki,Hikage:2003fc,Park:2005fk,10.1111/j.1365-2966.2008.14358.x,Gott:2008kk,Choi:2010sx,Zhang:2010tha,Blake:2013noa,Wiegand:2013xfa,Wang:2015eua,Buchert:2017uup,Hikage_2001}. In this work we study one particular Minkowski Functional; the genus. This statistic is a topological quantity, that has a simple and intuitive geometric interpretation and is relatively insensitive to non-linear physics. This makes it a valuable probe for cosmology.

Our focus is on the genus of the matter density field at redshifts $z < 0.6$, as traced by galaxies. By directly comparing the measured genus amplitude to its Gaussian expectation value, one can measure the cosmological parameters that dictate the shape of the matter power spectrum \citep{10.1143/PTP.76.952,1970Ap......6..320D,Adler,Gott:1986uz,Hamilton:1986,1989ApJ...345..618M}. By comparing the genus amplitude at high and low redshift, one can also infer the equation of state of dark energy $w_{\rm de}$ by exploiting the fact that this quantity is a standard ruler \citep{Park:2009ja}. Further information regarding the $N$-point functions ($N>2$) can be extracted from the shape of the genus curve \citep{Matsubara:1994wn,Pogosyan:2009rg,Gay:2011wz,Codis:2013exa}. 

Extracting information from large scale structure using the genus, and eliminating systematic effects such as shot noise, non-linear gravitational evolution and redshift space distortion, has been the subject of a series of recent works by the authors. In \citet{Appleby:2017ahh}, we used mock galaxy catalogs to study various systematic effects that could bias our reconstruction of cosmological parameters. In \citet{Appleby:2018jew} we applied these lessons to two-dimensional shells of mock galaxy lightcone data, to test the constraining power of the statistic. In this work we further refine our analysis, extract the two-dimensional genus from a  galaxy catalog and use this information for cosmological parameter estimation.  

In this work, we measure the genus of two-dimensional shells of the SDSS-III BOSS LOWZ and CMASS galaxy catalogs \citep{2015ApJS..219...12A}. After extracting the genus curves from the data, we compare the measurements to their theoretical expectation value. We use the fact that on quasi-linear scales, the genus amplitude is only weakly sensitive to non-linear gravitational collapse \citep{1988ApJ...328...50M,Matsubara:1995ns,Park:2009ja}. This allows us to use simple Gaussian statistics at relatively small scales, as we quantify in what follows. In a companion paper \citet{Appleby_inprep} we extract the genus amplitudes from a combination of BOSS data and the SDSS main galaxy sample, and use this quantity as a standard ruler to place constraints on the expansion history of the Universe. 

The paper will proceed as follows. In section \ref{sec:theory} we review the theory underpinning the genus of random fields. The data, mask, mock catalogs and construction of covariance matrices are described in section \ref{sec:obs}. Finally in section \ref{sec:constraints} we place constraints on cosmological parameters, then close with a discussion in section \ref{sec:discuss}.

\section{Genus -- Theory}
\label{sec:theory}

The theory underlying the Minkowski Functionals of random fields was derived in \citet{1970Ap......6..320D,Adler,10.1143/PTP.76.952,Hamilton:1986,Ryden:1988rk,
1987ApJ...319....1G,1987ApJ...321....2W}
 for Gaussian fields, and expanded in \citet{Matsubara:1994wn,2003ApJ...584....1M,Matsubara:1995dv,1988ApJ...328...50M,Matsubara:1995ns,10.1111/j.1365-2966.2008.12944.x,Pogosyan:2009rg,Gay:2011wz,Codis:2013exa} for non-Gaussian generalisations. In this paper we will be concerned with the three dimensional matter density field as traced by galaxies -- $\delta_{\rm 3D}$, and specifically two-dimensional slices\footnote{More precisely we generate shells centered on the observer. The BOSS data is sufficiently distant that we can use the distant observer approximation to predict the amplitude of the genus curve.} of this field in planes perpendicular to the line of sight --  $\delta_{\rm 2D}$. The two-dimensional power spectrum $P_{\rm 2D}$ of $\delta_{\rm 2D}$ is related to its three-dimensional counterpart $P_{\rm 3D}(k)$ as  

\begin{equation}\label{eq:p2d} P_{\rm 2D}(k_{\perp},z) = {2 \over \pi} \int dk_{\parallelsum} P_{\rm 3D}\left(k,z\right) {\sin^{2} \left( k_{\parallelsum} \Delta\right) \over k_{\parallelsum}^{2} \Delta^{2}} ,  \end{equation} 

\noindent where $k=\sqrt{\vec{k}^{2}_{\perp}+k^{2}_{\parallelsum}}$, and we have performed real space top hat smoothing along $x_{\parallelsum}$, where $\Delta$ is the thickness of the slice. Here, $\vec{k}_{\perp}$ and $k_{\parallelsum}$ are the Fourier modes perpendicular and parallel to $x_{\parallelsum}$.

For the low-redshift matter density that we will probe via galaxy catalogs, the underlying three-dimensional power spectrum is given by 

\begin{equation}\label{eq:p3df} P_{\rm 3D} (k,k_{\parallelsum},z) = b^{2} \left( 1 + \beta {k_{\parallelsum}^{2} \over k^{2}}\right)^{2} P_{\rm m}(z, k) + P_{\rm SN} , \end{equation}

\noindent where $P_{\rm m}(z,k)$ is the matter power spectrum at redshift $z$, $P_{\rm SN}$ is the shot noise power spectrum that we estimate as $P_{\rm SN} =1/\bar{n}$, where $\bar{n}$ is the number density of the galaxy catalog. $\beta = \Omega_{\rm m}^{\gamma}/b$ is the redshift space distortion parameter, $b$ is the linear galaxy bias and $\gamma \simeq 3(1-w_{\rm de})/(5-6w_{\rm de})$. The expression ($\ref{eq:p3df}$) accounts for linear redshift space distortion and shot noise. 

For two-dimensional slices of a three-dimensional Gaussian field,  the genus per unit area is given by \citep{Hamilton:1986}

\begin{eqnarray} \label{eq:gg2d} & &  g_{\rm 2D}(\nu) = {1 \over 2(2\pi)^{3/2}} {\sigma_{1}^{2} \over \sigma_{0}^{2}} \nu e^{-\nu^{2}/2} , \\
\nonumber & & \sigma_{0}^{2} = \langle \delta_{\rm 2D}^{2} \rangle  , \qquad  \sigma_{1}^{2} = \langle |\nabla \delta_{\rm 2D} |^{2} \rangle  , \end{eqnarray} 

\noindent where $\sigma_{0,1}$ are cumulants of the two dimensional field and $\nu$ is a constant-density threshold. For a Gaussian field the shape of the genus curve is fixed as a function of threshold $g_{\rm 2D} \sim \nu e^{-\nu^{2}/2}$, and only the amplitude -- 

\begin{eqnarray} 
\label{eq:ag} & & A_{\rm G}^{(\rm 2D)} \equiv  {1 \over 2(2\pi)^{3/2}} {\sigma_{1}^{2} \over \sigma_{0}^{2}}  \end{eqnarray} 

\noindent contains information. We can relate the cumulants to the power spectrum as

\begin{eqnarray}
\label{eq:s02} & & \sigma_{0}^{2} = {1 \over (2\pi)^{2}}\int d^{2} k_{\perp} e^{-k_{\perp}^{2}R_{\rm G}^{2}} P_{\rm 2D}(k_{\perp},z)   , \\
\label{eq:s12} & & \sigma_{1}^{2} = {1 \over (2\pi)^{2}}\int d^{2} k_{\perp} k_{\perp}^{2} e^{-k_{\perp}^{2}R_{\rm G}^{2}} P_{\rm 2D}(k_{\perp},z), 
  \end{eqnarray} 

\noindent where we have smoothed the two-dimensional field using a Gaussian kernel of width $R_{\rm G}$. 

The genus amplitude ($\ref{eq:ag}$) is proportional to the ratio of $\sigma_{1}, \sigma_{0}$ cumulants, and as such will be insensitive to the total amplitude of the power spectrum. This is a generic property of the Minkowski functionals.

For a weakly non-Gaussian field, one can perform an expansion of the genus in $\sigma_{0}$ as follows \citep{2003ApJ...584....1M} 

\begin{eqnarray} \nonumber & &   g_{\rm 2D}(\nu_{\rm A}) = A_{\rm G}^{(\rm 2D)} e^{-\nu_{\rm A}^{2}/2} \left[ H_{1}(\nu_{\rm A})+ \left[ {2 \over 3} \left( S^{(1)} - S^{(0)}\right) \times \right. \right. \\
\label{eq:mat1} & & \quad \left. \left. H_{2}(\nu_{\rm A}) +  
  {1 \over 3} \left(S^{(2)} - S^{(0)}\right)H_{0}(\nu_{\rm A}) \right] \sigma_{0} + {\cal O}(\sigma_{0}^{2}) \right] , \end{eqnarray} 

\noindent where $A_{\rm G}^{(\rm 2D)}$ is the Gaussian amplitude ($\ref{eq:ag}$) and the skewness parameters $S^{(0)}, S^{(1)}, S^{(2)}$ are related to the three point cumulants

\begin{eqnarray} \label{eq:sk1} & &  S^{(0)} = {\langle \delta^{3}_{\rm 2D} \rangle \over \sigma_{0}^{4}} , \\
 & & S^{(1)} = - {3 \over 4} {\langle \delta^{2}_{\rm 2D} (\nabla^{2} \delta_{\rm 2D}) \rangle \over \sigma_{0}^{2} \sigma_{1}^{2}} , \\ 
 & & S^{(2)} = -3  {\langle (\nabla \delta_{\rm 2D} \cdot  \nabla \delta_{\rm 2D})  (\nabla^{2} \delta_{\rm 2D}) \rangle \over \sigma_{1}^{4}} . \end{eqnarray}  

\noindent $H_{i}(x)$ are Hermite polynomials, the first few of which are given by $H_{0}(x) = 1$, $H_{1}(x)= x$, $H_{2}(x)= 1 - x^{2}$, $H_{3}(x) = x^{3}-3x$. We have defined $\nu_{A}$ as the density threshold such that the excursion set has the same area fraction as a corresponding Gaussian field - 

\begin{equation}\label{eq:afrac} f_{A} = {1 \over \sqrt{2\pi}} \int^{\infty}_{\nu_{A}} e^{-t^{2}/2} dt , \end{equation}

\noindent where $f_{A}$ is the fractional area of the field above $\nu_{A}$. This choice of $\nu_{\rm A}$ parameterization eliminates the non-Gaussianity in the one-point function \citep{1987ApJ...319....1G,1987ApJ...321....2W,1988ApJ...328...50M}. The expansion ($\ref{eq:mat1}$) has been continued to arbitrary order in \citet{Pogosyan:2009rg}. 

The amplitude of the genus, which is the coefficient of $H_{1}(\nu_{\rm A})$ in ($\ref{eq:mat1}$), is not modified by the non-Gaussian effect of gravitational collapse to leading order in the $\sigma_{0}$ expansion. We can therefore directly compare the measured genus amplitude to the expectation value ($\ref{eq:ag}$), after smoothing the field over suitably large scales. The quantity $A_{\rm G}^{(\rm 2D)}$ is defined as the ratio of $\sigma_{1}, \sigma_{0}$ cumulants and so will be a measure of the shape of the underlying linear matter power spectrum. It follows that the cosmological parameters to which this statistic will be sensitive are the dark matter fraction $\Omega_{\rm c}h^{2}$, primordial power spectral index $n_{\rm s}$ and also weakly to the baryon fraction $\Omega_{\rm b}h^{2}$. Conversely, it will be practically insensitive to the amplitude of the power spectrum and any linear bias factors\footnote{See appendix A for a more detailed discussion of this point.}. 

One important caveat associated with our approach is that we are dealing with a masked field, and hence a bounded domain. The genus of a field with boundary will not be represented by the expression ($\ref{eq:mat1}$), which was derived assuming an unbounded domain. In what follows, we measure the integrated Gaussian curvature per unit area of the masked sky, and assume that this result yields an unbiased estimate of the genus of an unbounded field. We use the method outlined in \cite{Schmalzing:1997uc} and applied to large scale structure in \cite{Appleby:2018jew}, which provides such an unbiased estimator.

We will measure genus curves from shells of galaxy data, and extract cosmological information by comparing the amplitude of the curves to their Gaussian expectation value. In the following sections we elucidate the galaxy catalogs used in this analysis, and the mock catalogs used to construct the covariance matrices required for statistical inference.

\section{Observational Data} 
\label{sec:obs}

To measure the genus over the redshift range $0.2 < z < 0.6$, we use the SDSS-III Baryon Oscillation Spectroscopic Survey (BOSS) \citep{2000AJ....120.1579Y}. The $12^{\rm th}$ release of the Sloan Digital Sky Survey (SDSS) \citep{2015ApJS..219...12A} imaged $9,376 {\rm deg^{2}}$ of the sky in the $\emph{ugriz}$ bands \citep{1996AJ....111.1748F}. The survey was performed with the 2.5m Sloan telescope \citep{2006AJ....131.2332G} at the Apache Point Observatory in New Mexico. The resulting extra-galactic catalog contains 1,372,737 unique galaxies, with redshifts measured using an automated pipeline \citep{2012AJ....144..144B}. 

The BOSS data is decomposed into two distinct catalogs. The LOWZ sample consists of galaxies at redshift $z<0.4$, and are selected using various color-magnitude cuts that are intended to match the evolution of a passively evolving stellar population. In this way, a bright and red ``low-redshift" galaxy population is selected with the intention of extending the SDSS-I and II Luminous Red Galaxies (LRGs) to higher redshift and increased number density.

\begin{table}
\begin{center}
\caption{\label{tab:ii}}
 \begin{tabular}{||c  c ||}
 \hline
 Parameter \, & \, Fiducial Value \\ [0.5ex] 
 \hline\hline
 $\Omega_{\rm m}$ \,  & \, $0.307$   \\ 
 $h$ \,  & \, $0.677$   \\
 $\Delta$ \, & \, $80 {\rm Mpc}$   \\
 $R_{\rm G}$ \,  & \, $20 {\rm Mpc}$ \\ 
 \hline 
\end{tabular}\label{tab:ini}
\end{center} 
Fiducial parameters used to fix the slice thickness, and the parameters used to calculate the genus in this work. $\Delta$ is the thickness of the two dimensional slices of the density field, and $R_{\rm G}$ is the Gaussian smoothing scale used in the two-dimensional planes perpendicular to the line of sight. 
\end{table}

The CMASS ``high-redshift" $0.4 < z < 0.7$ galaxies are selected using a set of color-magnitude cuts. $(g-r)$ and $(r-i)$ cuts are specified to segregate ``high-redshift" galaxies. However, the sample is not biased towards red galaxies as some of the colour limits imposed on the SDSS-I/II sample have been removed. The colour-magnitude cut is varied to ensure massive objects are sampled as uniformly as possible with redshift. We direct the reader to \cite{2016MNRAS.455.1553R} for further details of the galaxy samples, including details of targeting algorithms.

The CMASS and LOWZ samples are provided with the galaxy weights $w_{\rm cp}, w_{\rm noz}, w_{\rm systot}$ to account for observational systematics. $w_{\rm cp}$ represents the `close pairs' weight, which accounts for the subsample of galaxies that are not assigned a spectroscopic fibre due to fibre collisions. This sample is not random, as these missed galaxies must be within a fibre collision radius ($62 {\rm arcsec}$) of another target. This systematic is corrected by upweighting the nearest galaxy by $w=(1+n)$, where $n$ is the number of neighbours without a redshift. The spectroscopic pipeline failed to obtain a redshift for $1.8\%$ ($0.5\%$) of CMASS (LOWZ) targets. In the case of such a failure, a similar upweighting scheme is adopted as for $w_{\rm cp}$ -- the nearest neighbour of any failed redshift galaxy is upweighted by $w_{\rm noz}$. Note that failed redshift galaxies could be first upweighted by a factor $w_{\rm cp}$ -- in this case $w_{\rm cp}$ is added as a weight to its nearest neighbour (for $w_{\rm noz}$). The upweighted object must be classified as a `good' galaxy. The $w_{\rm systot}$ weight applies to the CMASS sample only, and is used to remove non-cosmological fluctuations in the CMASS target density due to stellar density and seeing. The LOWZ galaxies are generally bright compared to the CMASS galaxies and do not show significant density variations because of non-cosmological fluctuations from stellar density and seeing. Therefore, the LOWZ targets do not require the $w_{\rm systot}$ weight \citep{2016MNRAS.455.1553R}. The total galaxy weight adopted in this analysis is 

\begin{equation} w_{\rm tot} = w_{\rm systot}(w_{\rm cp} + w_{\rm noz} - 1) . \end{equation}

We measure the genus of two-dimensional shells of the BOSS LOWZ and CMASS data. To construct a set of two-dimensional galaxy number density fields, we first bin the galaxies into redshift shells. The redshift shell boundaries are fixed such that each shell has constant comoving thickness $\Delta = 80 {\rm Mpc}$, assuming a flat $\Lambda$CDM input cosmology with parameters presented in table \ref{tab:ii}. As the slice thickness is chosen in terms of comoving distance, we have introduced a cosmological parameter dependence -- if we select an incorrect cosmology our bins will not have uniform thickness. However, in Appendix B we vary the cosmology used to generate the slices and verify that our results are robust to this choice, for reasonable variations of the parameters $h$ and $\Omega_{\rm m}$. The redshift shells are concentric and non-overlapping over the range $0.25 < z < 0.6$. To generate a constant number density sample in each redshift shell, we apply a lower stellar mass cut in each shell to fix $\bar{n}_{\rm cut} = 6.5 \times 10^{-5}  ({\rm Mpc})^{-3}$. The galaxy stellar mass estimates are derived using the {\it Wisconsin PCA BC03} model \citep{Tinker:2016zpi}. With this choice, the shot noise contribution to the power spectrum ($\ref{eq:p3df}$) is approximately given by $P_{\rm SN} = 1/\bar{n}_{\rm cut}$, and is constant in each redshift shell.

\begin{figure}[h!]
  \includegraphics[width=0.5\textwidth]{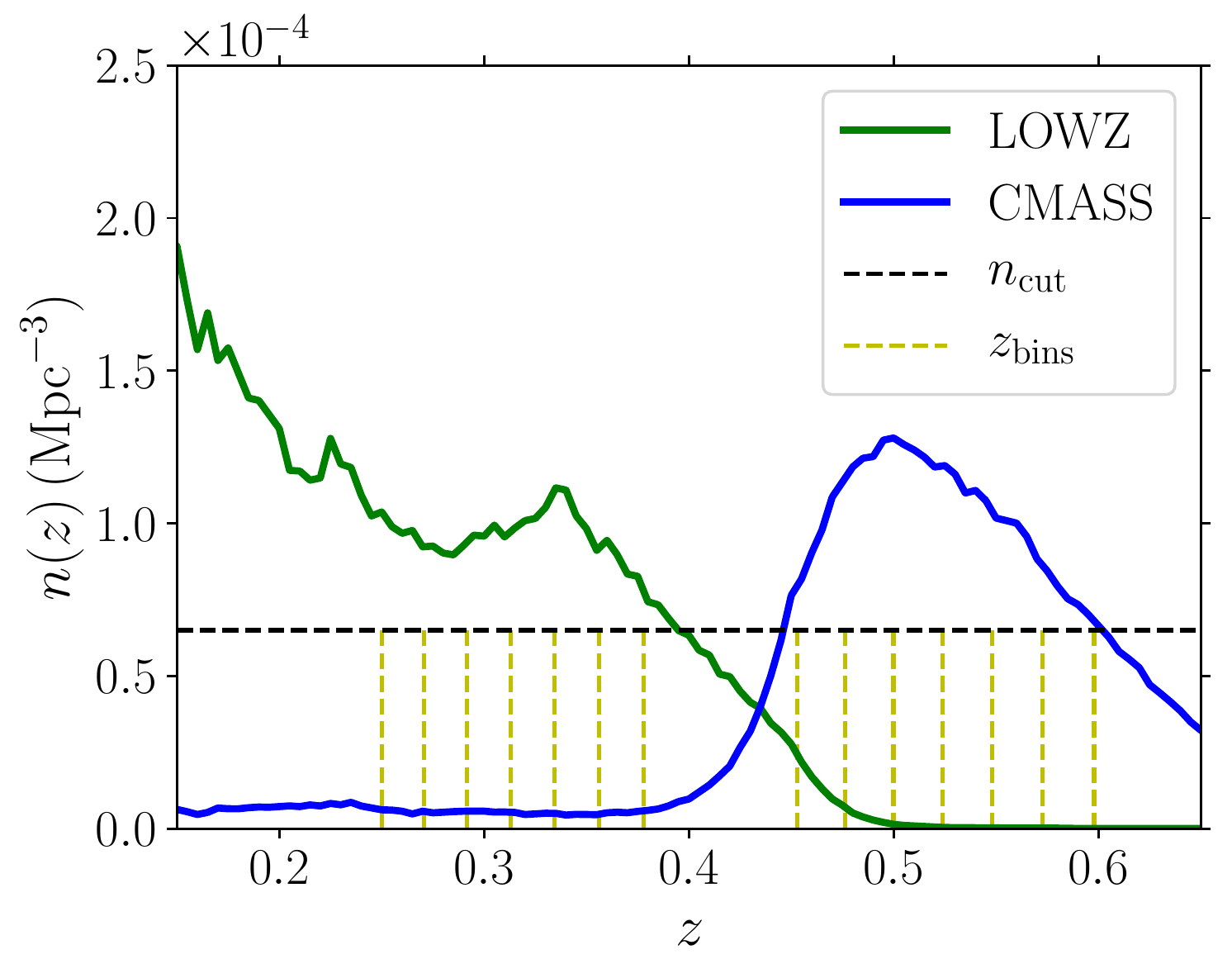}
  \caption{Number densities of the LOWZ and CMASS galaxy catalogs as a function of redshift. The black dashed line is the density cut that we apply in this work, and the yellow dashed lines represent the redshift bin limits used to generate the shells.}
  \label{fig:1}
\end{figure}

In Table \ref{tab:1} we present the redshift limits of the shells that we choose for the CMASS and LOWZ samples respectively. We select a total of $N_{\rm z} = 12$ slices ($6/6$ from the LOWZ/CMASS catalogs respectively). The full LOWZ and CMASS catalogs span the ranges $0.15 < z < 0.45$ and $0.45 < z < 0.7$ respectively. However, we only select LOWZ data for $0.25 < z < 0.38$ and CMASS $0.45 < z < 0.6$. For LOWZ data below $z<0.25$, the density field that we generate does not have sufficient volume to accurately reconstruct the genus curve which will affect our genus amplitude measurements. For LOWZ data at $z > 0.38$, and CMASS data at $z > 0.6$, the catalogs do not possess sufficient number density and the genus measurements would be significantly affected by shot noise. In Figure \ref{fig:1} we present the full redshift distribution of the LOWZ and CMASS samples. The number density of our sample $\bar{n}_{\rm cut}$ is exhibited as a dashed black horizontal line, and the redshift bin limits are presented as vertical yellow dashed lines.

\begin{table}
\begin{center}
\caption{\label{tab:1}}
 \begin{tabular}{|| c | c  ||}
 \hline
 LOWZ & CMASS  \\
 \hline
 $0.250 <  z \leq 0.271$ & $0.453 < z \leq 0.476$ \\
 $0.271 < z \leq 0.292$ & $0.476 < z \leq 0.500$ \\
 $0.292 < z \leq 0.313$ & $0.500 < z \leq 0.524$ \\
 $0.313 < z \leq 0.334$ & $0.524 < z \leq 0.548$ \\
 $0.334 < z \leq 0.356$ & $0.548 < z \leq 0.573$ \\
 $0.356 < z \leq 0.378$ & $0.573 < z \leq 0.598$  \\ 
 \hline
\end{tabular}
\end{center} 
 The redshift limits of the LOWZ and CMASS shells used in this work. \\
\end{table}

To generate two-dimensional density fields at each redshift, the galaxies are binned into a regular HEALPix\footnote{http://healpix.sourceforge.net} \citep{Gorski:2004by} lattice on the sphere according to their galactic latitude and longitude $(b,\ell)$. We use nearest pixel binning, applying the weight $w_{{\rm tot}, m}$ to the nearest pixel center to which the $m^{\rm th}$ galaxy belongs.  This generates a set of density fields $\delta_{i,j} \equiv (n_{i,j}-\bar{n}_{j})/\bar{n}_{j}$, where $1 \leq j \leq N_{\rm z}$ denotes the redshift bin (of which there are $N_{\rm z}=12$ in total) and $1 \leq i \leq N_{\rm pix}$ is the pixel identifier on the unit sphere. $\bar{n}_{j}$ is the mean number of galaxies contained within an unmasked pixel at each redshift shell, and $n_{i,j}$ is the number of galaxies contained within pixel $i$ in redshift slice $j$. We use $N_{\rm pix} = 12 \times 512^{2}$ pixels.

\subsection{\label{sec:mask}Two-Dimensional Masks}

The mask is an equal area pixel map of the same number of pixels as the galaxy maps. Each pixel is defined by a weight $\Theta_{i}$, $1 \leq i \leq N_{\rm pix}$, obtained by binning the survey angular selection function into the HEALPix basis. The resulting weights lie in the range $0 \leq \Theta_{i} \leq 1$.

In what follows, we apply a mask weight cut and only use pixels with $\Theta_{i} > \Theta_{\rm cut}$, with $\Theta_{\rm cut} =0.8$. We account for the angular selection function by directly weighting each galaxy with $w_{\rm tot}$ as described in section \ref{sec:obs}, so the mask $\Theta_{i}$ is converted to a binary map - $\Theta_{i} = 1$ if $\Theta_{i} > \Theta_{\rm cut}$ or $\Theta_{i}=0$ otherwise.

We apply the binary mask $\Theta_{i}$ to the galaxy fields $\delta_{i,j} \to \Theta_{i} \times \delta_{i,j}$, then smooth $\delta_{i,j}$ in harmonic space with a Gaussian kernel of width $\theta_{j} = R_{\rm G}/d_{\rm c}(z_{j})$, where $R_{\rm G}=20 \, {\rm Mpc}$ is a constant co-moving scale and $d_{\rm c}(z_{j})$ is the comoving distance to the center of the $j^{\rm th}$ redshift slice. We denote the smoothed density fields as $\tilde{\delta}_{i,j}$. We also smooth the mask $\Theta_{i}$ with the same angular scale $\theta_{j}$ at each redshift, generating a set of smoothed masks $\tilde{\Theta}_{i,j}$. We then apply a second cut to the density fields; $\tilde{\delta}_{i,j} = 0$ if $\tilde{\Theta}_{i,j} \leq \Theta_{\rm cut}$ and $\tilde{\delta}_{i,j} = \tilde{\delta}_{i,j}/\tilde{\Theta}_{i,j}$ if $\tilde{\Theta}_{i,j} > \Theta_{\rm cut}$. Finally, we re-apply the original, unsmoothed binary mask $\Theta_{i}$, as $\tilde{\delta}_{i,j} \to \Theta_{i}\tilde{\delta}_{i,j}$. This method removes regions of the density field in the vicinity of the mask boundary, where the true field is not accurately reproduced.

We extract the genus from each redshift shell using the method described in \citet{Appleby:2017ahh} and divide by the total co-moving area $A_{\rm c} = 4\pi f_{{\rm sky}, j} d^{2}_{\rm c}(z_{j})$, where $f_{{\rm sky}, j}$ is the fraction of sky that is unmasked in the $j^{\rm th}$ redshift slice. We measure the genus at $201$ values of the threshold $\nu_{\rm A}$, equi-spaced over the range $-2.5 < \nu_{\rm A} < 2.5$, then take the average over every four values to obtain $N_{\nu_{\rm A}}=50$ measurements. We label the measured values $g_{j}^{n}$, where $j$ runs over the redshift shells and $1 \leq n \leq N_{\nu_{\rm A}}$ over the $N_{\nu_{\rm A}}=50$, $\nu_{\rm A}$ thresholds. The two-dimensional density fields are presented in Figures \ref{fig:dens_LOWZ},\ref{fig:dens_CMASS} of Appendix B.

In Figure \ref{fig:4} we exhibit the genus curves measured from the $N_{\rm z} = 6$ shells of LOWZ [top panel] and CMASS [bottom panel] data. The curves extracted from the six LOWZ shells exhibit large scatter compared to the CMASS data, due to cosmic variance and the smaller volume available at low redshift. 

These genus curves will be used to extract cosmological information. However, before we can do so we must estimate the statistical uncertainty of the measurements. In the following section we describe the mock catalogs used to generate the relevant covariance matrices.

\subsection{Mock Catalogs}
\label{sec:mocks}

To estimate the covariance between the binned $g_{j}^{n}$ genus measurements, we use $N_{\rm r}=500$ Multidark patchy mocks \citep{2016MNRAS.456.4156K,2016MNRAS.460.1173R}. Full details of their creation can be found in \citet{2016MNRAS.456.4156K}. Briefly, the mocks were generated using an iterative procedure to reproduce a reference galaxy catalog using approximate gravity solvers and a statistical biasing model \citep{2014MNRAS.439L..21K}. The reference catalog arises from the Big-MultiDark N-body simulation, which used Gadget-2 \citep{Springel:2005mi} to evolve $3840^{3}$ particles in a $(2.5 h^{-1} {\rm Mpc})^{3}$ volume. Halo abundance matching was used to reproduce the clustering of the observational data. The Patchy code \citep{Kitaura:2013cwa,10.1093/mnras/stv645} is used to match the two- and three-point clustering statistics with the full reference simulation in different redshift bins. Stellar masses are assigned and mock lightcones are generated, accounting for the survey mask and selection effects. The resulting mock catalogs accurately reproduce the number density, two-point correlation function, selection function and survey geometry of the DR12 observational data. The simulated data was generated using a Planck cosmology with $\Omega_{\rm m} = 0.307$, $\Omega_{\rm b} = 0.048$, $n_{\rm s} = 0.961$, $H_{0} = 67.77 {\rm km s^{-1} Mpc^{-1}}$. 

\begin{figure}
  \includegraphics[width=0.5\textwidth]{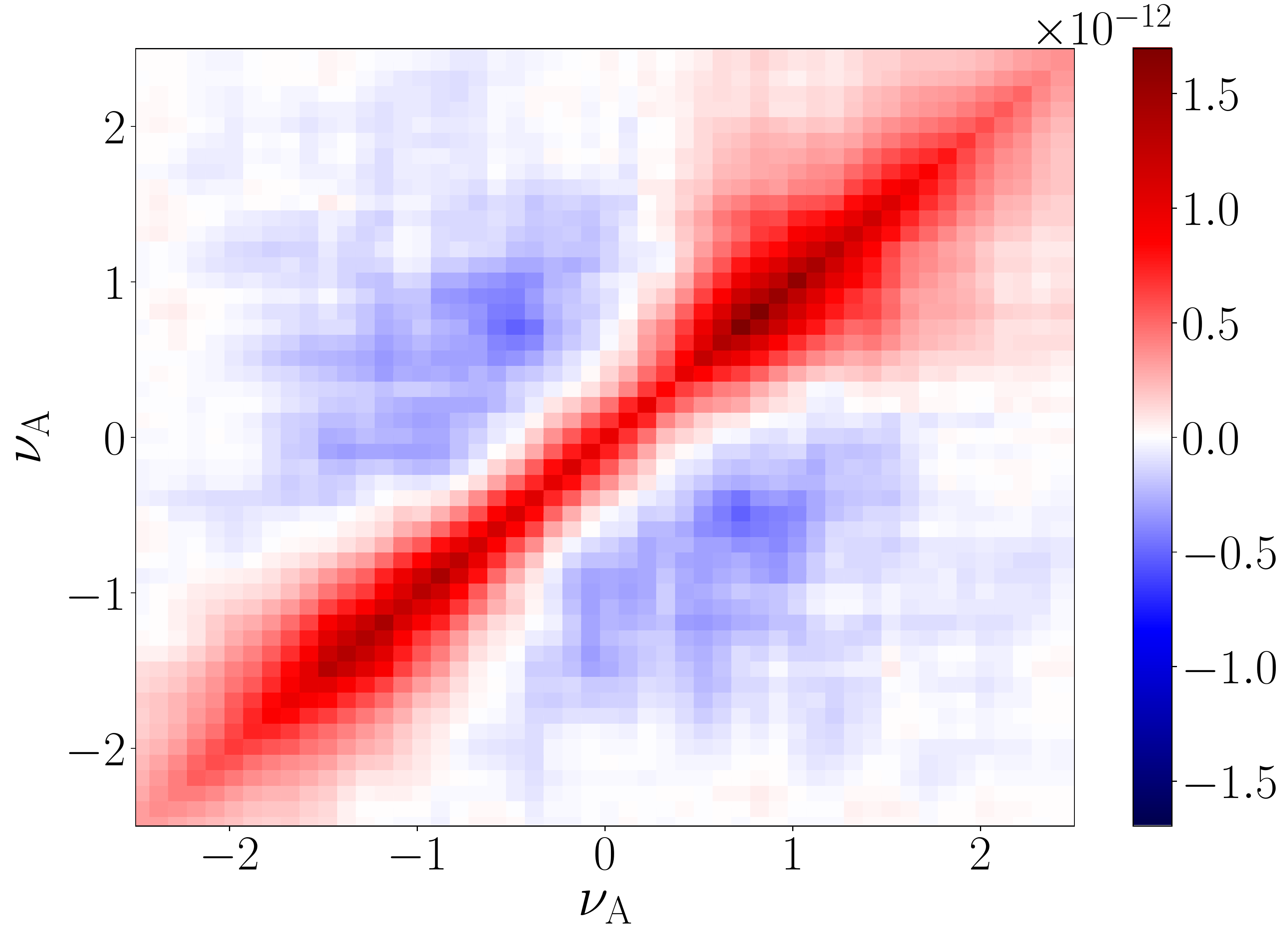} \\
  \caption{An example of one covariance matrix $\Sigma_{m,n}$ obtained from the mock catalogs. The shape is indicative of all twelve covariance matrices generated for the redshift shells. }
  \label{fig:covboss}
\end{figure}

For each mock catalog, we repeat our analysis - sort the galaxies into redshift shells, apply a mass cut then bin the surviving galaxies into pixels on the sphere. We then apply our masking and smoothing procedure and construct the genus curves $g_{p, j}^{n}$ where $p, j, n$ represent the $p^{\rm th}$ mock realisation, $j^{\rm th}$ redshift shell and $n^{\rm th}$ $\nu_{\rm A}$ threshold. As for the actual data, we measure the genus at $201$ values of $\nu_{\rm A}$ over the range $-2.5 < \nu_{\rm A} < 2.5$ then average every four points to obtain $N_{\nu_{\rm A}}=50$ values. From these measurements the twelve covariance matrices $\Sigma_{m,n} (z_{j})$ -- one for each redshift shell -- can be constructed as 

\begin{equation}\label{eq:cov2D} \Sigma_{m,n}(z_{j}) = {1 \over N_{\rm r}-1} \sum_{p=1}^{N_{\rm r}} \left( g_{p, j}^{n} - \langle g_{j}^{n} \rangle \right) \left( g_{p, j}^{m} - \langle g_{j}^{m} \rangle  \right) , \end{equation}

\noindent where $\langle g^{n}_{j} \rangle$ is the average value of the genus curve at the $n^{\rm th}$, $\nu_{{\rm A}, n}$ threshold and $j^{\rm th}$ redshift bin. We assume that the genus curves obtained at different redshift slices are uncorrelated. In Figure \ref{fig:covboss} we present one covariance matrix obtained from the patchy mocks, which is representative of all covariance matrices generated. One can observe strong positive correlation between $\nu_{\rm A}$ thresholds separated by $\Delta \nu_{\rm A} < 0.5$ (red) and a weaker, negative correlation between thresholds of larger separation (blue).

\begin{figure}
  \includegraphics[width=0.48\textwidth]{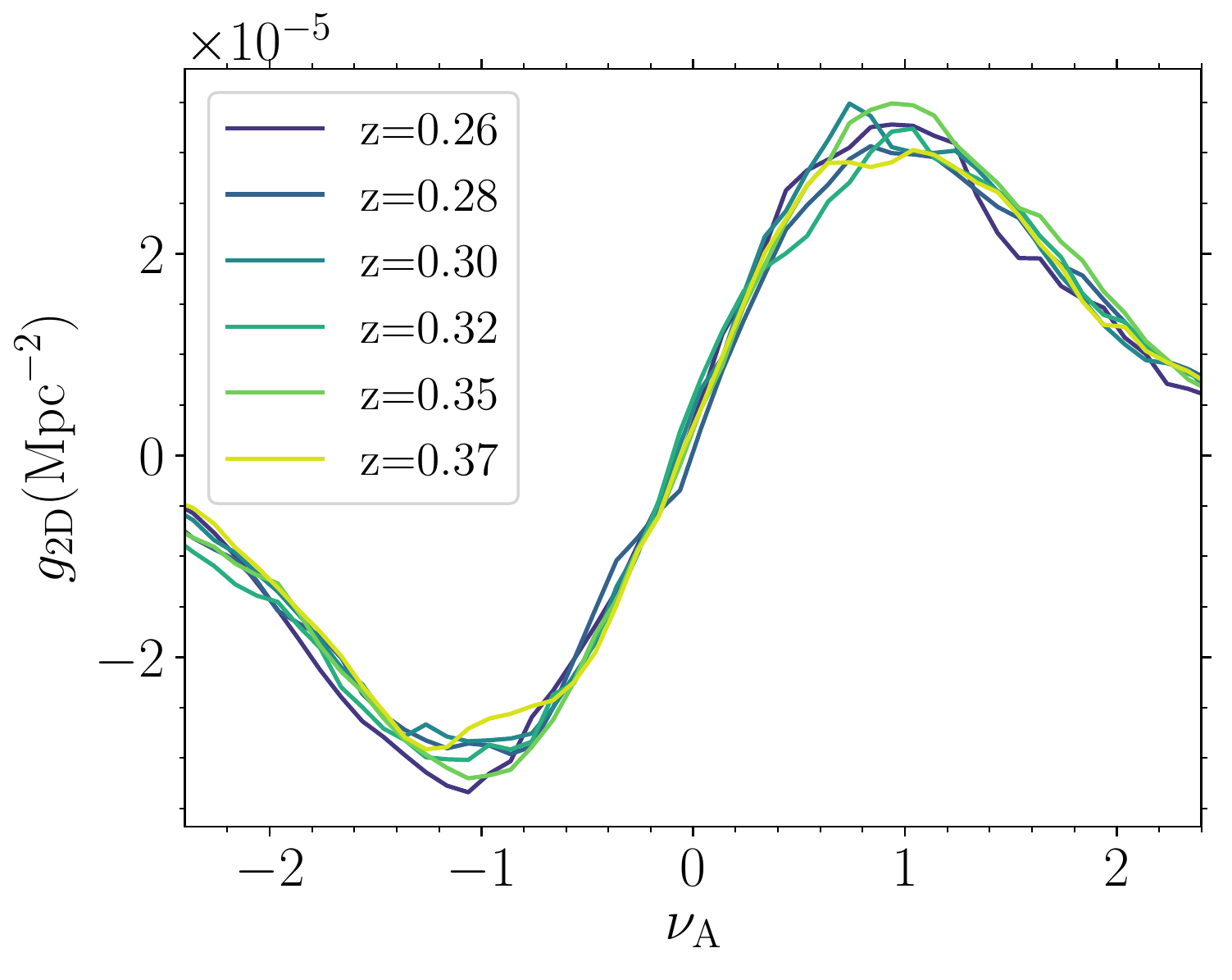}\\
    \includegraphics[width=0.48\textwidth]{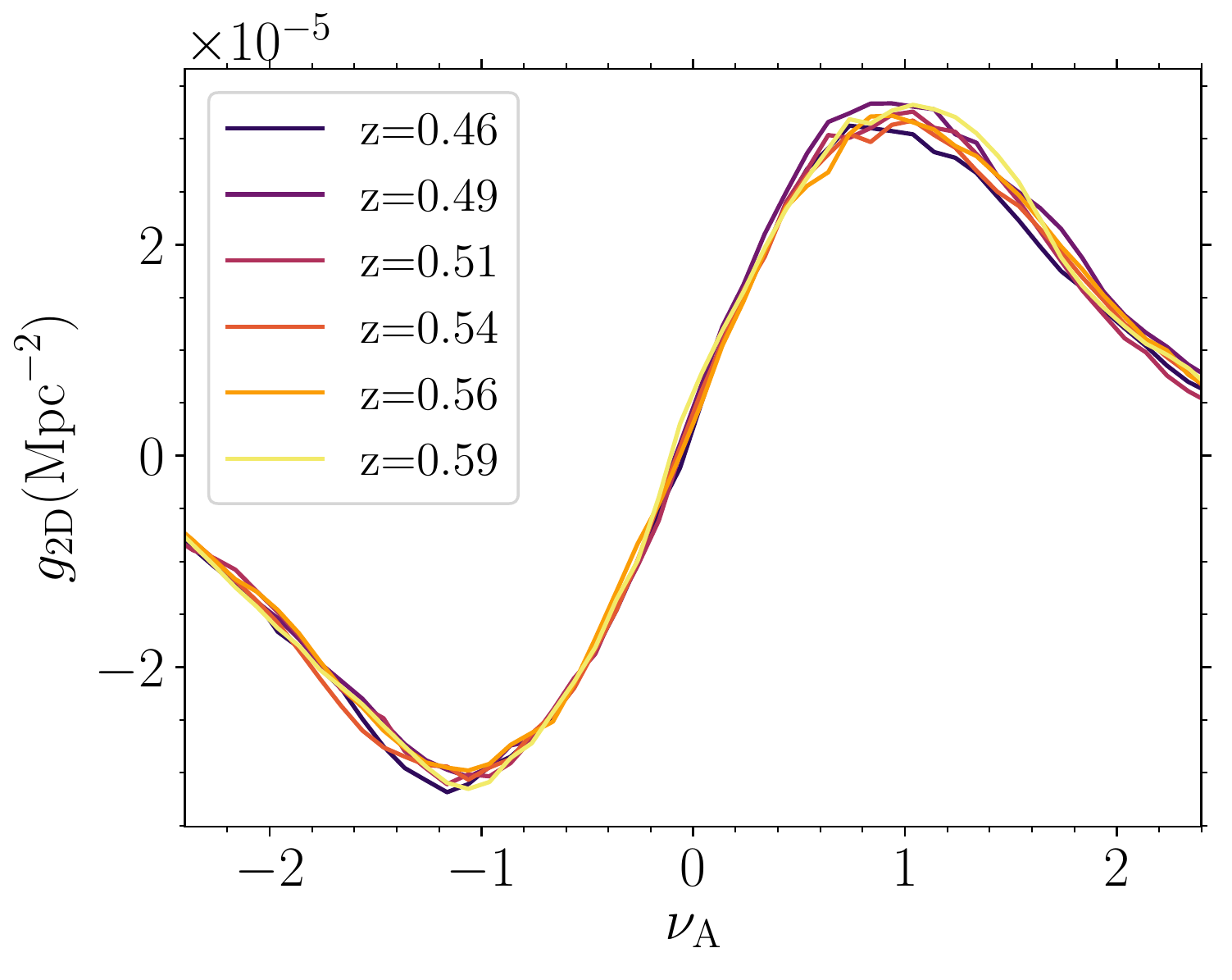}
  \caption{\label{fig:4} Six, two-dimensional genus curves obtained from the BOSS LOWZ [top panel] and CMASS [bottom panel] data. The curves are colour coded to the redshift shells, and no evolution with redshift is observed.}
\end{figure}

\section{Results - Cosmological Parameter Estimation }
\label{sec:constraints}

Finally, we use the measured genus curves $g^{n}_{j}$ and reconstructed covariance matrices $\Sigma_{m,n}(z_{j})$ to constrain cosmological parameters. In each redshift shell, we minimize the following $\chi^{2}_{j}$ function

\begin{equation}\label{eq:ch2d} \chi^{2}_{j} =  \sum_{n=1}^{N_{\nu_{\rm A}}}\sum_{m=1}^{N_{\nu_{\rm A}}} \Delta g^{n}_{j} \Sigma_{n,m}^{-1}(z_{j}) \Delta g^{m}_{j} , \end{equation} 

\noindent where 

\begin{eqnarray} \nonumber & &  \Delta  g^{n}_{j} = g_{j}^{n}  -   A^{(\rm 2D)}_{{\rm G},j} e^{-\nu_{{\rm A},n}^{2}/2} \left[ a_{0, j} H_{0}(\nu_{{\rm A},n}) + \right. 
\\ \label{eq:herm2d} & & \qquad \quad  \left.  H_{1}(\nu_{{\rm A},n}) + a_{2, j} H_{2}(\nu_{{\rm A},n}) + a_{3, j} H_{3}(\nu_{{\rm A},n}) \right]  . \end{eqnarray}

\noindent This functional form matches the theoretical expansion ($\ref{eq:mat1}$). The parameters varied are $a_{0,j}, a_{2,j}$, $a_{3,j}$, which are assumed to be arbitrary constants in each shell, and $\Omega_{\rm c}h^{2}$, $\Omega_{\rm b}h^{2}$, $n_{\rm s}$ ; the cosmological parameters that dictate the genus amplitude $A^{(\rm 2D)}_{{\rm G},j}$, which is defined in equation ($\ref{eq:ag}$) and related to the three dimensional matter power spectrum via equations ($\ref{eq:p2d}$,$\ref{eq:p3df}$). 

$A^{(\rm 2D)}_{{\rm G},j}$ is sensitive to the shape of the linear matter power spectrum, and hence to the parameters $\Omega_{\rm c}h^{2}$, $\Omega_{\rm b}h^{2}$, $n_{\rm s}$. Conversely, it is practically insensitive to the amplitude of the power spectrum, so we fix $\ln[10^{10}A_{\rm s}] = 3.089$ according to its Planck best fit \cite{Aghanim:2018eyx} and linear galaxy bias $b=2$, inferred from the mock catalogs. In appendix A we discuss the sensitivity of the genus to the amplitude of the power spectrum, and argue that for sparse galaxy data some residual sensitivity exists due to the presence of a shot noise contribution. For the smoothing scales used in this work, the sensitivity can be safely neglected.

In principle, the parameters $a_{0, 2}$ are sensitive to cosmology, as they are related to the three-point cumulants and hence shape of the bispectrum. However, in this work we do not utilise this information and treat these parameters as free over the range $-1 < a_{0}, a_{2} < 1$. $a_{0},a_{2}$ are the leading order corrections to the genus shape due to non-Gaussianity generated by gravitational collapse. The parameter $a_{3}$ should be of order $\sigma_{0}^{2}$ in the perturbative non-Gaussian expansion of the field, and we include it as a check that higher order terms remain negligible. We assign uniform priors of $0.5 < n_{\rm s} < 1.2$ and $0.05 < \Omega_{\rm c}h^{2} < 0.2$, $0.018 < \Omega_{\rm b}h^{2} < 0.026$ on the cosmological parameters.

In Figures \ref{fig:all_parms_lowz},\ref{fig:all_parms_cmass} in Appendix C we exhibit the two dimensional marginalised contours obtained from each individual LOWZ and CMASS slice for the parameters $\Omega_{\rm c}h^{2},\Omega_{\rm b}h^{2}, n_{\rm s}, a_{0}, a_{2}$. Each coloured contour corresponds to the result from a particular redshift slice. Although the parameter uncertainties are large when only using individual shells, we find some general trends. The sensitivity of $A_{{\rm G}, j}^{(\rm 2D)}$ to $\Omega_{\rm b}h^{2}$ is extremely weak -- we obtain no significant constraint within the prior range selected. The parameters $a_{0,2}$ are effectively independent to $n_{\rm s}$ and $\Omega_{\rm c}h^{2}$; this is due to the fact that $a_{0,2}$ are coefficients of even Hermite polynomials, whereas the genus amplitude is the coefficient of the odd polynomial $H_{1}(\nu_{\rm A})$. Conversely, there exists a strong correlation between $a_{0}, a_{2}$ for the same reason (both are even polynomial coefficients). We do not plot $a_{3}$, as it is included simply as a consistency check. This parameter is present at the $\sim 1\%$ level, but does not significantly impact our results. It is one of multiple terms that would be induced at order ${\cal O}(\sigma_{0}^{2})$. 

For the cosmological parameters, there is a strong negative degeneracy between $n_{\rm s}$ and $\Omega_{\rm c}h^{2}$. Both parameters can increase the amount of small scale power, by tilting the power spectrum and shifting the peak respectively.

\begin{figure*}
  \includegraphics[width=0.48\textwidth]{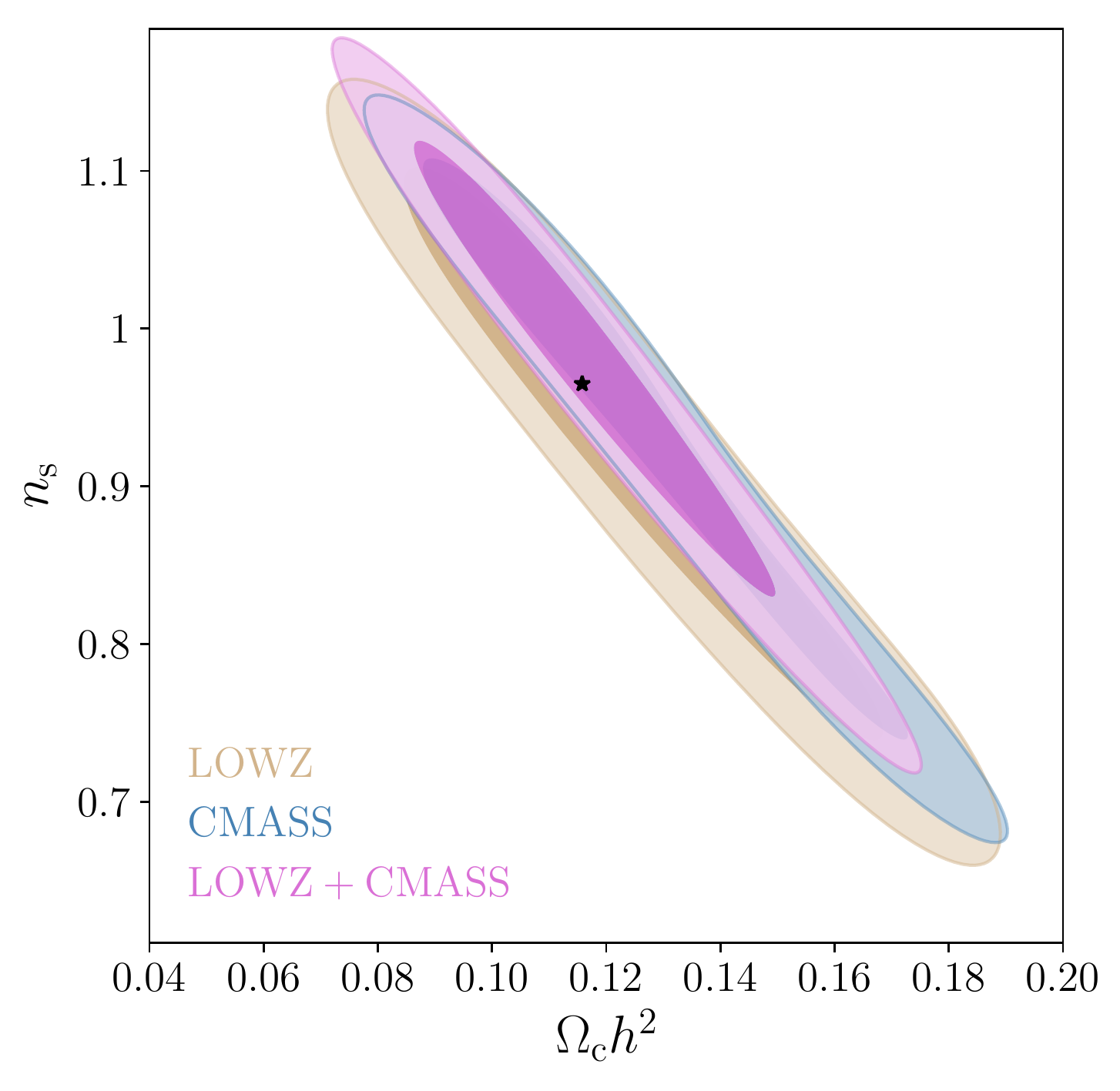}
  \includegraphics[width=0.48\textwidth]{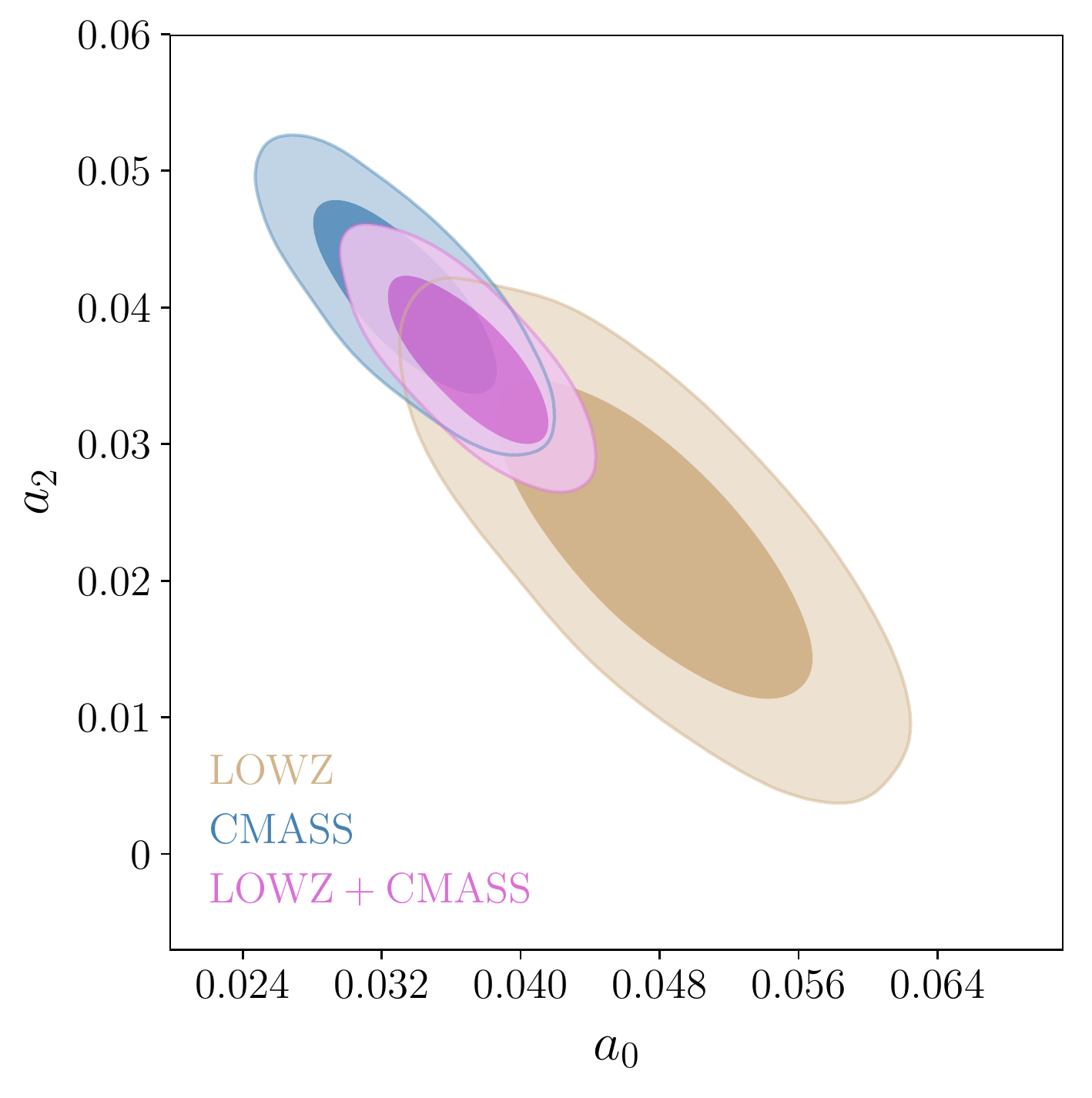}
  \caption{ [Left panel] Marginalised $68/95\%$ contours for the parameters $\Omega_{\rm c}h^{2}$ and $n_{\rm s}$, obtained from the combined LOWZ (brown), CMASS (blue) and all twelve shells of BOSS data (purple).  The black star is the Planck best fit value assuming a flat $\Lambda$CDM model  [Right panel] Two-dimensional $68/95\%$ contours for the parameters $a_{0}$, $a_{2}$, with colour scheme identical to the left panel. We observe a degeneracy between the two parameters.  }
  \label{fig:parm}
\end{figure*}

To improve the constraining power of the statistic, in Figure \ref{fig:parm} we present the combined constraint on $\Omega_{\rm c}h^{2}$, $n_{\rm s}$ [left panel], and $a_{2},a_{0}$ [right panel], obtained by combining all six CMASS shells [blue contour],  the combined LOWZ data [brown contour], and all twelve shells combined [purple contour]. Specifically we fit a single set of parameters $\Omega_{\rm c}h^{2}$, $n_{\rm s}$, $a_{2},a_{0}$, $a_{3}$ separately to all six LOWZ and CMASS genus curves, and also to the entire set of twelve curves. We assume that the genus measurements at each redshift are independent and so sum the $\chi^{2}_{j}$ contributions from each redshift shell and minimize the following $\chi^{2}$ functions

\begin{eqnarray}   \chi_{\rm lowz}^{2} &=& \sum_{j=1}^{6} \chi^{2}_{j},  \\ 
 \chi_{\rm cmass}^{2} &=& \sum_{j=7}^{12} \chi^{2}_{j}, \\ 
 \chi_{\rm all}^{2} &=& \sum_{j=1}^{12} \chi^{2}_{j}
\end{eqnarray}

\noindent  We  include the parameter $\Omega_{\rm b}h^{2}$ and marginalise over it, despite this quantity being effectively unconstrained within our prior range. We only present two pairs of contours in the main body of the text, because all other combinations are not informative. For completeness we provide the full corner plot in Figure \ref{fig:full_corner} in Appendix C. The CMASS data provides a tighter constraint compared to the LOWZ data, as expected due to the larger volume being probed at high redshift. All data are self-consistent and also in agreement with Planck measurements of these parameters (black star). The parameters $a_{0}$, $a_{2}$ are correlated and represent $\sim {\cal O}(5\%)$ corrections to the shape of the genus curve, relative to its Gaussian form. This indicates that the non-Gaussian perturbative expansion in $\sigma_{0}$ is valid at the scales being probed in this work.

\begin{figure}
  \includegraphics[width=0.45\textwidth]{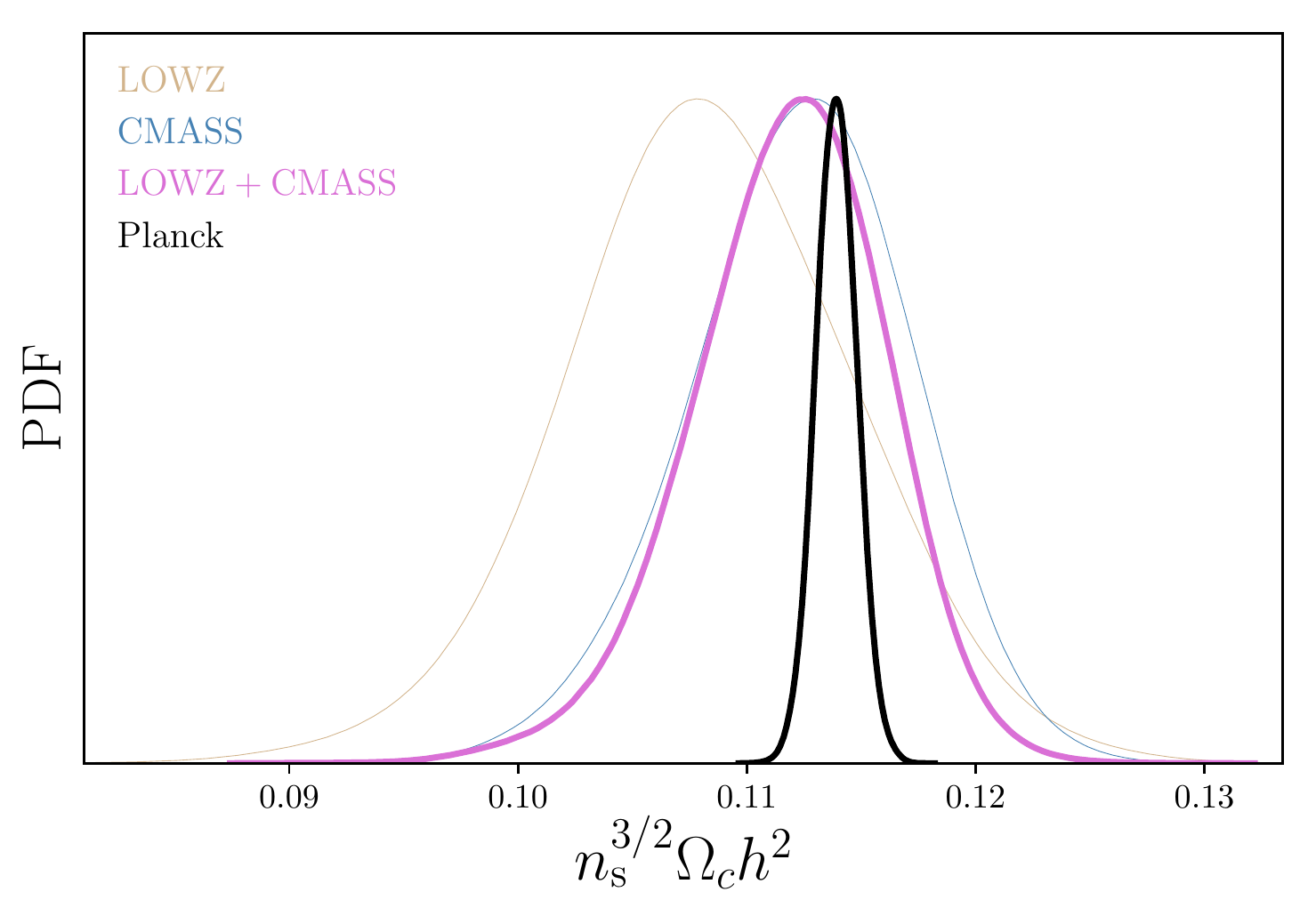}\\
    \includegraphics[width=0.49\textwidth]{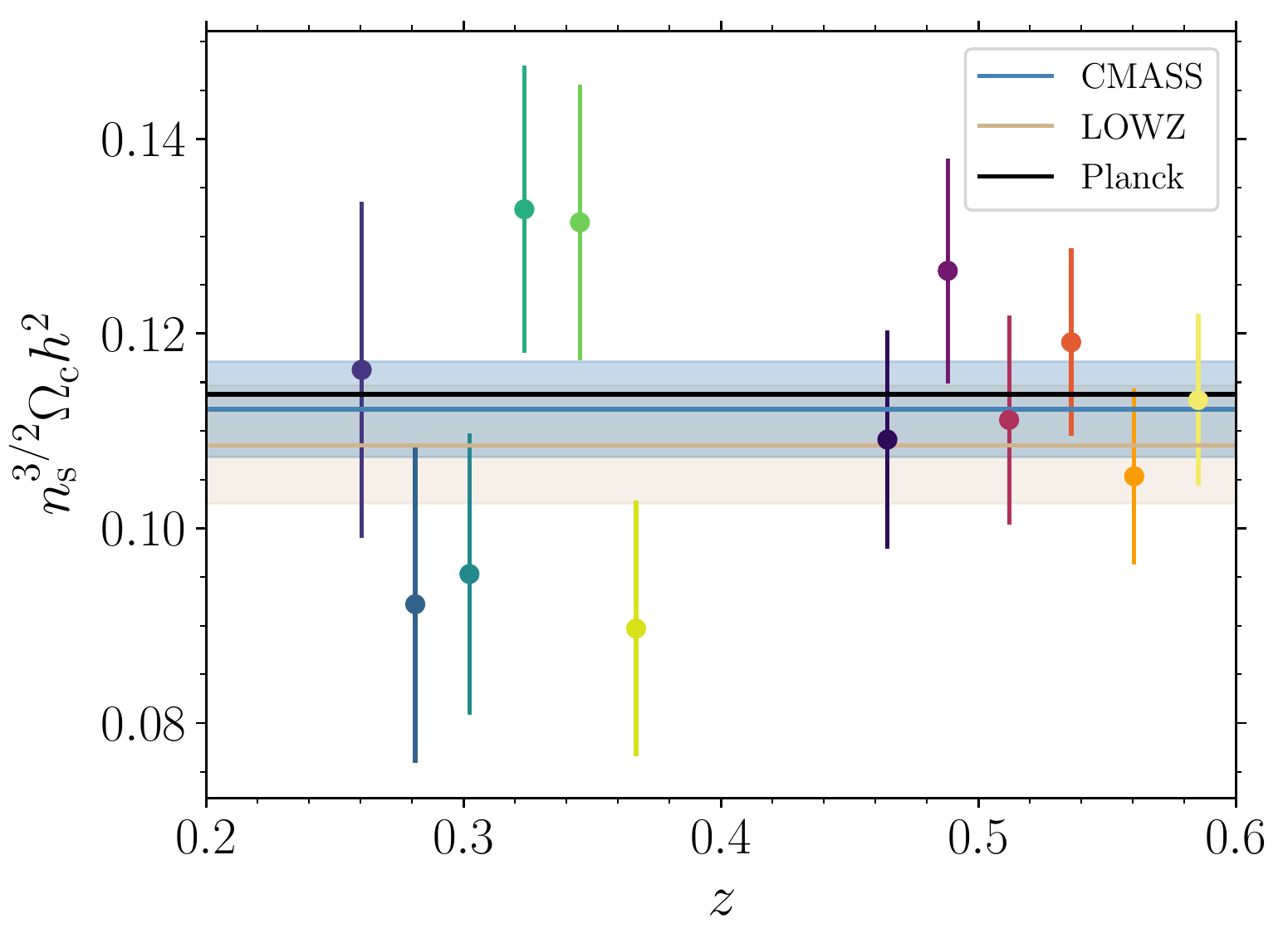}\\
  \caption{[Top panel] One-dimensional marginalised posterior likelihood for the parameter combination $n_{\rm s}^{3/2} \Omega_{\rm c}h^{2}$. The blue/brown distributions correspond to the CMASS and LOWZ data respectively, the purple distribution the combined result of all CMASS and LOWZ shells. The black distribution is the Planck posterior for this parameter combination, assuming a flat $\Lambda$CDM model [Bottom panel] The one-dimensional marginalised best fit and $1-\sigma$ uncertainty on $n_{\rm s}^{3/2}\Omega_{\rm c}h^{2}$ from each LOWZ and CMASS shell (points and error bars). The solid brown/blue lines and shaded areas are the best fit and $1-\sigma$ uncertainty from the combined LOWZ/CMASS shells, and the black solid line is the Planck best fit for this parameter combination. }
  \label{fig:parm_f}
\end{figure}

Due to the strong degeneracy between $n_{\rm s}$ and $\Omega_{\rm c}h^{2}$, we cannot simultaneously constrain these parameters using the genus amplitude alone. However, a tight constraint on the particular combination $n_{\rm s}^{3/2} \Omega_{\rm c}h^{2}$ can be derived from the data. If we rotate the contours into the $n_{\rm s}^{3/2} \Omega_{\rm c}h^{2}$ - $n_{\rm s}$ plane, we effectively obtain a one-dimensional constraint on $n_{\rm s}^{3/2} \Omega_{\rm c}h^{2}$. In Figure \ref{fig:parm_f} we present the marginalised one-dimensional posterior likelihood for this parameter combination, for the LOWZ (brown), CMASS (blue), combined LOWZ and CMASS (purple) and Planck (black) data. For the Planck constraint we use the publicly available baseline $\Lambda$CDM MCMC chains. The BOSS data is fully consistent with the Planck result, indicating that the shape of the linear matter power spectrum is consistent between $z \lesssim 1$ and $z \sim 1000$ over the scales probed in this analysis.

Finally, in Figure \ref{fig:parm_f} (bottom panel) we present the best fit and $1-\sigma$ uncertainties of the combination $n_{\rm s}^{3/2} \Omega_{\rm c}h^{2}$ as a function of redshift, obtained from the twelve LOWZ and CMASS redshift shells individually (coloured points/error bars). The Planck best fit is shown as a solid black line, and the combined result from all LOWZ and CMASS slices are presented as blue/brown lines and solid bands (the bands indicate $1-\sigma$ limits). The results are self-consistent, and fully consistent with the Planck result. In table \ref{tab:parms_f} we present our results in tabulated form. 

Our results provide a tight constraint on the combination  $n_{\rm s}^{3/2} \Omega_{\rm c}h^{2}$, which dictates the shape of the linear matter power spectrum. The genus amplitude is consistent with early Universe measurement of the power spectrum, indicating conservation of $P_{\rm 3D}(z,k)$ with redshift. This is expected for the $\Lambda$CDM model, for which the linear power spectrum shape is conserved from the last scattering surface to the present time, and only the amplitude varies. We emphasize that the amplitude cannot be measured efficiently using topological statistics.

\begin{table}[b!]
\caption{\label{tab:parms_f}}
 \begin{tabular}{||c  c  c  c  c ||}
 \hline
 Data  & $n_{\rm s}^{3/2} \Omega_{\rm c}h^{2}$ & $a_{0}$ & $a_{2}$ & $a_{3}$  \\ [0.5ex]
 \hline\hline
 & & & &  \\
 LOWZ  & $0.108^{+0.006}_{-0.006}$ & $0.048^{+0.006}_{-0.006}$ &  $0.023^{+0.008}_{-0.008}$ & $-0.003^{+0.004}_{-0.005}$   \\ 
  & & & &  \\
  CMASS   & $0.112^{+0.005}_{-0.005}$  &  $0.033^{+0.004}_{-0.004}$ &  $0.041^{+0.005}_{-0.005}$ & $0.005^{+0.003}_{-0.003}$   \\
   & & & &  \\
  ALL   & $0.112^{+0.004}_{-0.004}$  &  $0.037^{+0.003}_{-0.003}$ &  $0.036^{+0.004}_{-0.004}$ & $0.003^{+0.003}_{-0.003}$   \\
   & & & &  \\
 \hline 
\end{tabular}
\\ Marginalised best fit and $1-\sigma$ uncertainty on the parameter combination $n_{\rm s}^{3/2} \Omega_{\rm c}h^{2}$, and the Hermite polynomial coefficients $a_{0,2,3}$ from the combined LOWZ and CMASS and Combination of all twelve shells used (`ALL').
\end{table}

\section{Discussion}
\label{sec:discuss}

In this work we have measured the two-dimensional genus of shells of BOSS LOWZ and CMASS data. After extracting the  genus curves, we used them to place cosmological parameter constraints using the amplitude of the curve. The genus amplitude provides a measure of the shape of the underlying linear matter power spectrum; hence we were able to constrain  $\Omega_{\rm c}h^{2}$, $n_{\rm s}$. The parameters $\Omega_{\rm c}h^{2}$ and $n_{\rm s}$ present negative correlation as both can act to affect the slope of the power spectrum on the smoothing scales adopted in this study. We found that the genus amplitude is effectively insensitive to the baryon fraction. We were able to place a tight constraint on $n_{\rm s}^{3/2} \Omega_{\rm c}h^{2} = 0.108 \pm 0.006$ and $n_{\rm s}^{3/2} \Omega_{\rm c}h^{2} = 0.112 \pm 0.005$ for the LOWZ and CMASS data sets respectively, with a total constraint $n_{\rm s}^{3/2} \Omega_{\rm c}h^{2} = 0.112 \pm 0.004$ after combining all data. Our constraints are completely consistent with the Planck best fit values for a flat $\Lambda$CDM model.

Our results are practically insensitive to reasonable variations of $A_{\rm s}$ and linear galaxy bias $b$. However, the Minkowski functionals can be sensitive to these quantities for sparse galaxy samples, as the relative amplitude between the matter power spectrum and the shot noise contribution modifies the genus amplitude. We discuss this caveat further in Appendix A. Shot noise is the dominant issue in the reconstruction of Minkowski functionals from a continuous field inferred from a point distribution. It modifies the genus amplitude, but also the shape of the Minkowski functionals as the noise can be non-Gaussian. When shot noise is a significant contributor to the field, we lose the interpretation that the genus amplitude is a measure of the slope of the matter power spectrum. 

Given that our analysis is applied to a spectroscopic galaxy catalog, one might question the logic of using two-dimensional slices of data when we have access to accurate redshift information. The reasoning is two-fold. First, the galaxy catalog is sparse, and we mitigate this issue by taking thick slices along the line of sight. Binning galaxies in this way is simply a smoothing choice, so we can interpret our decision as anisotropic smoothing perpendicular and parallel to the line of sight. Smoothing on larger scales parallel to the line of sight has advantages, such as allowing us to use linear redshift space distortion physics. Second, in future work we wish to compare our results with higher redshift photometric redshift catalogs, which will require us to bin galaxies into thick shells. An understanding of how photometric redshift uncertainty modifies our analysis must be further explored before this comparison can be made. 

Our analysis can be refined in a number of ways. We have only used information extracted from the amplitude of the genus curve. For a non-Gaussian field, the shape of the genus contains information on the three-point function of the density field -- by relating the Hermite polynomial coefficients $a_{0,2}$ to the three-point cumulants one can extract information on the shape of the galaxy bispectrum. We intend to perform this comparison in future work. 

The redshift space distortion correction to the genus has been calculated at linear order in \citet{Matsubara:1995wj}. It would be of interest to study the non-linear effects of the velocity field \citep{Codis:2013exa}, all the way down to the small scale finger of god effect. Better understanding of this systematic can be used to reduce the uncertainty of our measurements, as it would allow us to smooth the density field on smaller scales parallel to the line of sight. 

Finally, it would be of interest to consider methods by which we can break the parameter degeneracy between $\Omega_{\rm c}h^{2}$ and $n_{\rm s}$. One method would be to combine measurements of the genus at different smoothing scales. On small scales we can expect to be predominantly sensitive to $n_{\rm s}$, whereas by using a large smoothing scale one will be increasingly sensitive to the peak position of the power spectrum. This dependence will rotate the two-dimensional contour in the $n_{\rm s}$-$\Omega_{\rm c}h^{2}$ plane. To perform this test we must understand the covariance between genus measurements at different scales. We can also use overlapping redshift bins, and measure the two-dimensional genus as a continuous function of $z$. Finally the two- and three- dimensional genus amplitudes will also be sensitive to the power spectrum slope at different scales. In future work we will combine these measurements to simultaneously constrain the parameters governing the shape of the matter power spectrum.

\section*{Acknowledgement}

SA is supported by a KIAS Individual Grant QP055701 via the Quantum Universe Center at Korea Institute for Advanced Study. SEH was supported by Basic Science Research Program through the National Research Foundation of Korea funded by the Ministry of Education (2018\-R1\-A6\-A1\-A06\-024\-977). SA would like to thank Christophe Pichon for helpful discussions during the preparation of this manuscript. We thank the Korea Institute for Advanced Study for
providing computing resources (KIAS Center for Advanced
Computation Linux Cluster System).

Funding for SDSS-III has been provided by the Alfred
P. Sloan Foundation, the Participating Institutions,
the National Science Foundation, and the U.S. Department
of Energy Office of Science. The SDSS-III web
site is http://www.sdss3.org/. SDSS-III is managed by
the Astrophysical Research Consortium for the Participating
Institutions of the SDSS-III Collaboration including
the University of Arizona, the Brazilian Participation
Group, Brookhaven National Laboratory, Carnegie Mellon
University, University of Florida, the French Participation
Group, the German Participation Group, Harvard
University, the Instituto de Astrofisica de Canarias, the
Michigan State/Notre Dame/JINA Participation Group,
Johns Hopkins University, Lawrence Berkeley National
Laboratory, Max Planck Institute for Astrophysics, Max
Planck Institute for Extraterrestrial Physics, New Mexico
State University, New York University, Ohio State
University, Pennsylvania State University, University of
Portsmouth, Princeton University, the Spanish Participation
Group, University of Tokyo, University of Utah,
Vanderbilt University, University of Virginia, University
of Washington, and Yale University.

The massive production of all MultiDark-Patchy mocks for the BOSS Final Data Release has been performed at the BSC Marenostrum supercomputer, the Hydra cluster at the Instituto de Fısica Teorica UAM/CSIC, and NERSC at the Lawrence Berkeley National Laboratory. We acknowledge support from the Spanish MICINNs Consolider-Ingenio 2010 Programme under grant MultiDark CSD2009-00064, MINECO Centro de Excelencia Severo Ochoa Programme under grant SEV- 2012-0249, and grant AYA2014-60641-C2-1-P. The MultiDark-Patchy mocks was an effort led from the IFT UAM-CSIC by F. Prada’s group (C.-H. Chuang, S. Rodriguez-Torres and C. Scoccola) in collaboration with C. Zhao (Tsinghua U.), F.-S. Kitaura (AIP), A. Klypin (NMSU), G. Yepes (UAM), and the BOSS galaxy clustering working group.

Some of the results in this paper have been derived using the healpy and HEALPix package

\newpage

\section*{Appendix A -- Sensitivity of Genus to Galaxy Bias and Power Spectrum Amplitude }
\label{sec:th}

In appendix A we discuss the extent to which the genus amplitude is sensitive to the linear galaxy bias and primordial power spectrum amplitude $A_{\rm s}$. We begin by repeating the definition of the two-dimensional genus amplitude as the ratio of cumulants --

\begin{equation}\label{eq:agaussapp} A^{(\rm 2D)}_{{\rm G}} = {1 \over 2(2\pi)^{3/2}} { \int d k_{\perp} k^{3}_{\perp} e^{-k_{\perp}^{2}R_{\rm G}^{2}} P_{\rm 2D}(k_{\perp},z)  \over \int d k_{\perp}  k_{\perp} e^{-k_{\perp}^{2}R_{\rm G}^{2}} P_{\rm 2D}(k_{\perp},z)  } , \end{equation}

\noindent where the two-dimensional power spectrum $P_{\rm 2D}$ is related to the three dimensional matter power spectrum according to 

\begin{equation}\label{eq:p2d3} P_{\rm 2D}(k_{\perp},z_{j}) = {2 \over \pi} \int dk_{\parallelsum}  P_{\rm 3D}\left(k,k_{\parallelsum},z \right) {\sin^{2} [\Delta k_{\parallelsum}] \over (\Delta k_{\parallelsum})^{2}} , \end{equation} 

\noindent where $k = \sqrt{k_{\perp}^{2} + k_{\parallelsum}^{2}}$. When extracting the genus from a galaxy catalog, we are probing the matter field measured in redshift space, reconstructed from a discrete point distribution. For such a field the underlying three-dimensional power spectrum in  ($\ref{eq:p2d3}$) can be approximated on large scales as --

\begin{equation}\label{eq:p3dfapp} P_{\rm 3D} (k,k_{\parallelsum},z) = b^{2} \left( 1 + \beta {k_{\parallelsum}^{2} \over k^{2}}\right)^{2} P_{\rm lin}(z, k) + P_{\rm SN} , \end{equation}

\noindent where $P_{\rm lin}(z,k)$ is the linear matter power spectrum at redshift $z$, $P_{\rm SN}$ is the shot noise power spectrum that we take as $P_{\rm SN} = 1/\bar{n}_{\rm cut}$, $\beta = \Omega_{\rm m}^{\gamma}/b$, $b$ is the linear galaxy bias and $\gamma \simeq 3(1-w_{\rm de})/(5-6w_{\rm de})$. 

\begin{figure}
  \includegraphics[width=0.45\textwidth]{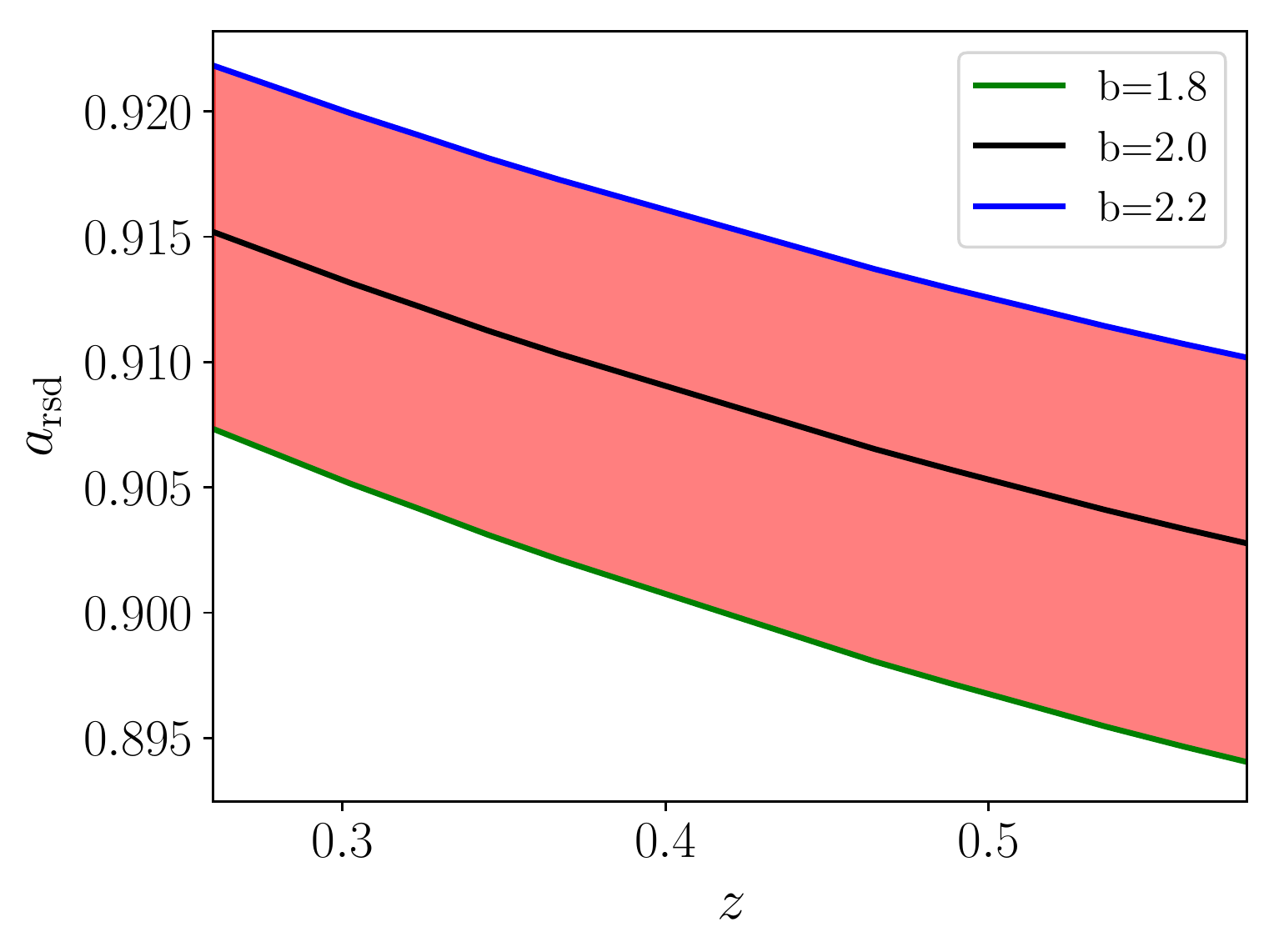}
  \caption{The ratio of the genus amplitude in redshift and real space, $a_{\rm rsd}$, as a function of redshift. The green/black/blue solid lines correspond to linear bias $b=1.8, 2.0, 2.2$ respectively. The effect of redshift space distortion decreases the genus amplitude, introduces mild redshift dependence and creates weak sensitivity to galaxy bias (cf. red filled region). }
  \label{fig:rsdth}
\end{figure}

The conventional wisdom in topological analysis is that the Minkowski functionals are insensitive to the amplitude of the power spectrum, and hence also any linear bias factors. However, galaxy bias and power spectrum amplitude enter into the genus amplitude in two ways. First, the presence of the redshift space distortion parameter $\beta \sim b^{-1}$ in ($\ref{eq:p3dfapp}$) introduces weak dependence on the galaxy bias. In Figure \ref{fig:rsdth} we present the dimensionless ratio $a_{\rm rsd} \equiv  A^{(\rm 2D)}_{{\rm G, rs}}/  A^{(\rm 2D)}_{{\rm G, re}}$, where $ A^{(\rm 2D)}_{{\rm G, rs}},  A^{(\rm 2D)}_{{\rm G, re}}$ are the genus amplitudes in redshift and real space respectively. We have fixed all cosmological parameters to their Planck best fit values, and used $\beta = 0$ in ($\ref{eq:p3dfapp}$) to calculate $A^{(\rm 2D)}_{{\rm G, re}}$. We have calculated $a_{\rm rsd}$ for three different values of the linear galaxy bias $b=1.8, 2, 2.2$, which are shown as green, black and blue lines in the figure. The net effect of redshift space distortion is to decrease the genus amplitude by roughly $9\%$. The red solid area indicates the sensitivity of the redshift space distortion effect due to galaxy bias -- varying the galaxy bias over the range $1.8 < b < 2.2$ introduces a weak $\sim {\cal O}(1\%)$ variation in the genus amplitude. 

The second effect, which generates sensitivity to the amplitude of the power spectrum $b^{2}A_{\rm s}$, is shot noise. The fact that we are attempting to infer the properties of a continuous fluid from a discrete point distribution introduces a noise contribution to the total measured power spectrum -- we have approximated this effect via the term $P_{\rm SN} = 1/\bar{n}_{\rm cut}$ in ($\ref{eq:p3dfapp}$), where $\bar{n}_{\rm cut}$ is the galaxy number density, selected to be constant in each redshift shell. 

 We see that the genus amplitude is actually a measurement of the sum of two power spectra -- the underlying matter power spectrum and $P_{\rm SN}$. Generically, the fact that $P_{\rm SN} \neq 0$ implies that the relative amplitudes of $b^{2} P_{\rm lin}$ and $P_{\rm SN}$ will impact the genus amplitude. This was observed in \citet{Kim:2014axe,Appleby:2017ahh}, where the genus amplitude was found to be a function of galaxy number density and sampling selection (or bias). This effect is minimized by Gaussian smoothing in the plane, which suppresses the contribution of $P_{\rm SN}$, and also selecting as high number density sample as possible. If the number density of the galaxy catalog is sufficiently large, the effect of shot noise can be safely neglected and the genus amplitude becomes insensitive to scale-independent bias factors $b(z)$ and also the primordial amplitude $A_{\rm s}$. However, for a sparse galaxy sample we can expect some sensitivity to both.

\begin{figure}
  \includegraphics[width=0.45\textwidth]{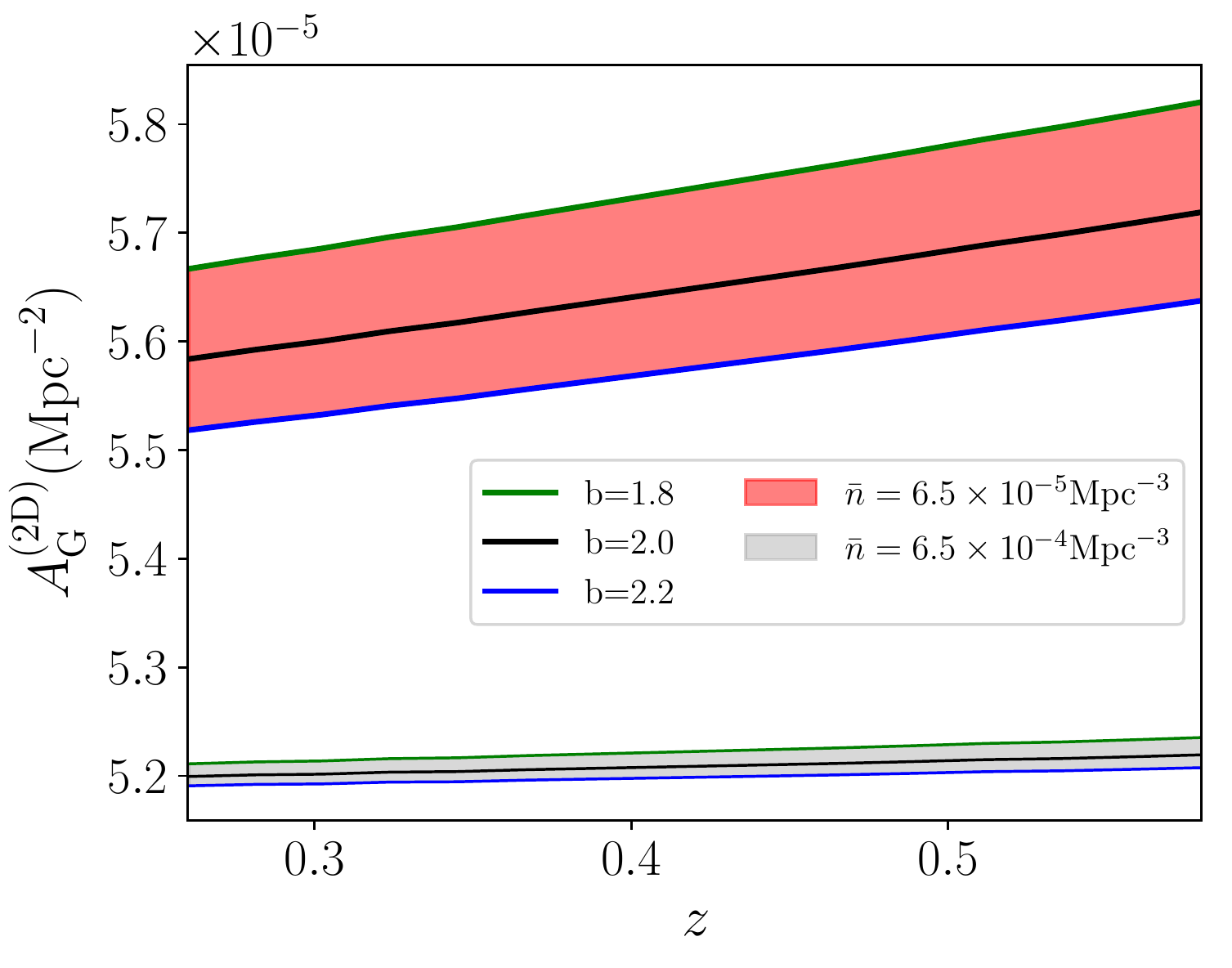}
  \caption{The Genus amplitude $A^{(\rm 2D)}_{\rm G}$ in real space, fixing cosmological parameters and varying the number density of tracer particles $\bar{n}$ and bias $b$. The red/grey solid regions cover the range $1.8 < b < 2.2$, for fixed number density $\bar{n} = 6.5 \times 10^{-5} {\rm Mpc}^{-3}$ (red) and $\bar{n} = 6.5 \times 10^{-4} {\rm Mpc}^{-3}$ (grey). The more sparse galaxy sample is more sensitive to galaxy bias (more specifically, the amplitude of the matter power spectrum). }
  \label{fig:sn}
\end{figure}

In Figure \ref{fig:sn} we exhibit $A^{(\rm 2D)}_{\rm G}$ -- equation ($\ref{eq:agaussapp}$) -- using the power spectrum ($\ref{eq:p3dfapp}$) assuming cosmological parameters and smoothing scales $\Delta$, $R_{\rm G}$ in table \ref{tab:ii}. To avoid conflating different systematic effects, we focus on the real space genus amplitude and set $\beta=0$. We keep the cosmological parameters fixed and vary the number density $\bar{n}_{\rm cut}$ and bias $b$. The red and grey filled regions cover the area between two limiting bias values $1.8 < b < 2.2$. The red region corresponds to our fiducial number density choice $\bar{n}_{\rm cut} = 6.5 \times 10^{-5} {\rm Mpc}^{-3}$ for the BOSS galaxy sample, and the grey region corresponds to a more dense sample $\bar{n}_{\rm cut} = 6.5 \times 10^{-4} {\rm Mpc}^{-3}$. The green, black and blue lines correspond to bias factors $b=1.8,2.0,2.2$ respectively. 

Two conclusions can be drawn from Figure \ref{fig:sn}. First, shot noise modifies the genus amplitude, and a more sparse galaxy sample (cf red region) will generate a larger genus amplitude than a denser sample (grey region). Second, if the galaxy sample is sufficiently sparse, then the genus amplitude becomes sensitive to galaxy bias (or more precisely the combination $b^{2} A_{\rm s}$) -- the width of the filled regions indicates the sensitivity of $A_{\rm G}^{(\rm 2D)}$ to bias, for given number density and fixing all other cosmological parameters. For our fiducial number density $\bar{n} = 6.5 \times 10^{-5} {\rm Mpc}^{-3}$ a $\sim 10\%$ variation in the galaxy bias generates a $\sim 0.5\%$ uncertainty in the genus amplitude (cf. red shaded area). One can also observe a faint redshift evolution in the red shaded region -- this is due to the amplitude of the matter power spectrum decreasing with redshift relative to the shot noise term, which we have fixed to be constant at each redshift. However, if we increase the number density of the galaxy sample by an order of magnitude (gray shaded region), the sensitivity of the genus amplitude to the bias becomes negligible -- as the shot noise decreases the genus amplitude becomes insensitive to the amplitude of the power spectrum. The redshift evolution is also suppressed by selecting a more dense sample. 

A more general word of caution is required -- the shot noise contribution is not perfectly represented by a white noise power spectrum with $P_{\rm SN} = 1/\bar{n}$. In fact, as we stray into scales at which shot noise affects our results, then the expansion of the genus in Hermite polynomials will not provide a good representation of the Poissonian signal, as the Poisson distribution has a different moment generating function. $R_{\rm G} > \bar{r}$, where $\bar{r}$ is the mean galaxy separation of the catalog, is an important condition on our analysis. Shot noise has been discussed in \cite{Appleby:2017ahh} and further in \cite{Appleby_inprep}.

We note that both the redshift space distortion effect and shot noise introduce a redshift dependence to the genus amplitude. For the smoothing scales and redshifts used in this work, the redshift space distortion effect systematically decreases the genus amplitude with increasing redshift. Conversely, the shot noise contribution increases the genus amplitude with redshift. Both effects are $\sim {\cal O}(1\%)$ and will act to practically cancel one another.

To completely suppress the shot noise and bias effects, the condition $\bar{r} \ll R_{\rm G}$ is required. For the smoothing scales and number densities considered in this work, the genus amplitude is only weakly sensitive to the combination of $b$, $A_{\rm s}$ and $P_{\rm SN}$. Nevertheless, we must include the $P_{\rm SN}$ contribution to the power spectrum to avoid systematic bias in our cosmological parameter reconstruction.

\section*{Appendix B -- Effect of variation of cosmological parameters on measured genus amplitudes} 

At numerous points in our analysis when measuring the genus curves from the BOSS galaxy catalog, we have been forced to fix the distance redshift relation. Specifically, we smoothed the field with constant comoving scale $R_{\rm G} = 20 \, {\rm Mpc}$, which corresponds to $\theta_{\rm G} = R_{\rm G}/d_{\rm c}(z_{j})$ angular smoothing on the unit sphere. We measured the genus per unit area, so divided the genus by the total area $A_{j} = 4\pi f_{{\rm sky}, j} d_{\rm c}^{2}(z_{j})$ of each data shell. We also selected redshift shells of thickness $80 {\rm Mpc}$, and fixed a constant number density of $\bar{n}_{\rm cut} = 6.5 \times 10^{-5} {\rm Mpc}^{-3}$. Each of these dimension-full operations has forced us to select a distance-redshift relation, which we have taken throughout to be the Planck best fit, flat $\Lambda$CDM cosmology presented in table \ref{tab:ii}. In appendix B we consider how robust our measurements of the genus curve are to such a choice.

To test this, we take an all-sky mock galaxy catalog generated from a known cosmological parameter set, and repeat our analysis according to the main body of the paper. We then repeat our analysis again for three different, incorrect cosmological models. For each model, we completely repeat our analysis, and compare the resulting genus amplitude measurements.

The data that we use for this test is the Horizon Run 4 all sky mock galaxy lightcone \cite{Kim:2015yma}. Horizon Run 4 is a dense, cosmological scale dark matter simulation in which $N = 6300^{3}$ particles in a volume of $V = (3150 {\rm Mpc}/h)^{3}$ are gravitationally evolved. The simulation uses a modified GOTPM code and the initial conditions are estimated using second order Lagrangian perturbation theory \cite{L'Huillier:2014dpa}. The cosmological parameters used are $h=0.72$, $n_{\rm s} = 0.96$, $\Omega_{\rm m} = 0.26$, $\Omega_{\rm b} = 0.048$. A single all-sky mock galaxy lightcone out to $z = 0.7$ is used in this work. Details of the numerical implementation, and the method by which mock galaxies are constructed can be found in \cite{Hong:2016hsd}. The mock galaxies are defined using the most bound halo particle galaxy correspondence scheme, and the survival time of satellite galaxies post merger is estimated via the merger timescale model described in \cite{Jiang:2007xd}. 

We begin by repeating the analysis of the paper. Using the correct cosmological parameters $h=0.72$, $\Omega_{\rm m} = 0.26$ to infer the distance-redshift relation, we bin the galaxies into redshift shells of width $\Delta = 80 {\rm Mpc}$ and pixels on the sphere. We apply a mass cut to fix the galaxy number density $\bar{n} = 6.5 \times 10^{-5} {\rm Mpc}^{-3}$ in each shell. As for the BOSS data, we take six redshift shells over the range $0.25 < z < 0.4$ and six over the range $0.45 < z < 0.6$, and measure the genus curves from these shells. Finally, we extract the genus amplitude $A^{(\rm 2D)}$ from these curves, using the method described in \cite{Appleby:2018jew}.

\begin{figure}
  \centering
  \includegraphics[width=0.45\textwidth]{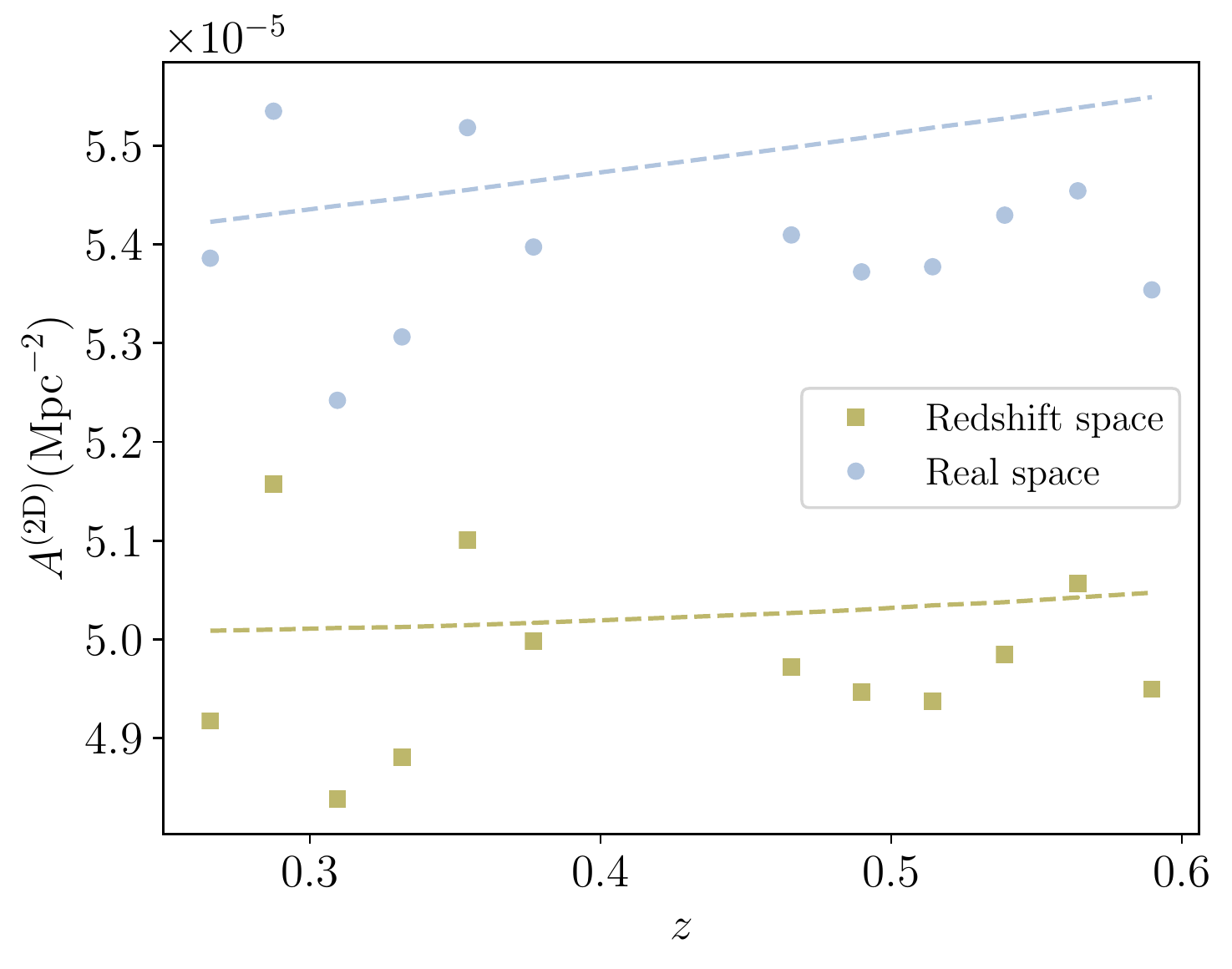}
  \caption{The measured genus amplitudes from the all-sky mock galaxy sample extracted from Horizon Run 4. The blue/brown solid points are the measured amplitudes in real/redshift space, one point for each redshift shell. The blue/brown dashed line is the Gaussian expectation value of $A^{(\rm 2D)}_{\rm G}$ in real/redshift space.}
  \label{fig:real_rsd_HR4}
\end{figure}

In Figure \ref{fig:real_rsd_HR4} we present the genus amplitudes extracted from the mock galaxy lightcone in real (blue squares) and redshift (yellow squares) space, for the twelve shells. The blue and yellow dashed lines correspond to the Gaussian theoretical prediction ($\ref{eq:agaussapp}$), with $\beta = 0$ (blue dashed) and $\beta = \Omega_{\rm m}^{6/11}/b$ (yellow dashed), taking $b=2$. 

Next, we repeat our analysis, using three different incorrect cosmological models to infer the distance-redshift relation. The specific models used are labeled ${\rm II}, {\rm III}, {\rm IV}$ in table \ref{tab:wrong}. For each cosmology the entire process of redshift binning, mass cut, smoothing and extracting the genus curve and amplitude is repeated. We arrive at a set of twelve genus amplitudes for each cosmological model, which we compare to those obtained using the correct cosmology. 

\begin{figure}
    \includegraphics[width=0.45\textwidth]{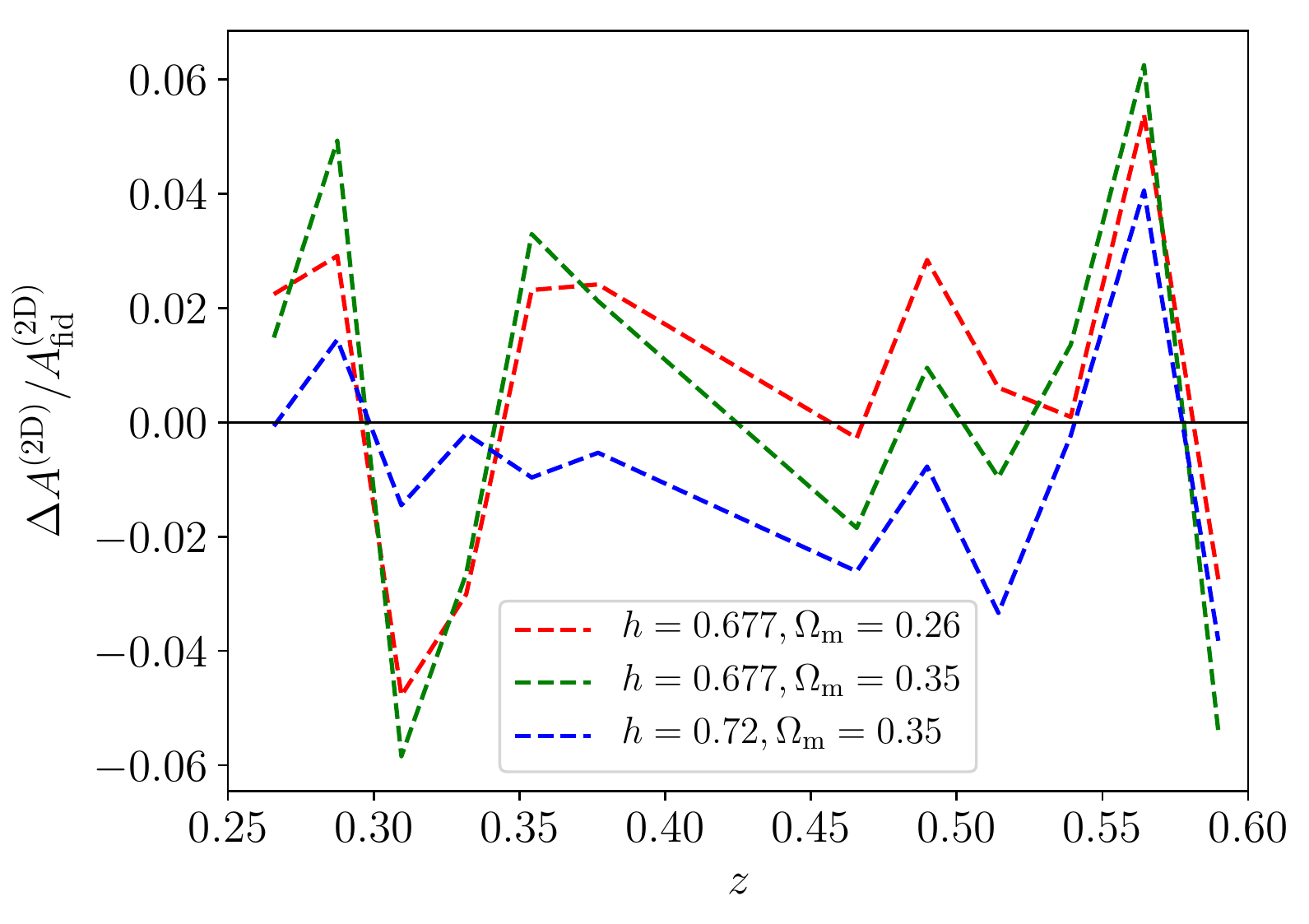}
  \caption{The fractional difference between the genus amplitude extracted from the mock data using the correct cosmological model $A_{\rm fid}^{(\rm 2D)}$ and the same quantity extracted using an incorrect cosmology to infer the distance-redshift relation - $\Delta A^{(\rm 2D)} = A^{(\rm 2D)}(h,\Omega_{\rm m}) - A_{\rm fid}$.  }
  \label{fig:wrong}
\end{figure}

In Figure \ref{fig:wrong} we present the fractional difference between the genus amplitudes from the wrong cosmology and the `correct' values inferred from the fiducial cosmological parameter set - $\Delta A^{(\rm 2D)}/A^{(\rm 2D)}_{\rm fid} =  (A^{(\rm 2D)} - A^{(\rm 2D)}_{\rm fid}) /A^{(\rm 2D)}_{\rm fid}$, where $A_{\rm fid}$ are the correct values. Although one can observe a mild systematic change with cosmology, the effect is at the $\sim 1\%$ level. Specifically, fixing $h=0.72$ and varying $\Omega_{\rm m}$ generates a marginally lower genus amplitude (cf. blue curve). Varying $h$ but selecting the correct value of $\Omega_{\rm m}=0.26$ does not produce a definite systematic bias (red curve). In all cases, the statistical scatter of the measurements dominates.

\begin{table}
\begin{center}
\caption{\label{tab:wrong}}
\centering 
 \begin{tabular}{||  c  c  c  ||}
 \hline
 Model \,  & \, $\Omega_{\rm m}$ \, & \, $h$   \\ [0.5ex] 
 \hline\hline
 Fid \, & \, $0.26$ \, & \, $0.72$  \\
 II \, & \, $0.35$ \, & \, $0.72$  \\
 III \, & \, $0.35$ \, & \, $0.677$  \\
 IV \, & \, $0.26$ \, & \, $0.677$  \\
 \hline 
\end{tabular}
\end{center}
 Different models that we adopt to infer the distance redshift relation, which is then used to generate the BOSS slices, and reconstruct the genus amplitude. `Fid' denotes the correct cosmological model used in the simulation.   \\
\end{table}

\section*{Appendix C -- Ancillary Results}

In appendix C we provide supporting results for completeness. The two dimensional marginalised $68/95\%$ contours for the parameter set $\Omega_{\rm c}h^{2}, n_{\rm s}, \Omega_{\rm b}h^{2}, a_{0}, a_{2}$  is presented in Figures \ref{fig:all_parms_lowz} (the six LOWZ shells) and \ref{fig:all_parms_cmass} (six CMASS shells). $\Omega_{\rm b}h^{2}$ is effectively unconstrained over the prior range taken, and we observe no significant correlation between $a_{0}, a_{2}$ and $n_{\rm s}, \Omega_{\rm c}h^{2}$. The full corner plot for the combined LOWZ and CMASS data is presented in Figure \ref{fig:full_corner}.

The twelve density fields used in our analysis are presented in Figures \ref{fig:dens_LOWZ},\ref{fig:dens_CMASS}. These maps have been smoothed with angular scale $\theta_{\rm G} = R_{\rm G}/d_{\rm c}(z,\Omega_{\rm m},h)$, where $R_{\rm G} = 20 \, {\rm Mpc}$ and we have used the Planck cosmological parameters to infer the comoving distance $d_{\rm c}$ to the center of each slice, and projected onto a Cartesian background. The left/right columns correspond to the North/South Galactic data respectively, and the maps are arrayed in ascending order of redshift. The genus curves extracted from these maps are exhibited in Figure \ref{fig:4} for the LOWZ (top panel) and CMASS (middle panel) shells.

\clearpage

\begin{widetext}

\begin{figure}
  \centering
  \includegraphics[width=0.95\textwidth]{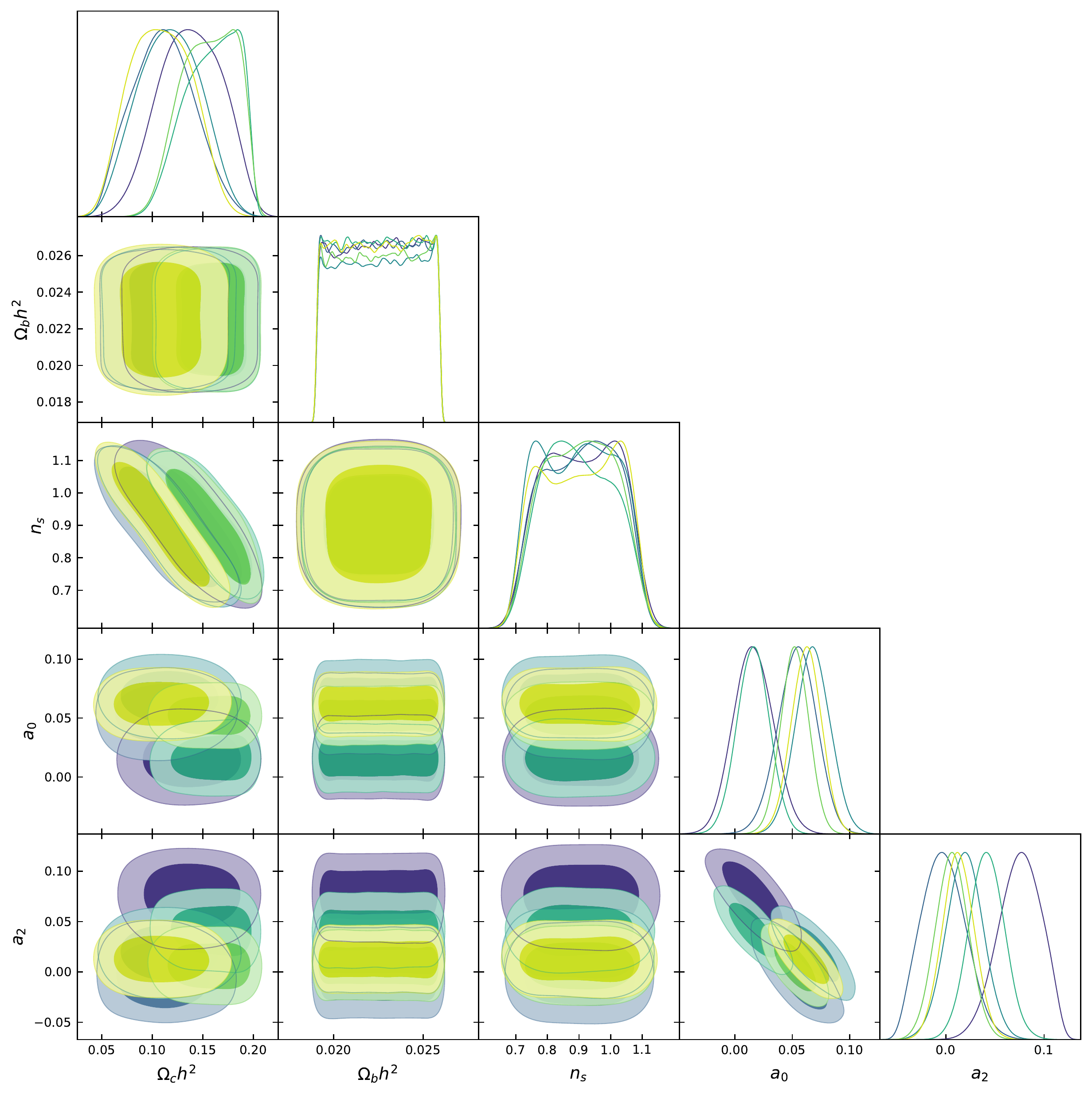}
  \caption{Corner plot of the parameters varied in this work. $\Omega_{\rm b}h^{2}$ is effectively unconstrained over its prior range, and $a_{0}$, $a_{2}$ are uncorrelated with the cosmological parameters $\Omega_{\rm c}h^{2}$, $n_{\rm s}$. The colours of the contours match Figure \ref{fig:4} in the main body of the paper -- we plot six contours which represent the results of each LOWZ shell.}
  \label{fig:all_parms_lowz}
\end{figure}

\begin{figure}
  \centering
  \includegraphics[width=0.95\textwidth]{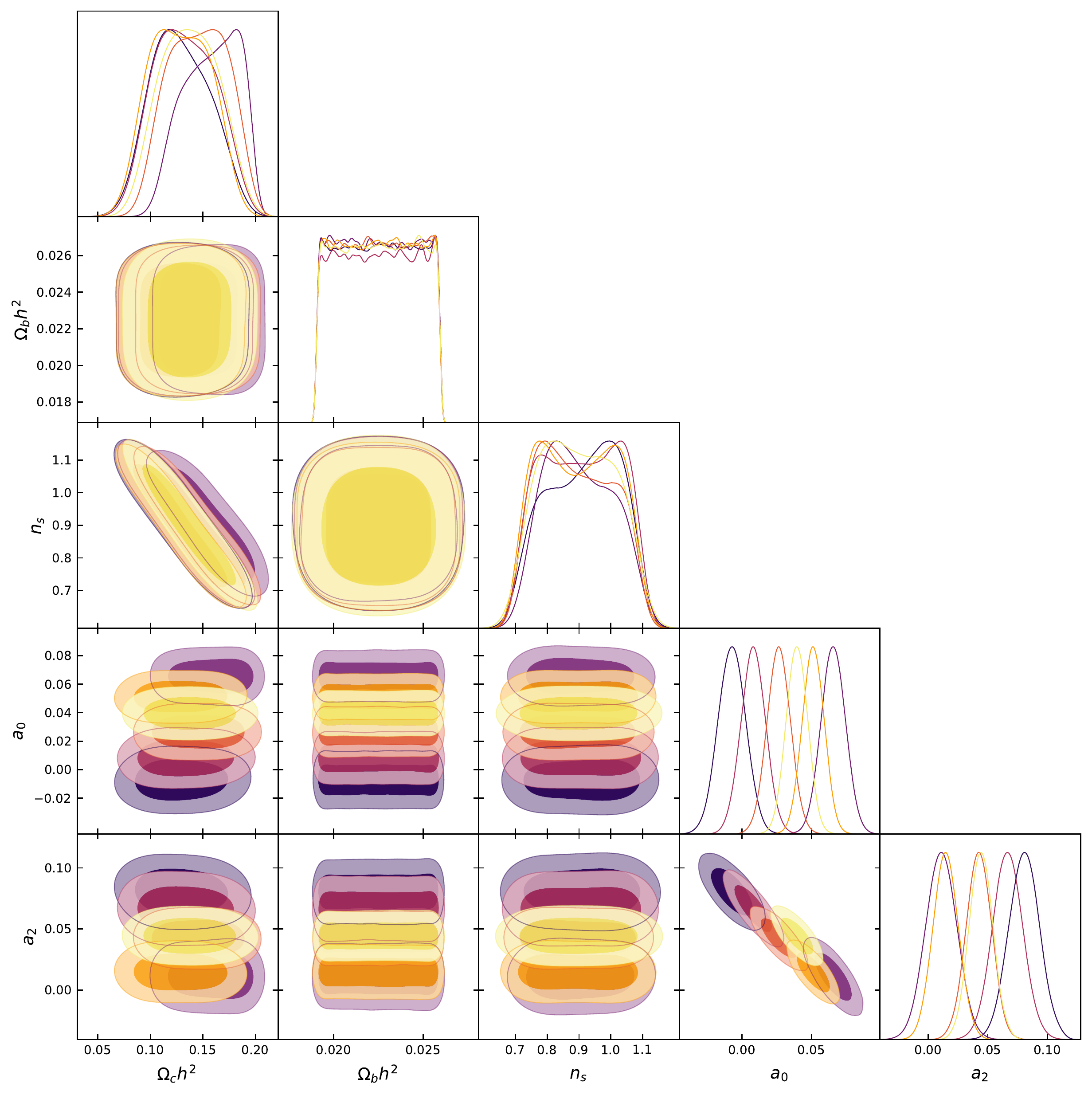}
  \caption{The same as Figure \ref{fig:all_parms_lowz} but for the six CMASS data shells.}
  \label{fig:all_parms_cmass}
\end{figure}

\begin{figure}
 \centering 
  \includegraphics[width=0.95\textwidth]{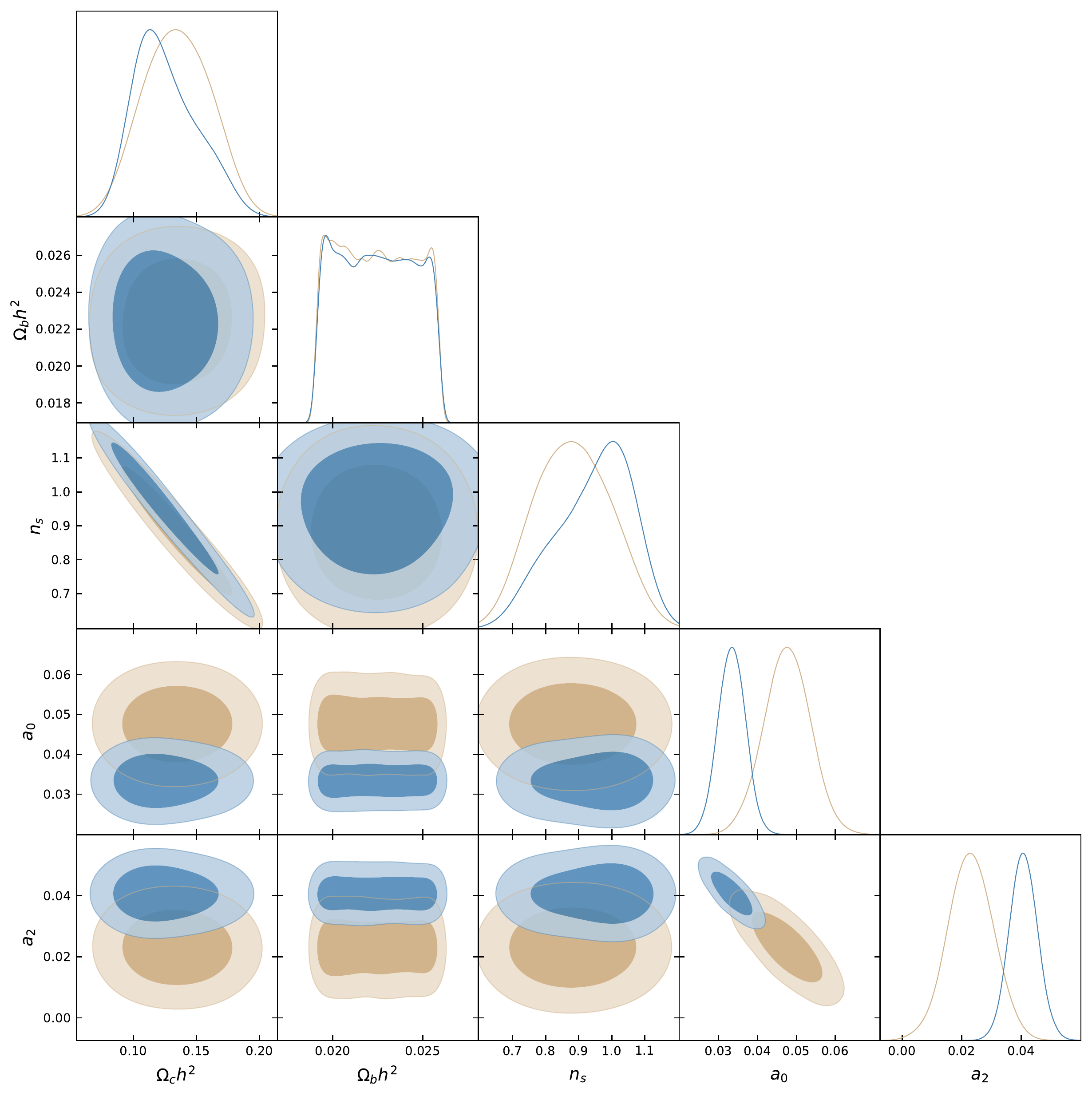}
  \caption{After combining all six LOWZ (brown contours) and CMASS (blue contours) shells, Two dimensional marginalised contours, fitting a single set of parameters $\Omega_{\rm c}h^{2}$, $\Omega_{\rm b}h^{2}$, $n_{\rm s}$, $a_{0,2}$ to all LOWZ (brown contours) and CMASS (blue contours) shells.}
  \label{fig:full_corner}
\end{figure}

\begin{figure}[h!]
 \centering 
  \includegraphics[width=0.38\textwidth]{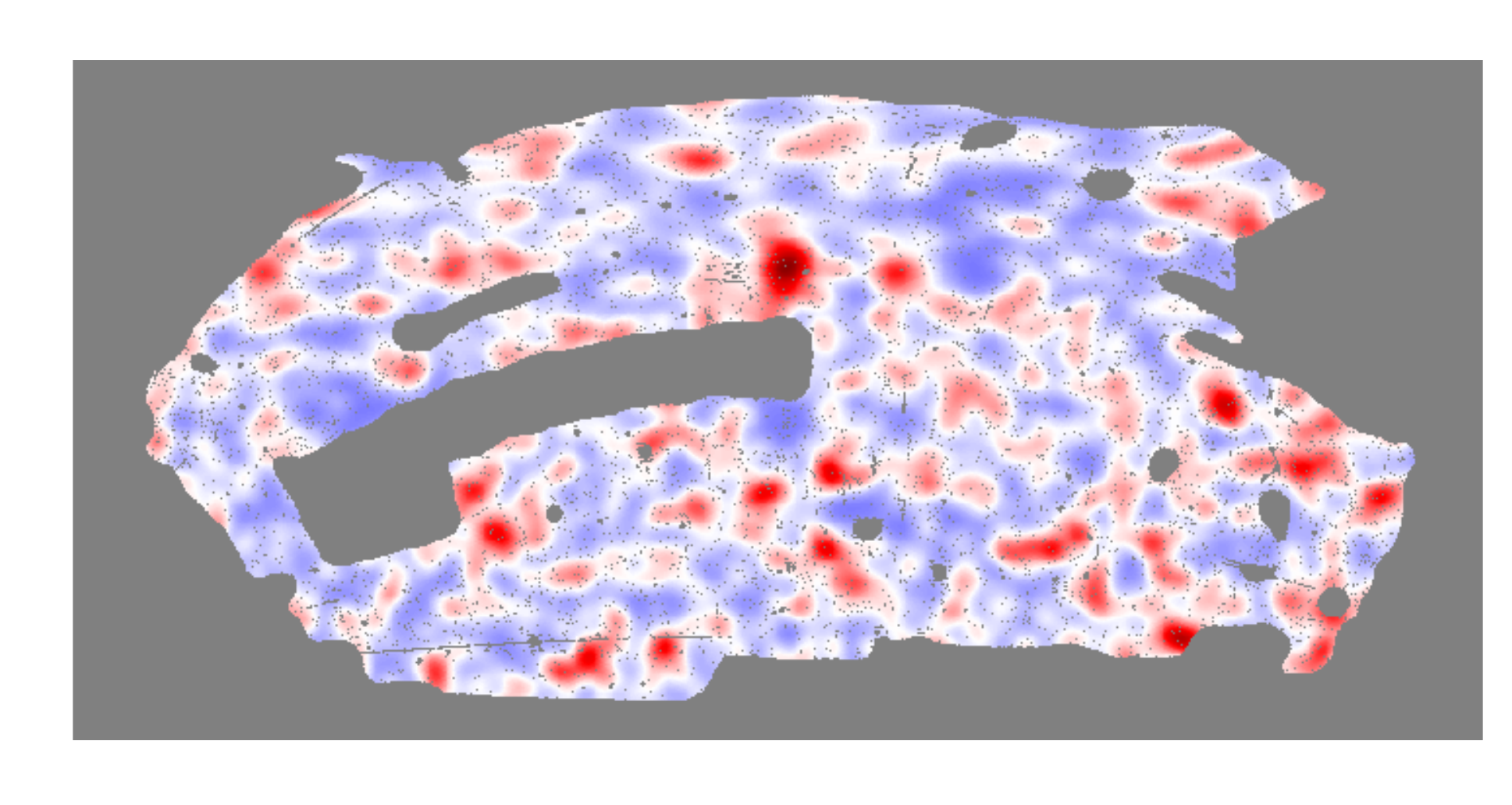} 
  \includegraphics[width=0.40\textwidth]{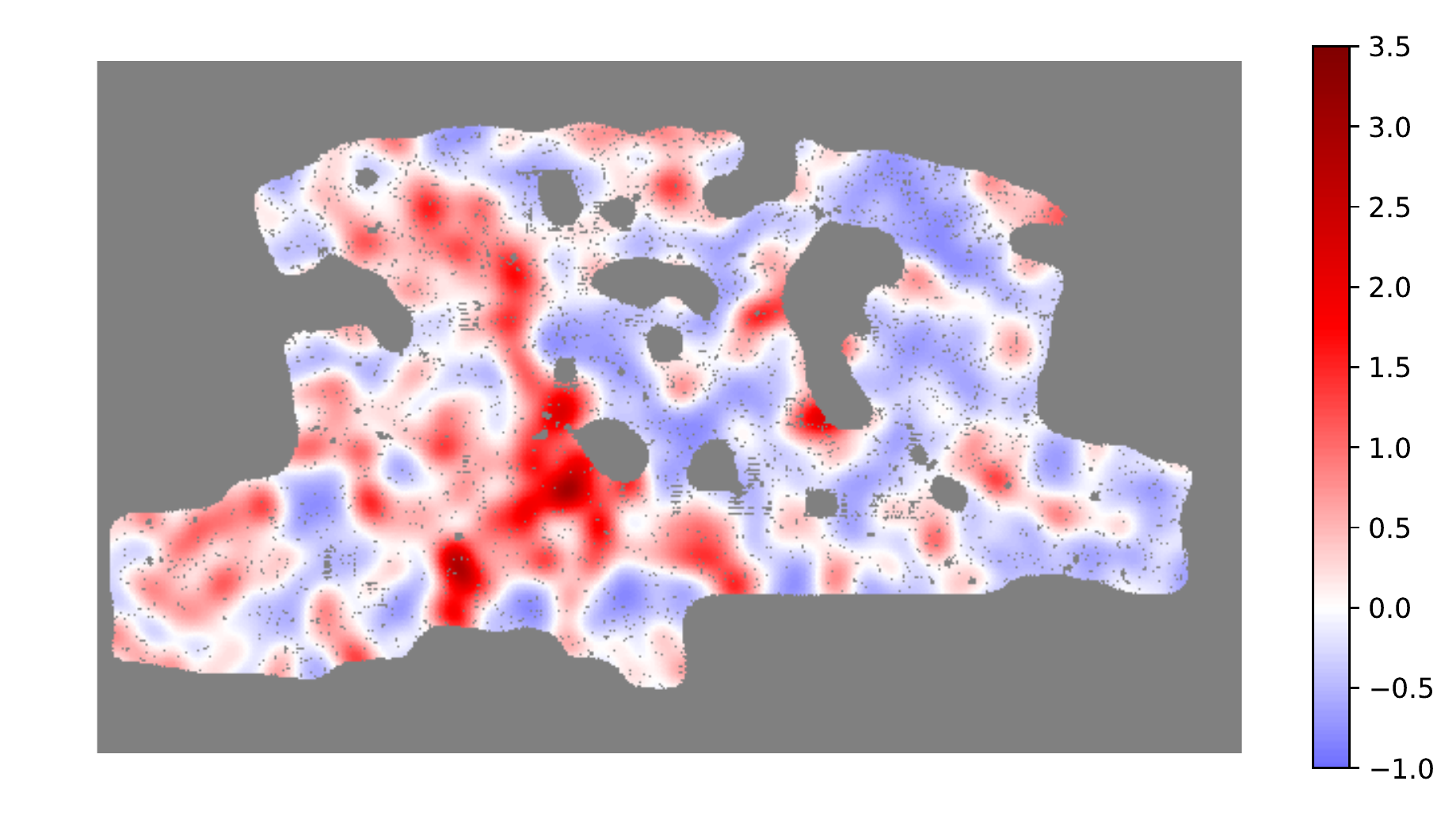} \\
    \includegraphics[width=0.38\textwidth]{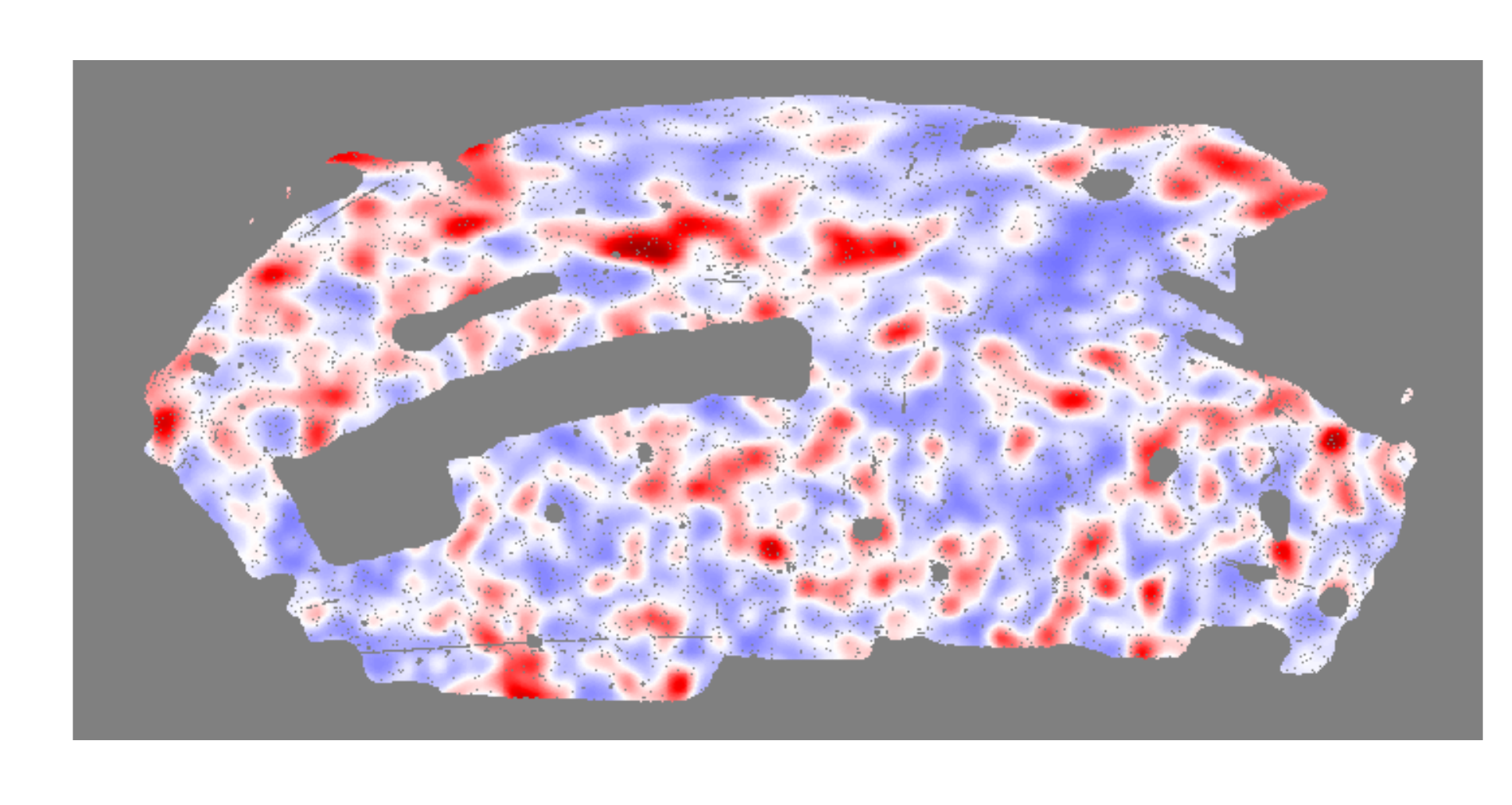} 
  \includegraphics[width=0.40\textwidth]{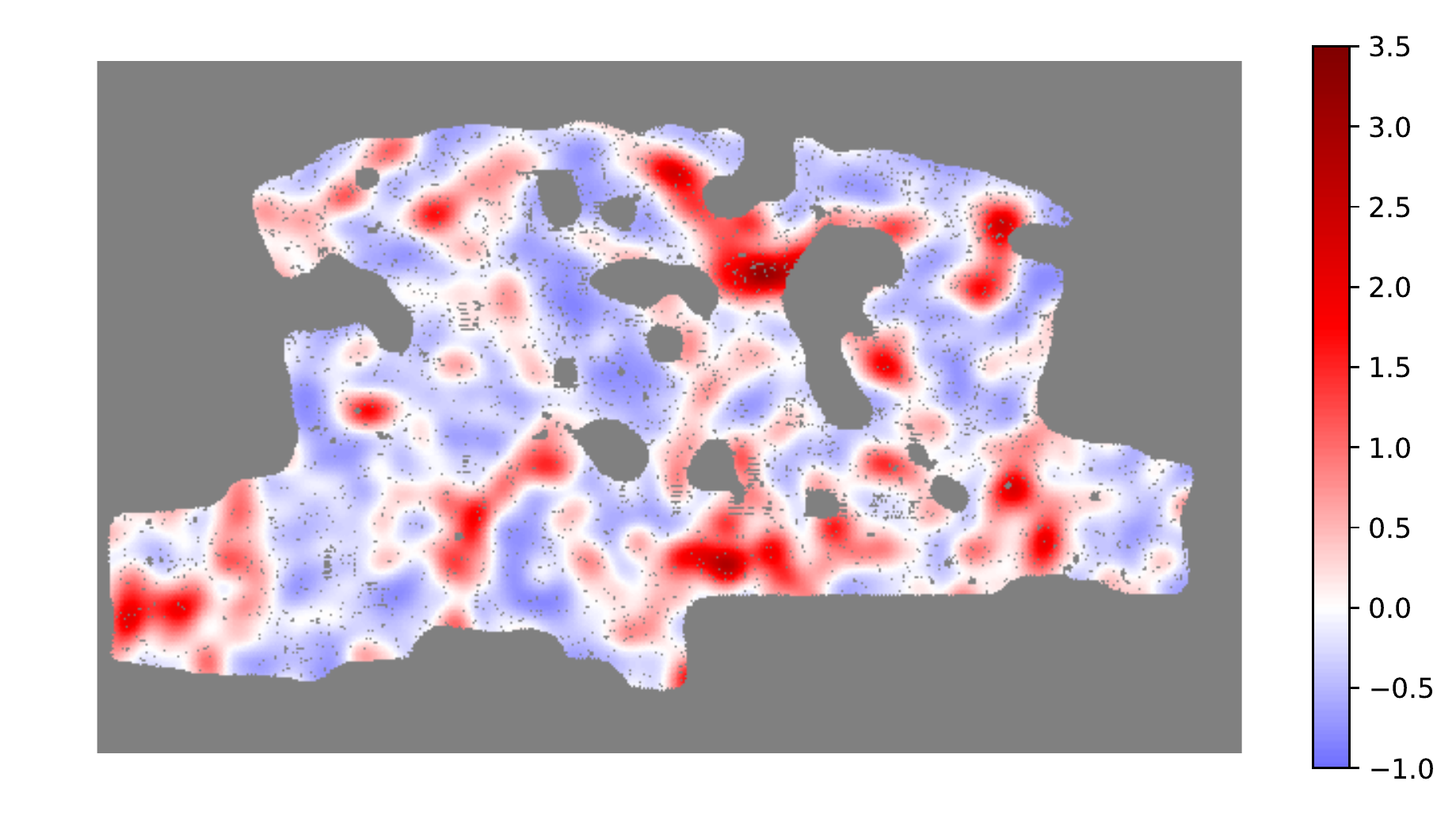} \\
    \includegraphics[width=0.38\textwidth]{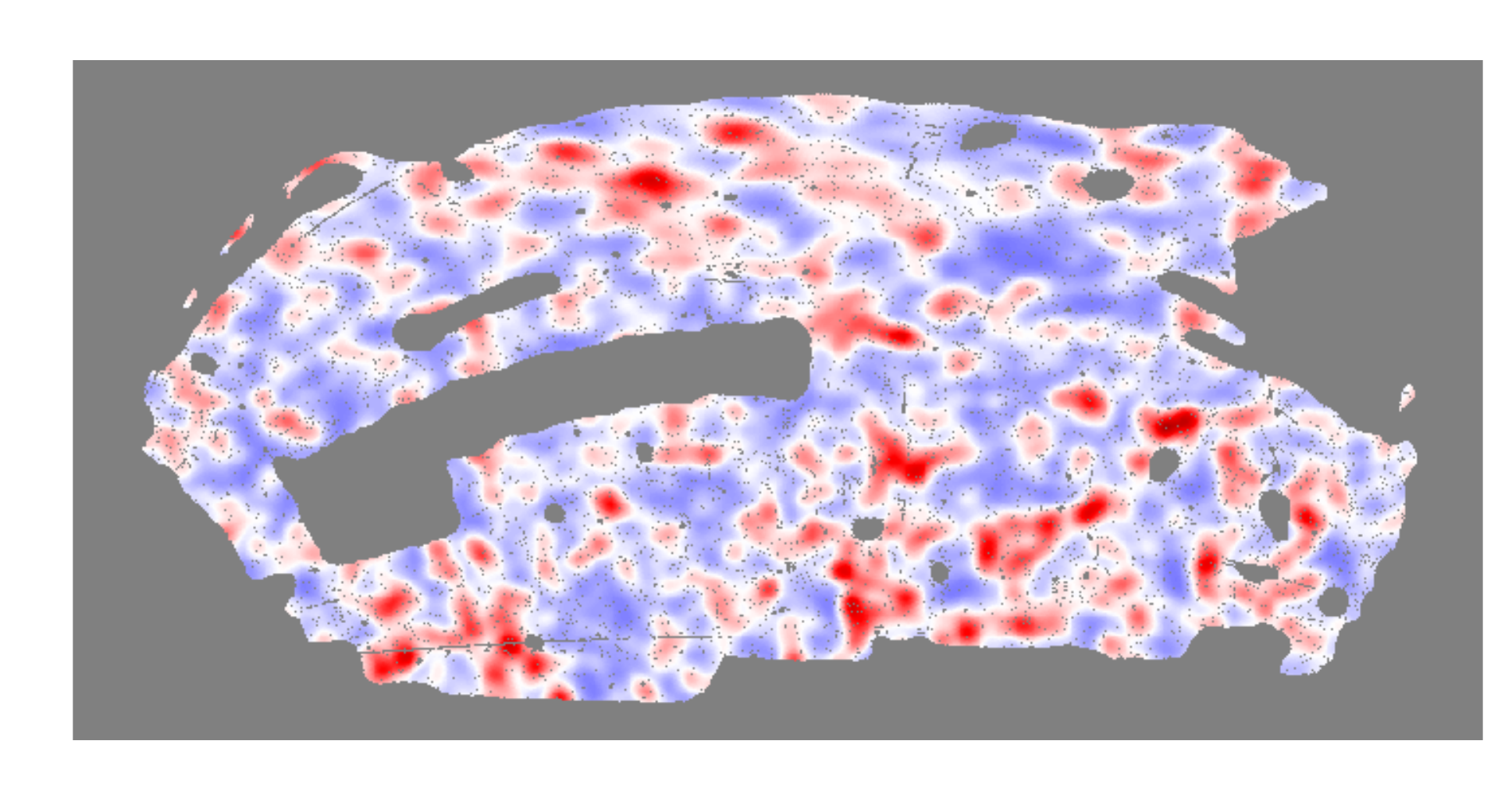} 
  \includegraphics[width=0.40\textwidth]{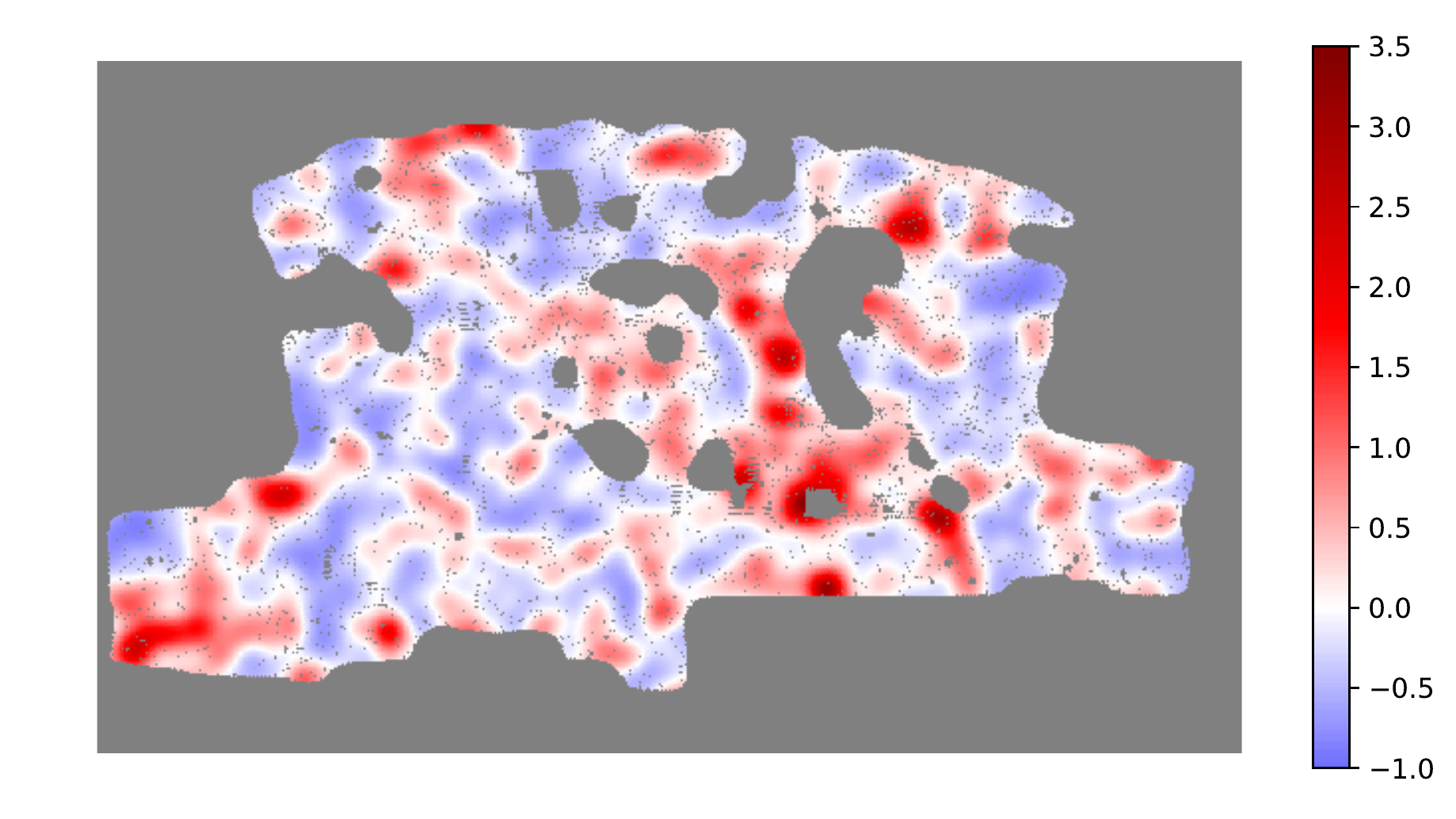} \\
    \includegraphics[width=0.38\textwidth]{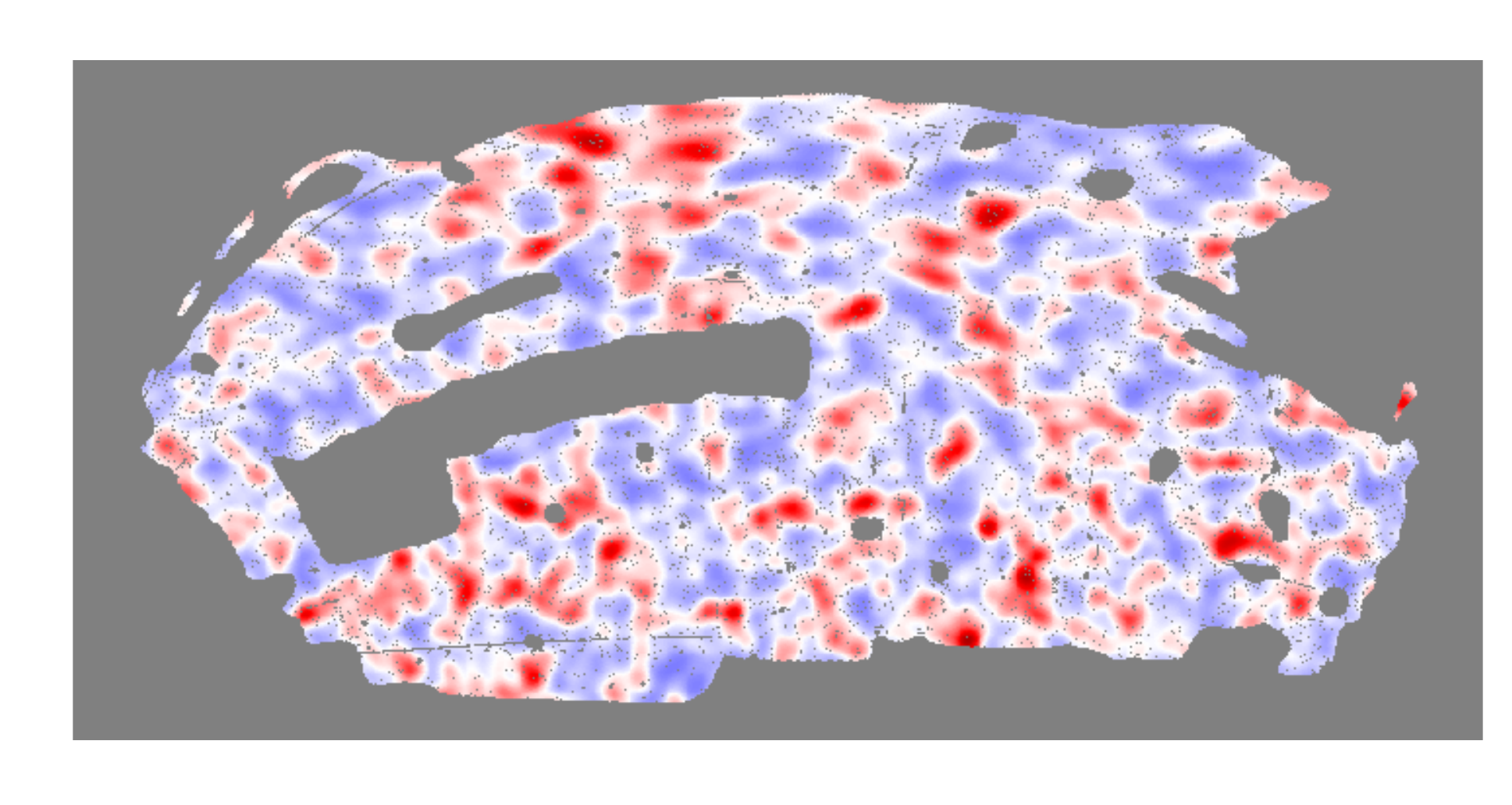} 
  \includegraphics[width=0.40\textwidth]{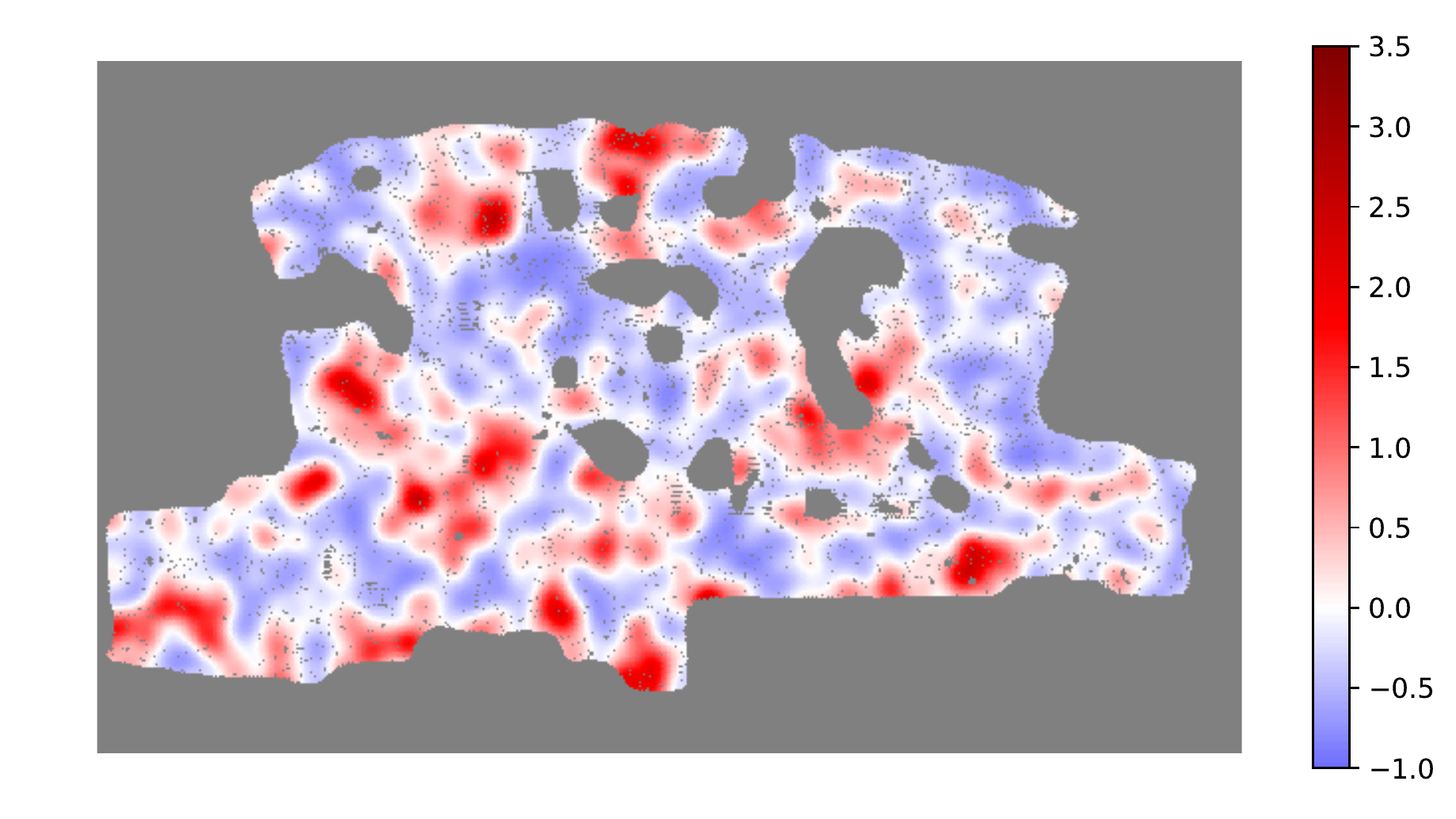} \\
    \includegraphics[width=0.38\textwidth]{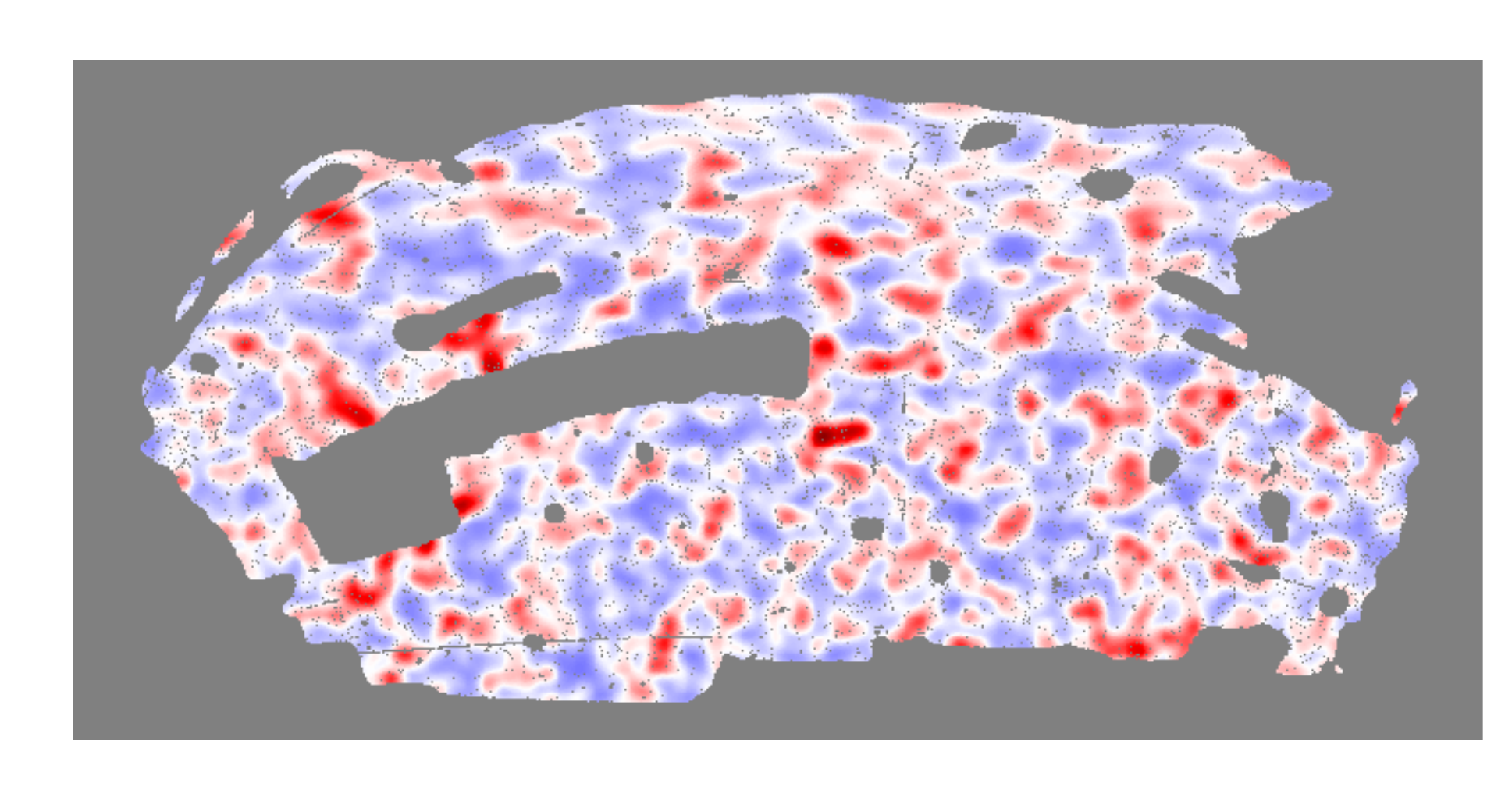} 
  \includegraphics[width=0.40\textwidth]{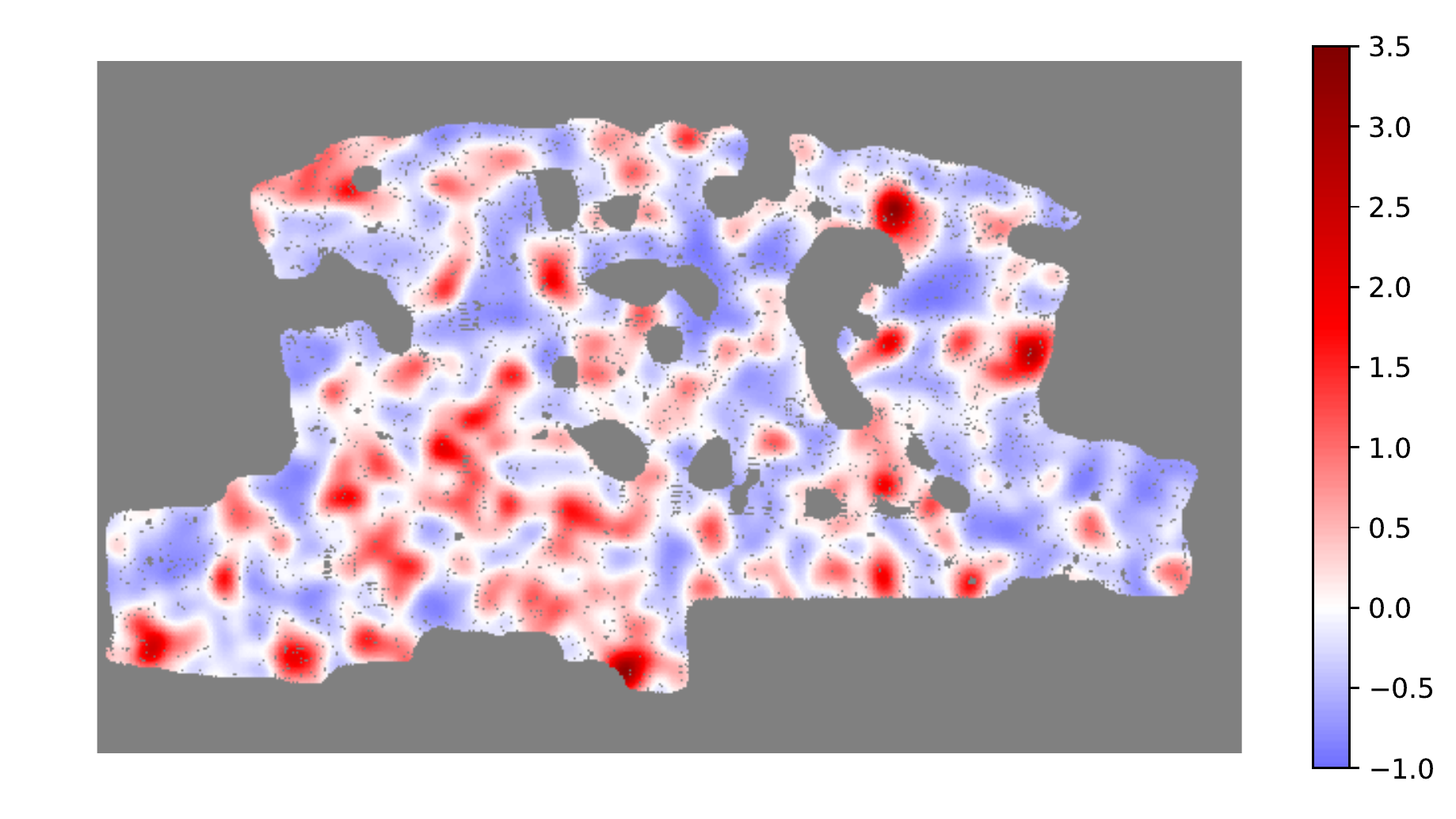} \\
    \includegraphics[width=0.38\textwidth]{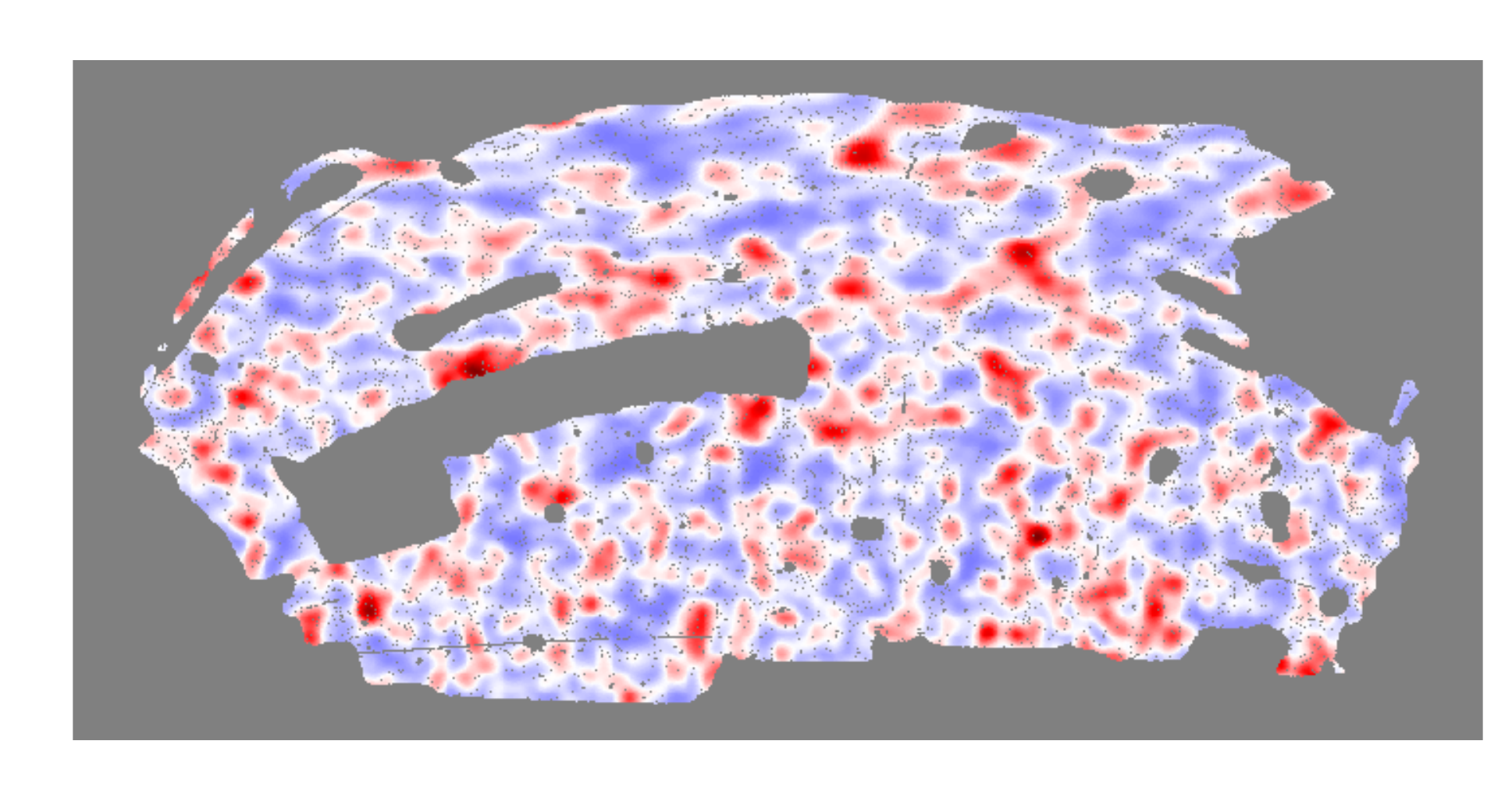} 
  \includegraphics[width=0.40\textwidth]{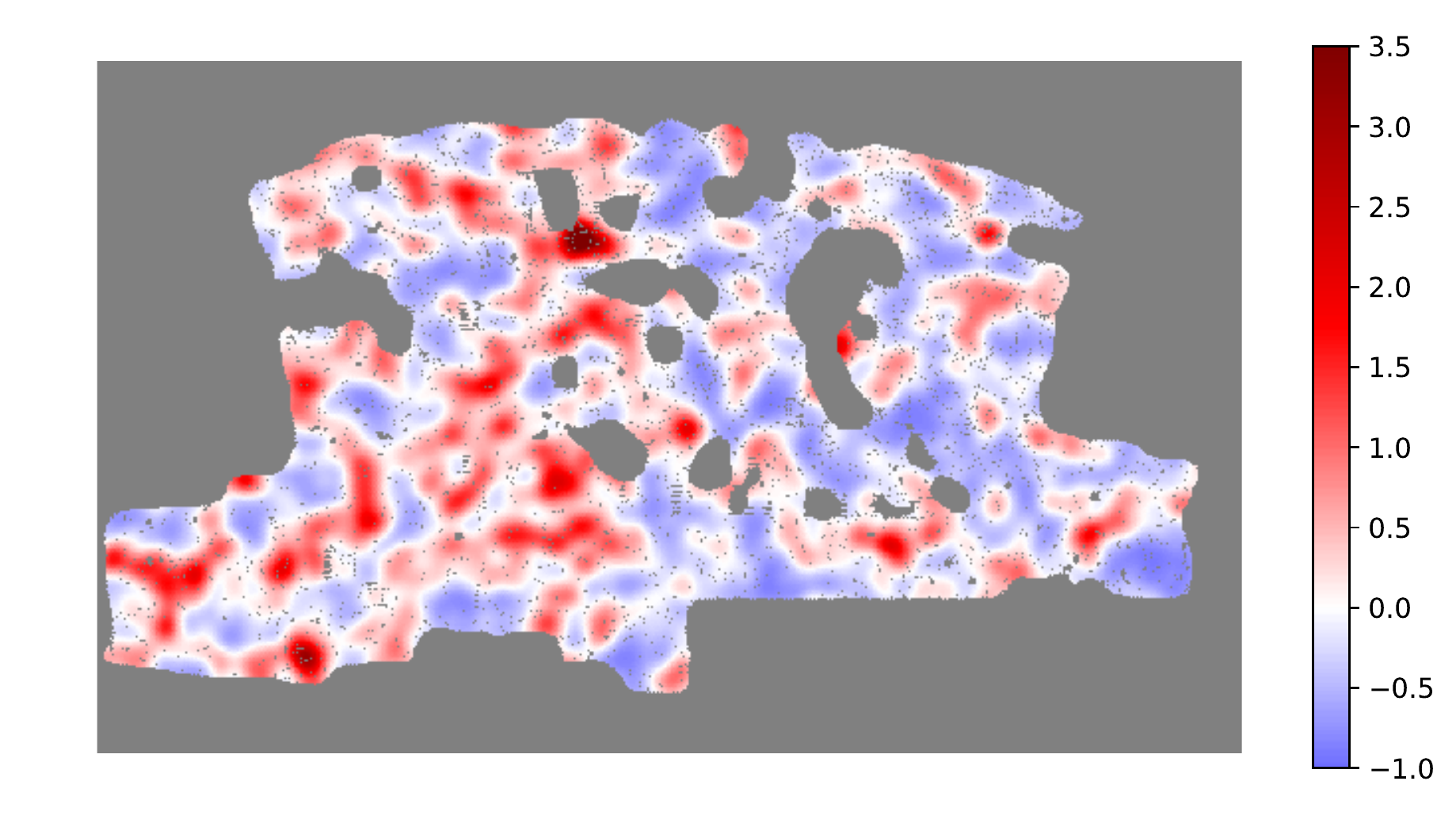} \\
  \caption{The density fields of the LOWZ shells in the North galactic plane (left panels) and South galactic plane (right panels), in ascending redshift order. All maps have been smoothed with Gaussian kernel of width $R_{\rm G}=20 {\rm Mpc}$. The spherical shells have undergone Cartesian projection, and the grey pixels have been masked. }
  \label{fig:dens_LOWZ}
\end{figure}

\newpage

\begin{figure}[h!]
  \centering
  \includegraphics[width=0.38\textwidth]{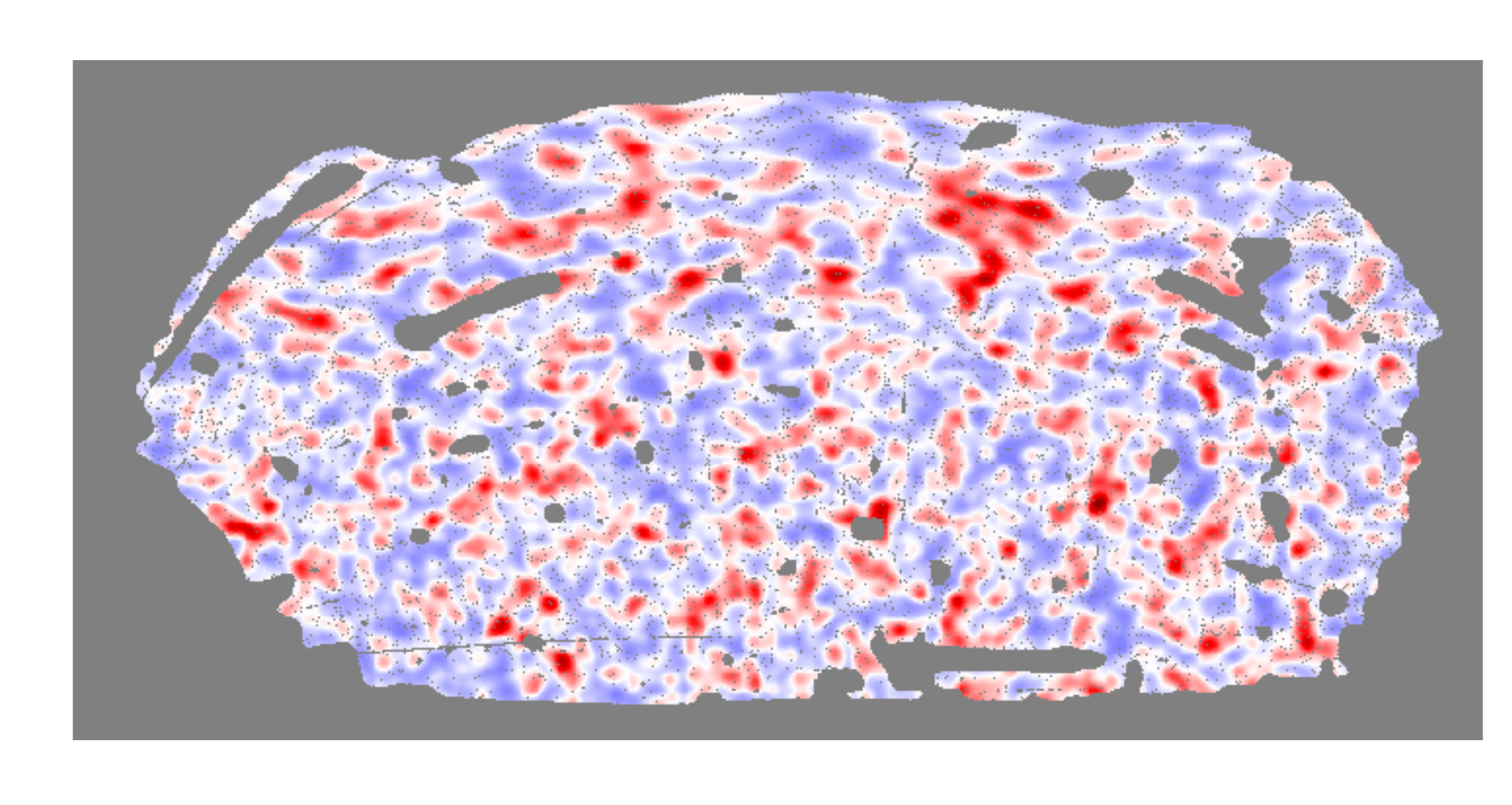} 
  \includegraphics[width=0.40\textwidth]{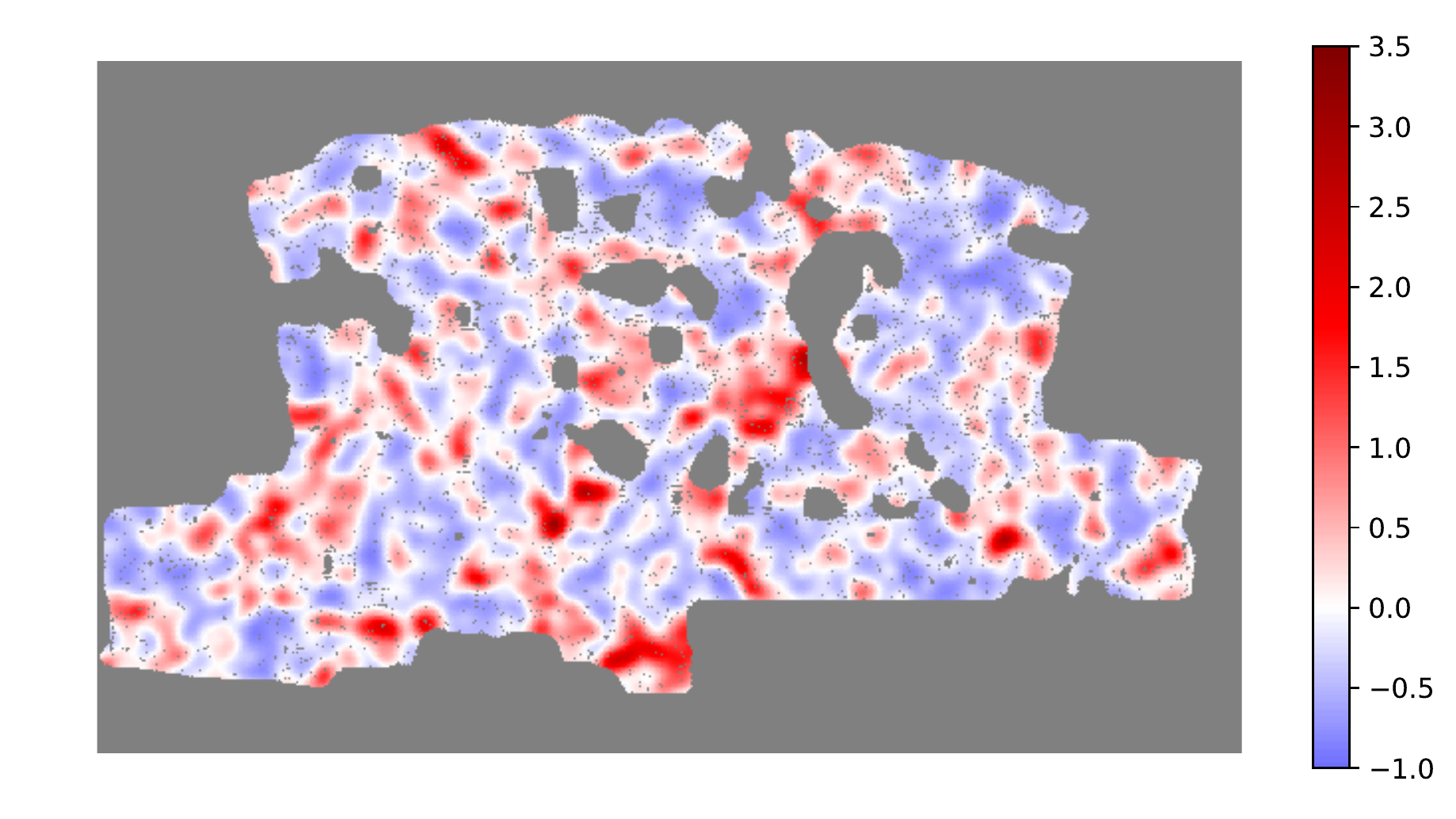} \\
    \includegraphics[width=0.38\textwidth]{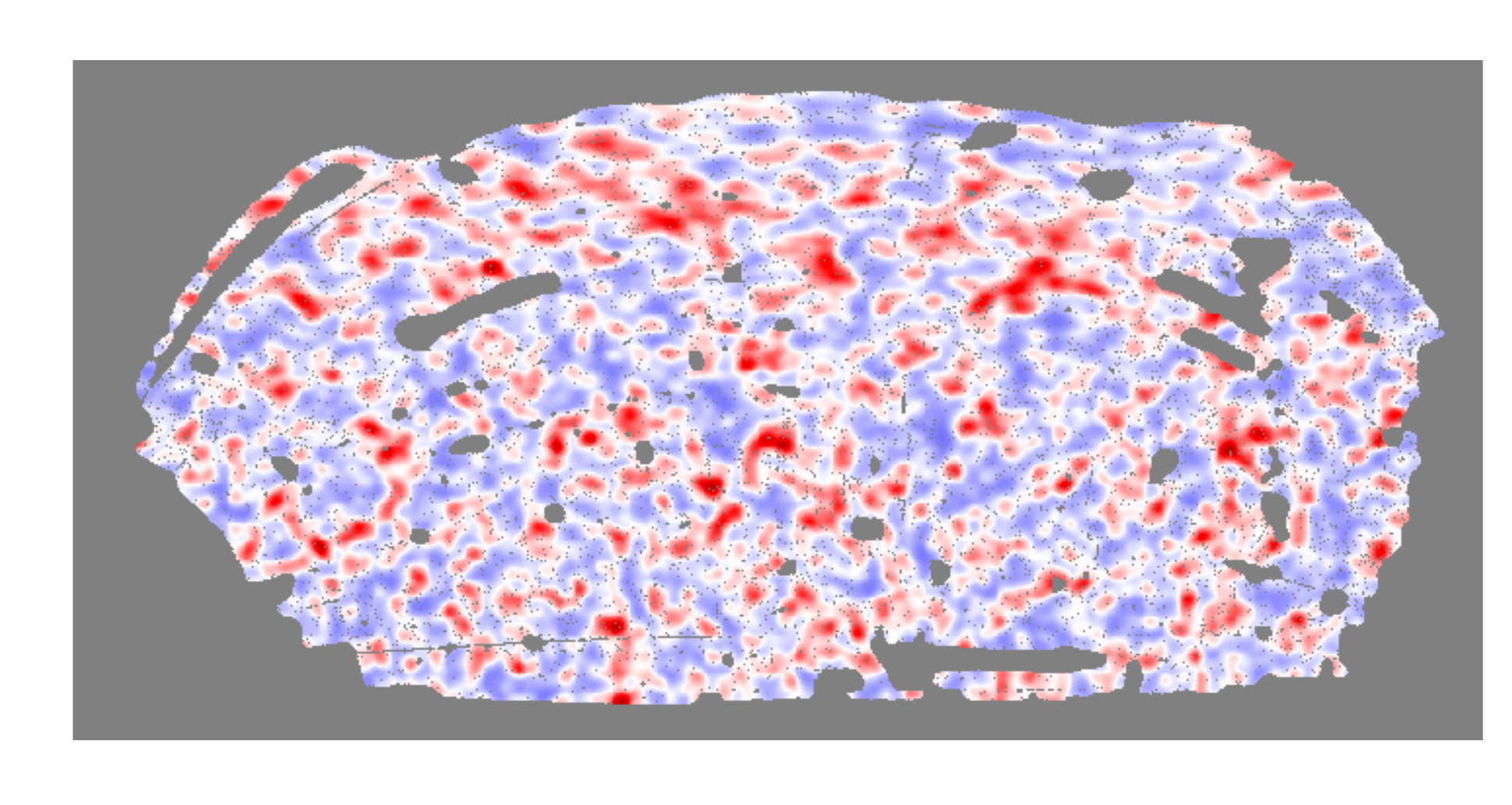} 
  \includegraphics[width=0.40\textwidth]{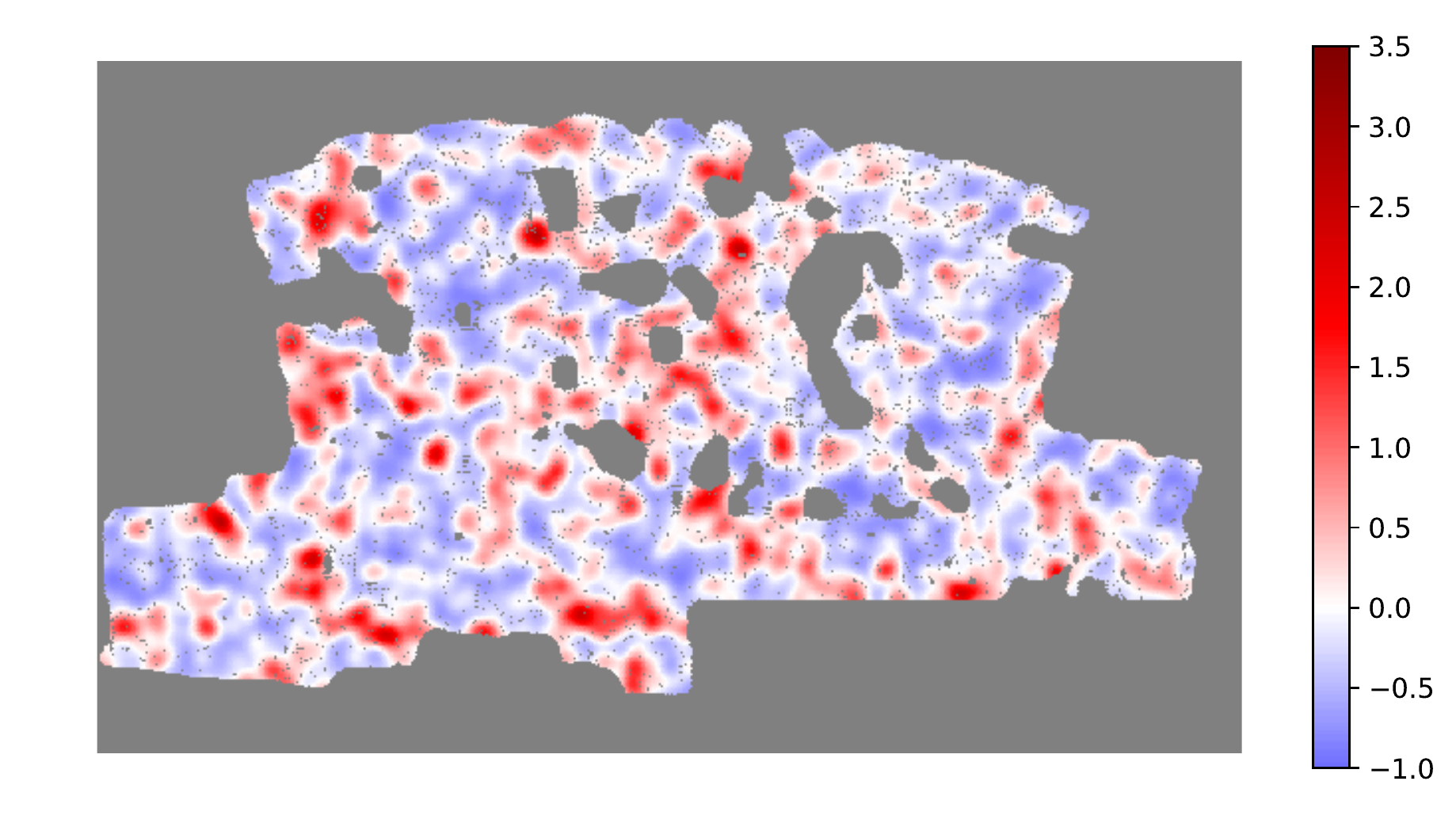} \\
    \includegraphics[width=0.38\textwidth]{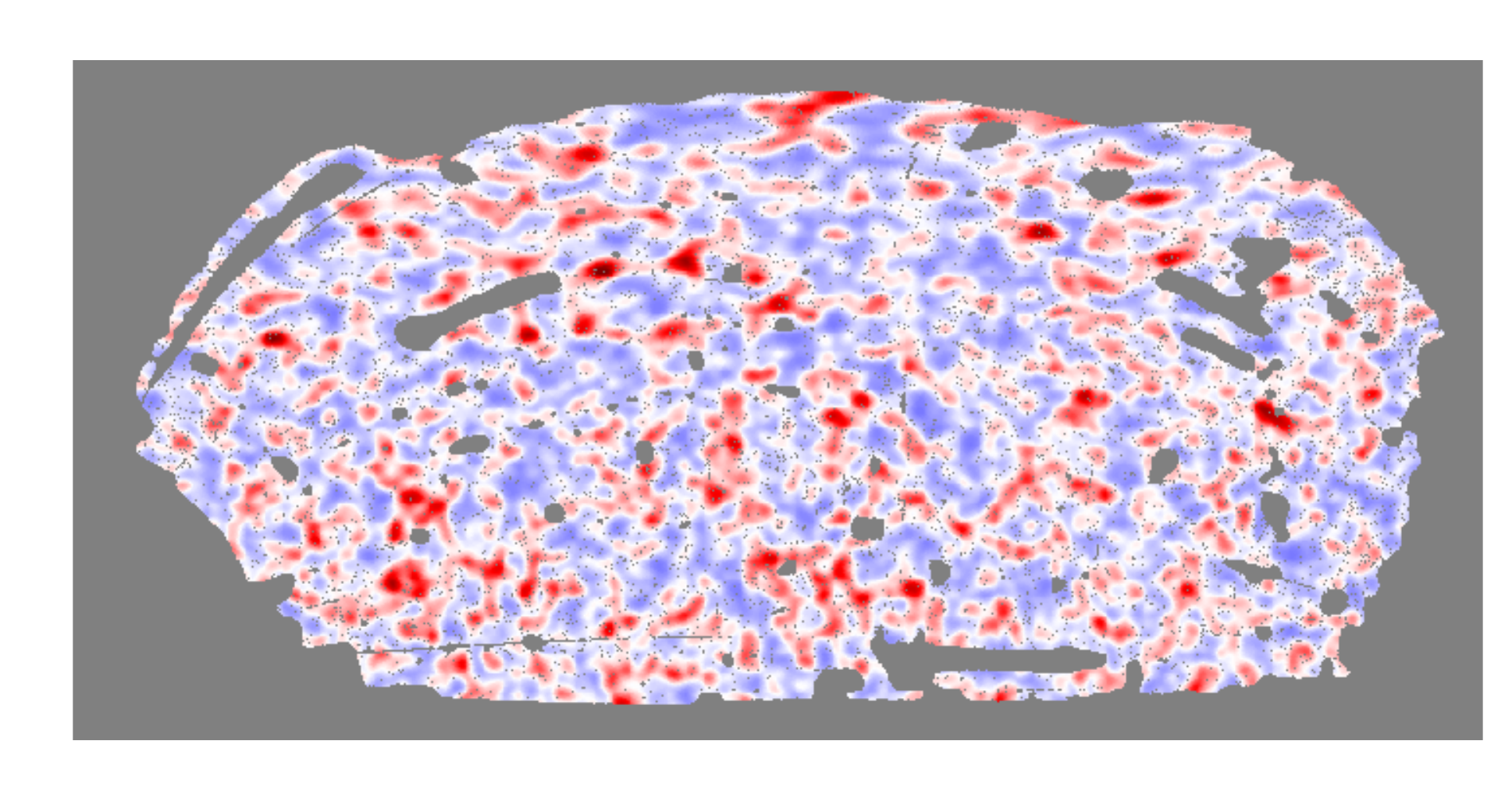} 
  \includegraphics[width=0.40\textwidth]{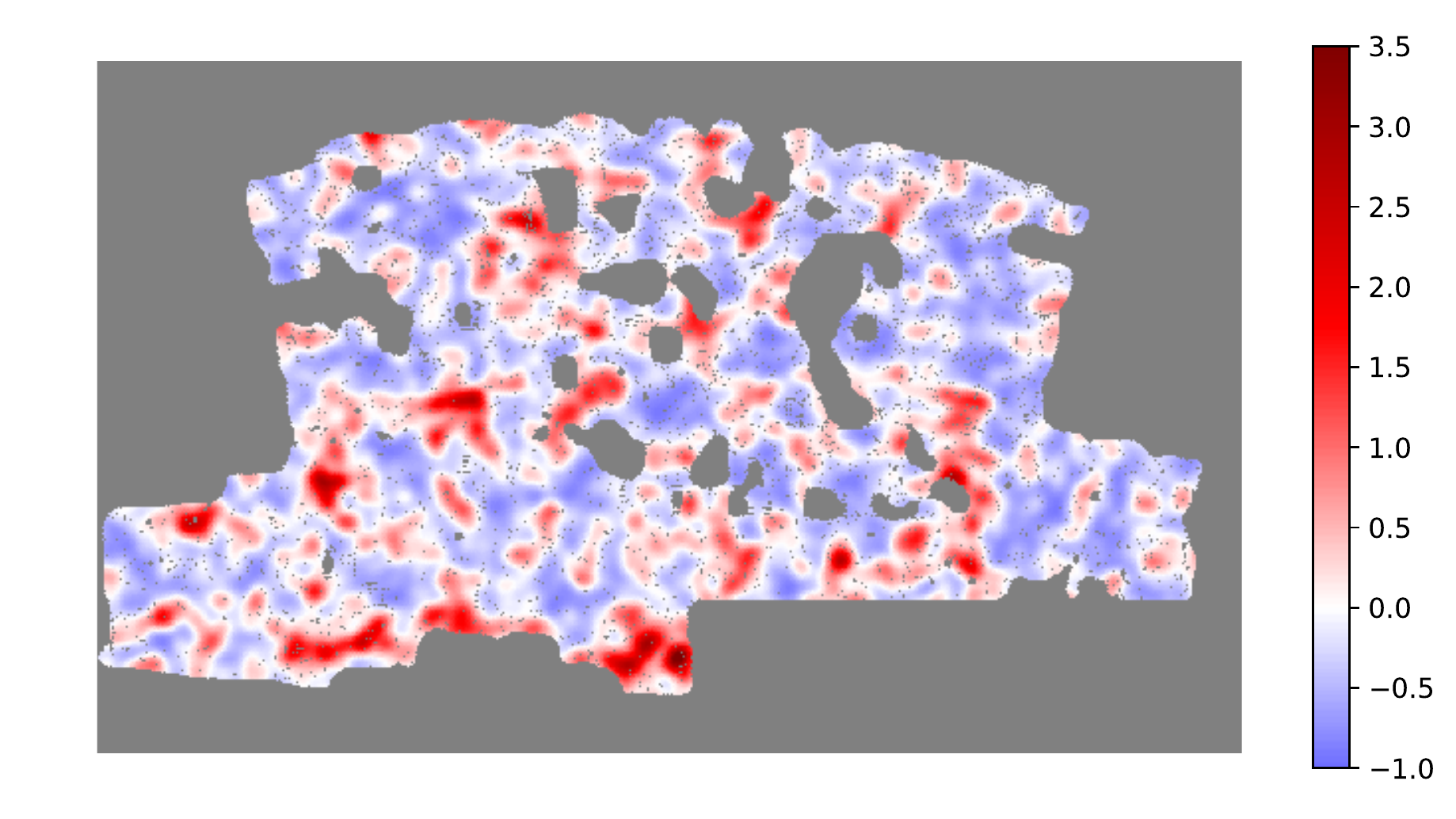} \\
    \includegraphics[width=0.38\textwidth]{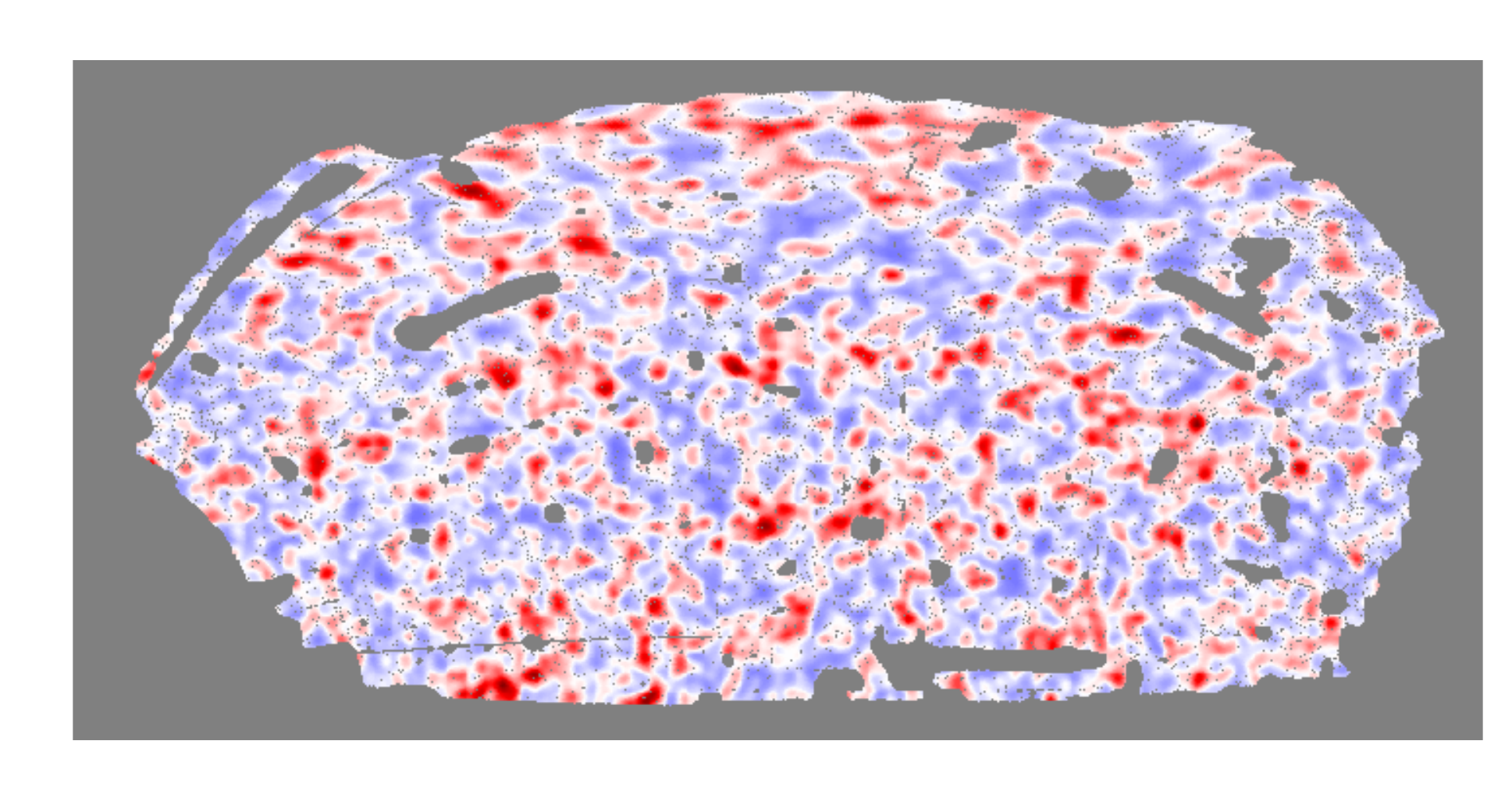} 
  \includegraphics[width=0.40\textwidth]{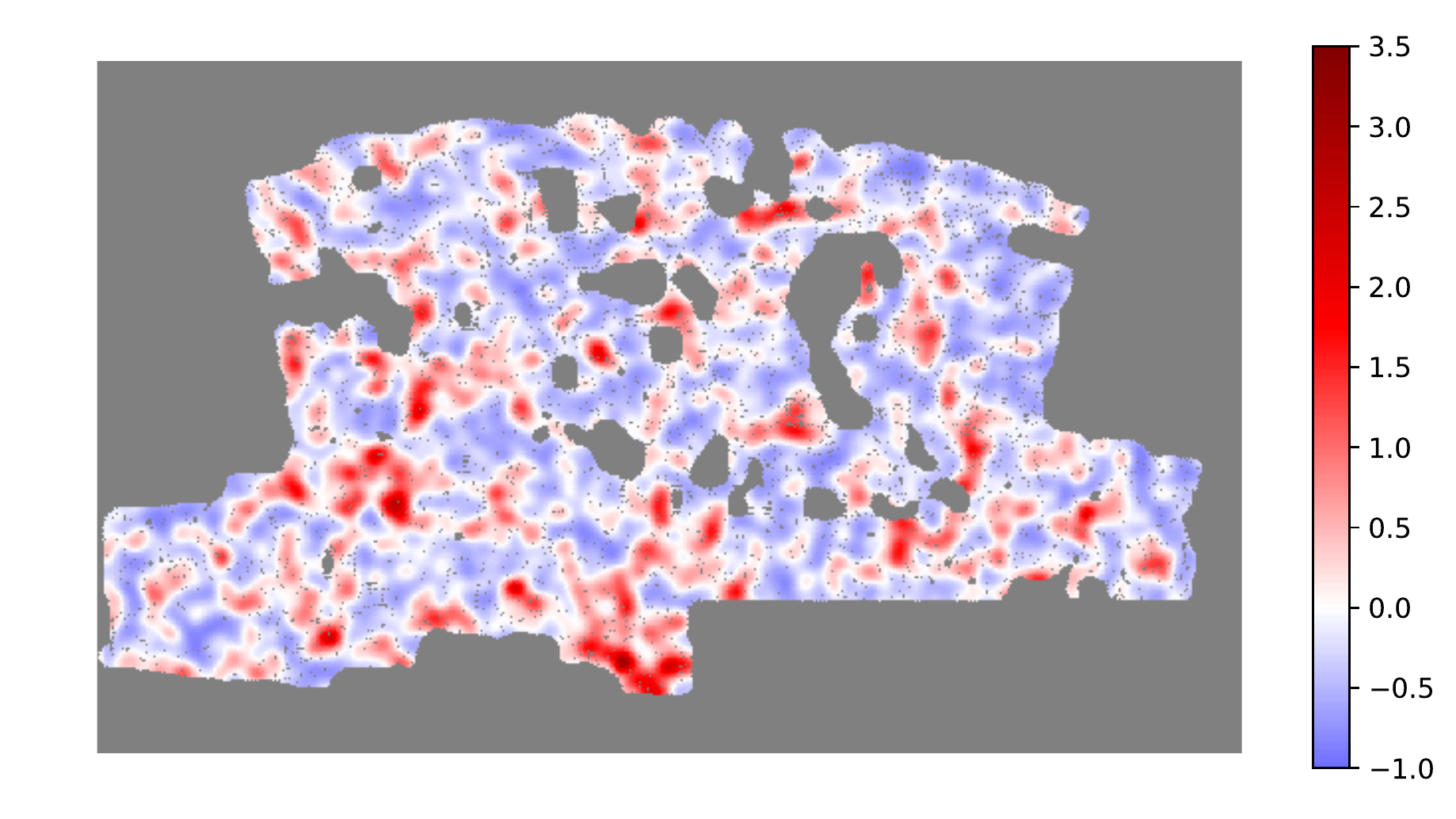} \\
    \includegraphics[width=0.38\textwidth]{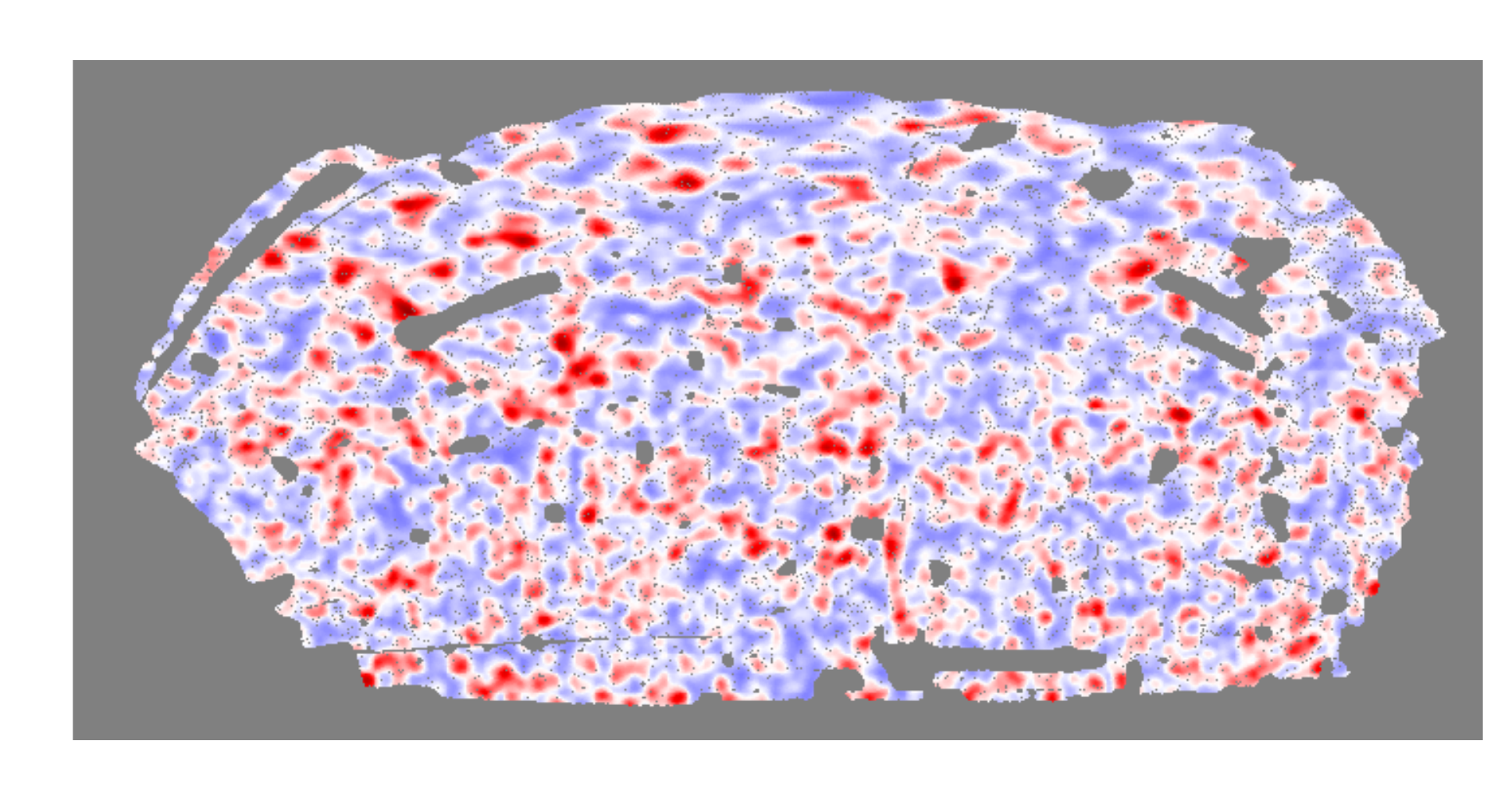} 
  \includegraphics[width=0.40\textwidth]{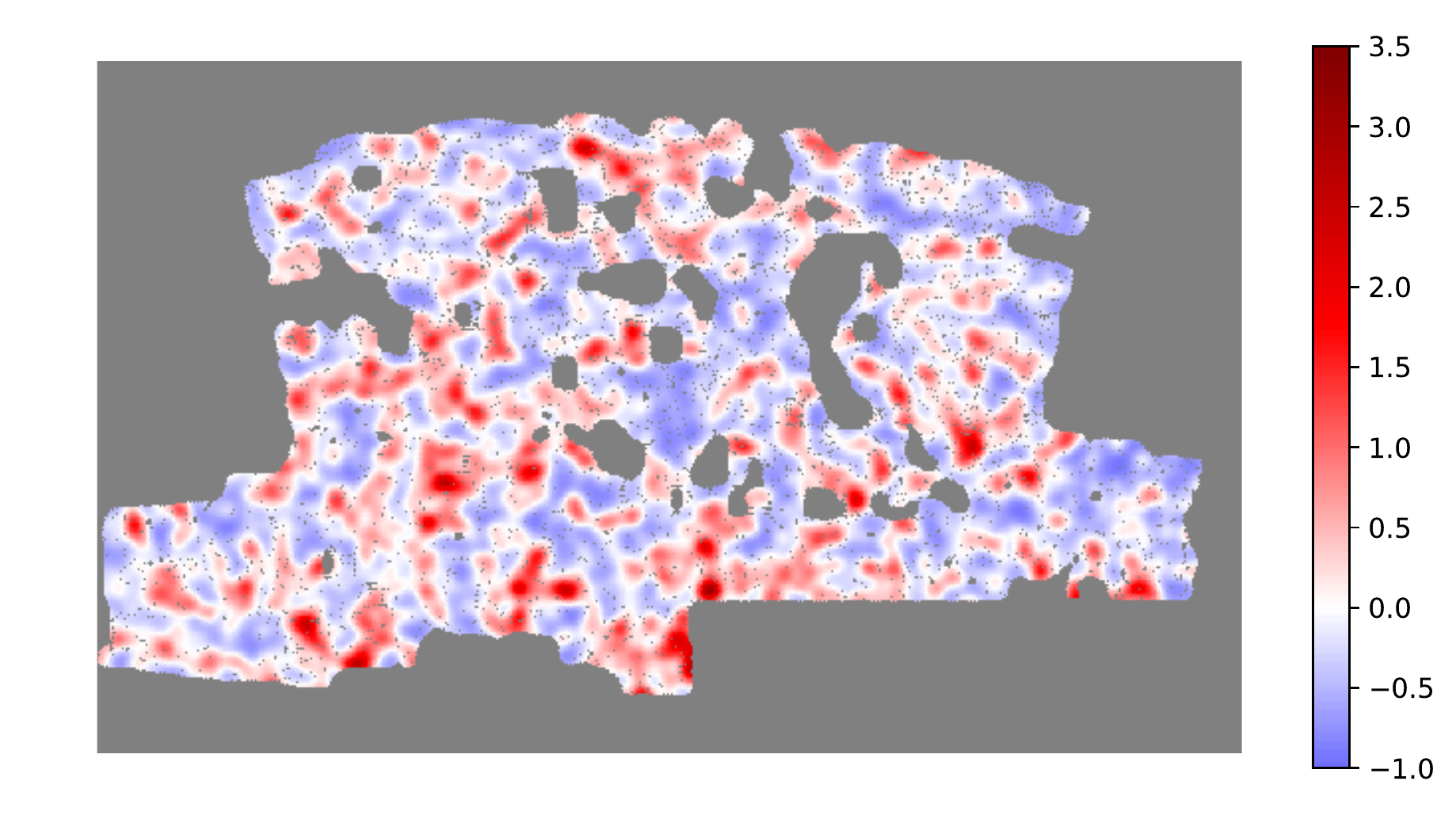} \\
    \includegraphics[width=0.38\textwidth]{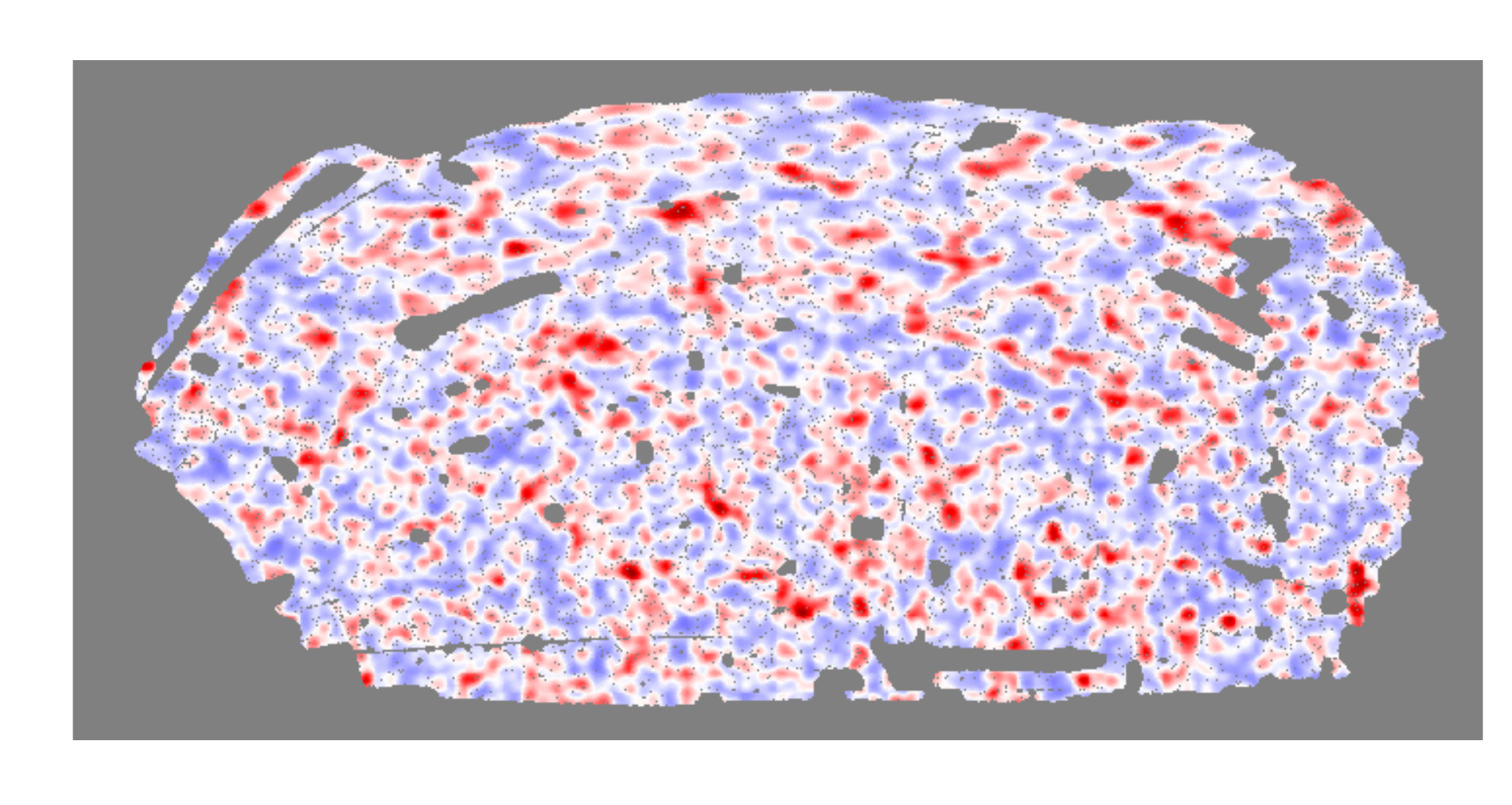} 
  \includegraphics[width=0.40\textwidth]{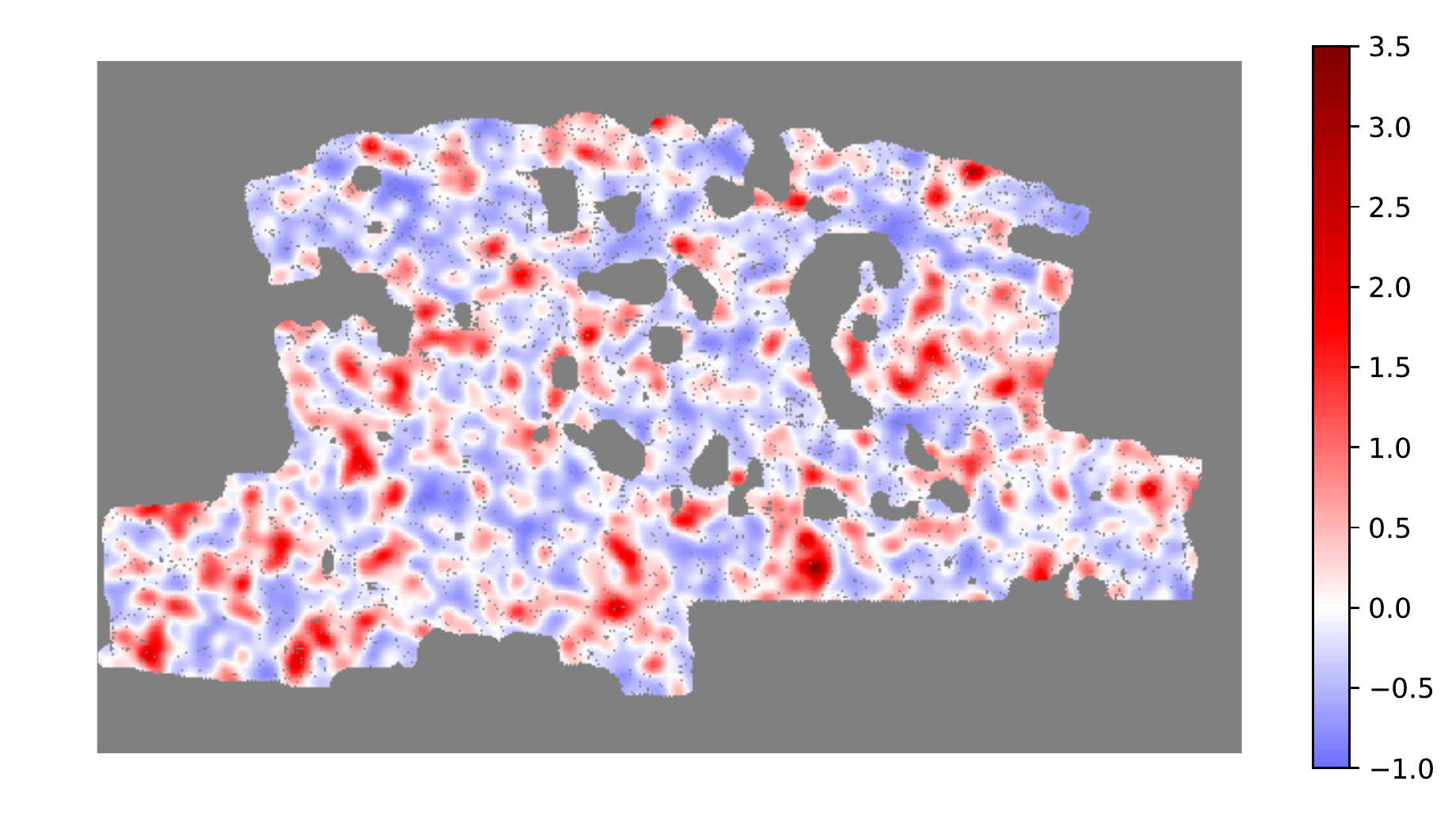} \\
  \caption{As in figure \ref{fig:dens_LOWZ}, but for the CMASS shells.}
  \label{fig:dens_CMASS}
\end{figure}

\end{widetext}

\clearpage

\bibliographystyle{ApJ}
\bibliography{biblio}{}

\end{document}